\documentclass[11pt]{article}
\usepackage[margin=1in]{geometry}
\usepackage{booktabs}
\usepackage{array}
\usepackage{multirow}
\usepackage{multicol}

\usepackage[
    backend=biber,
    style=numeric-comp,
    sortcites,
    sorting=none,
    giveninits=true,
    natbib,
    hyperref=true,
    maxbibnames=99,
    doi=false,isbn=false,url=false,eprint=false
]{biblatex}

\usepackage[hidelinks]{hyperref}

\title{A Locally Adaptive Algorithm for Multiple Testing \\with Network Structure}

\usepackage{subfigure}
\usepackage{placeins}
\usepackage{graphicx, color}
\usepackage{dsfont}
\usepackage{amssymb}
\usepackage{amsfonts}
\usepackage{url}
\usepackage{amsbsy}
\usepackage{setspace}
\usepackage{bm}
\usepackage{textcomp}
\usepackage{booktabs}
\usepackage{rotating}

\usepackage{algorithm, algorithmic}

\usepackage{amsmath,amsthm}

\usepackage{graphicx}
\usepackage{amsbsy}
\usepackage{mathtools}
\usepackage{bbm}

\usepackage{soul}
\usepackage{cleveref}

\usepackage[shortlabels]{enumitem}

\newcommand{\beq}{\begin{equation}}
\newcommand{\eeq}{\end{equation}}
\newcommand{\beas}{\begin{eqnarray*}}
\newcommand{\eeas}{\end{eqnarray*}}
\newcommand{\bea}{\begin{eqnarray}}
\newcommand{\eea}{\end{eqnarray}}
\newcommand{\bei}{\begin{itemize}}
\newcommand{\eei}{\end{itemize}}
\newcommand{\ben}{\begin{enumerate}}
\newcommand{\een}{\end{enumerate}}
\newcommand{\bet}{\begin{theorem}}
\newcommand{\eet}{\end{theorem}}
\newcommand{\bel}{\begin{lemma}}
\newcommand{\eel}{\end{lemma}}
\newcommand{\bep}{\begin{proposition}}
\newcommand{\eep}{\end{proposition}}
\newcommand{\bed}{\begin{definition}}
\newcommand{\eed}{\end{definition}}
\newcommand{\bec}{\begin{corollary}}
\newcommand{\eec}{\end{corollary}}
\newcommand{\bex}{\begin{example}}
\newcommand{\eex}{\end{example}}

\newcommand{\EE}{\mathbb E}

\newcommand{\ep}{\epsilon}

\def\limsup{\mathop{\overline{\rm lim}}}

\def\T{{ \mathsf{\scriptscriptstyle T} }}

\def\0{\boldsymbol{0}}

\def\be{\boldsymbol{\beta}}

\def\eps{\boldsymbol{\epsilon}}
\def\X{\boldsymbol{X}}
\def\R{\boldsymbol{R}}
\def\Y{\boldsymbol{Y}}
\def\Z{\boldsymbol{Z}}

\DeclarePairedDelimiter\paren{(}{)}
\DeclarePairedDelimiter\brac{[}{]}
\DeclarePairedDelimiter\cbrac{\{}{\}}
\DeclarePairedDelimiter\abs{|}{|}

\newbox\TempBox \newbox\TempBoxA

\def\pr{\mathbb{P}} 
\def\ep{\mathbb{E}} 
\def\Var{\text{Var}} 
\def\Corr{\textsf{Corr}}
\def\cH{\mathcal{H}}

\def\0{\boldsymbol{0}}

\def\be{\boldsymbol{\beta}}

\def\eps{\boldsymbol{\epsilon}}
\def\X{\boldsymbol{X}}
\def\Y{\boldsymbol{Y}}
\def\Z{\boldsymbol{Z}}

\def\cN{\mathcal{N}}
\def\cD{\mathcal{D}}

\def\II{{\mathbb I}}
\def\PP{{\mathbb P}}

\newcommand{\expect}{\operatorname{\mathbb{E}}\nobar}
\newcommand{\prob}{\operatorname{\mathbb{P}}\expectarg}
\newcommand{\var}{\operatorname{Var}\expectarg}

\DeclarePairedDelimiterX{\nobar}[1]{(}{)}{%
  \ifnum\currentgrouptype=16 \else\begingroup\fi
  #1
  \ifnum\currentgrouptype=16 \else\endgroup\fi
}
\DeclarePairedDelimiterX{\expectarg}[1]{(}{)}{%
  \ifnum\currentgrouptype=16 \else\begingroup\fi
  \activatebar#1
  \ifnum\currentgrouptype=16 \else\endgroup\fi
}
\DeclarePairedDelimiterX{\parenthesis}[1]{\{}{\}}{%
  \ifnum\currentgrouptype=16 \else\begingroup\fi
  \activatebar#1
  \ifnum\currentgrouptype=16 \else\endgroup\fi
}
\DeclarePairedDelimiterX{\bracket}[1]{[}{]}{%
  \ifnum\currentgrouptype=16 \else\begingroup\fi
  \activatebar#1
  \ifnum\currentgrouptype=16 \else\endgroup\fi
}

\providecommand{\customgenericname}{}

\newtheorem{innercustomgeneric}{\customgenericname}

\newcommand{\innermid}{\nonscript\;\delimsize\vert\nonscript\;}
\newcommand{\activatebar}{%
  \begingroup\lccode`\~=`\|
  \lowercase{\endgroup\let~}\innermid 
  \mathcode`|=\string"8000
}

\usepackage[dvipsnames]{xcolor}
\usepackage{caption}
\usepackage{subcaption}

\newtheorem{theorem}{Theorem}
\newtheorem{corollary}{Corollary}
\newtheorem{example}{Example}
\newtheorem{proposition}{Proposition}

\newtheorem{lemma}{Lemma}
\newtheorem{definition}{Definition}
\newtheorem*{remark}{Remark}

\author{Ziyi Liang\textsuperscript{1}, T. Tony Cai\textsuperscript{2}, Wenguang Sun\textsuperscript{3}, and Yin Xia\textsuperscript{4}}

\begin{document}

\maketitle
\let\thefootnote\relax
\footnotetext{\hspace{-0.62cm}\textsuperscript{1}Department of Statistics, University of California Irvine, Irvine, CA, USA.\\
  \textsuperscript{2}Department of Statistics and Data Science, University of Pennsylvania, Philadelphia, PA, USA.\\
  \textsuperscript{3}School of Management and Center for Data Science, Zhejiang University, Zhejiang, China. \\
  \textsuperscript{4}Department of Statistics and Data Science, Fudan University, Shanghai, China. \\}

\maketitle


\begin{abstract}
Incorporating auxiliary information alongside primary data can significantly enhance the accuracy of simultaneous inference. However, existing multiple testing methods face challenges in efficiently incorporating complex side information, especially when it differs in dimension or structure from the primary data, such as network side information. This paper introduces a locally adaptive structure learning algorithm (LASLA), a flexible framework designed to integrate a broad range of auxiliary information into the inference process. Although LASLA is specifically motivated by the challenges posed by network-structured data, it also proves highly effective with other types of side information, such as spatial locations and multiple auxiliary sequences. LASLA employs a $p$-value weighting approach, leveraging structural insights to derive data-driven weights that prioritize the importance of different hypotheses. Our theoretical analysis demonstrates that LASLA asymptotically controls the false discovery rate (FDR) under independent or weakly dependent $p$-values, and achieves enhanced power in scenarios where the auxiliary data provides valuable side information. Simulation studies are conducted to evaluate LASLA's numerical performance, and its efficacy is further illustrated through two real-world applications.

\end{abstract}

\vspace{9pt}

\section{Introduction} \label{sec:intro}

\subsection{Motivating application for network data} \label{sec:intro-network}
Statistical analysis of network-structured data is an important topic with a wide range of applications. Our study is motivated by genome-wide association studies (GWAS), where a primary objective is to identify disease-associated single-nucleotide polymorphisms (SNPs) across diverse populations. Previous studies have indicated that linkage analysis can provide insights into the genetic basis of complex diseases. Particularly, SNPs in linkage disequilibrium (LD) can jointly contribute to the representation of the disease phenotype \citep{2012schaub, Joiret2019-sl}. 
However, existing research in GWAS has often overlooked or underutilized the LD network information, representing a significant limitation. 
Therefore, it is essential to develop new integrative analytical tools that can effectively combine GWAS data with auxiliary data from linkage analysis.
 
\begin{figure}[ht]
	\centering
	\includegraphics[width=1\linewidth]{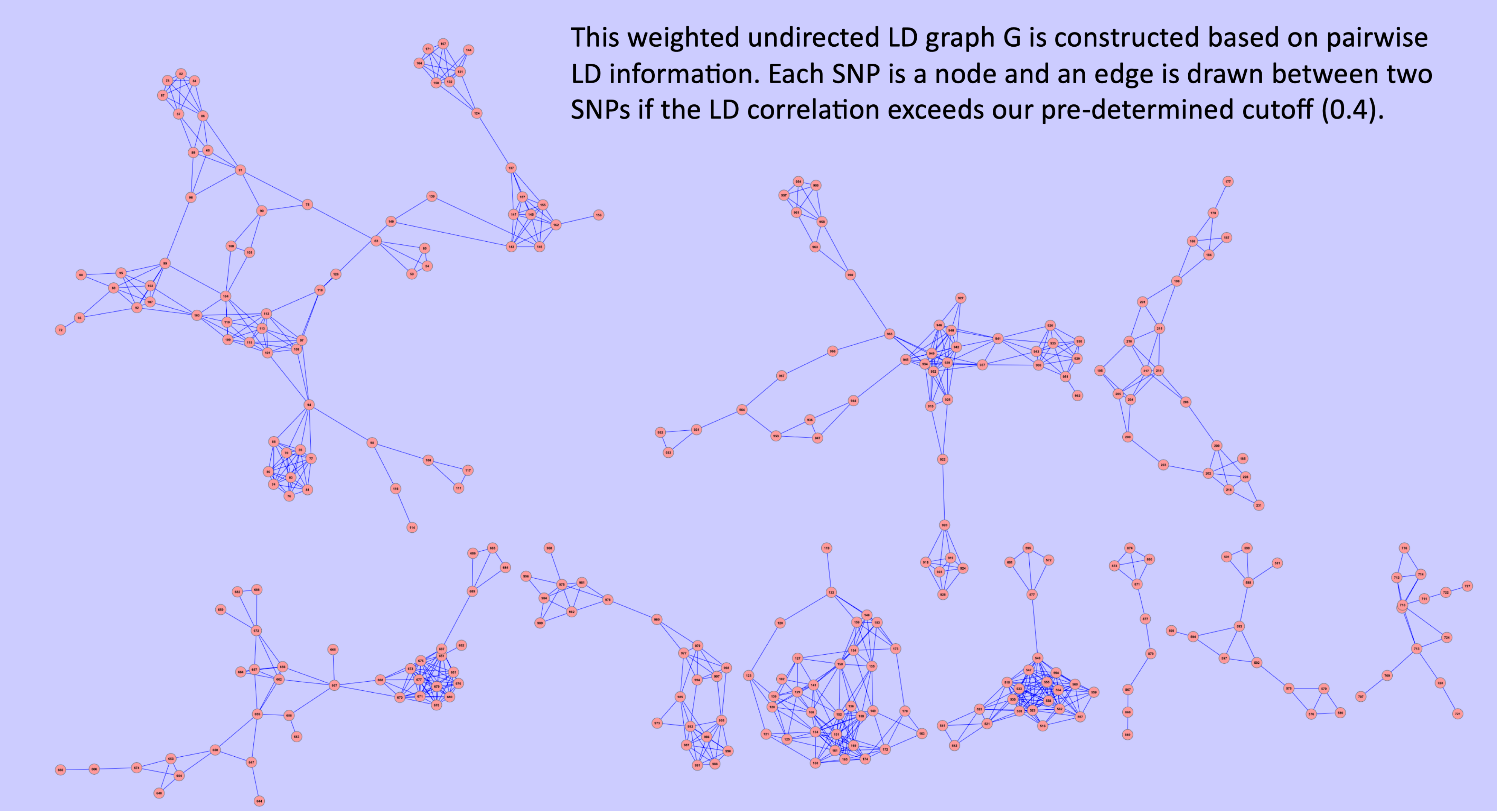}
	\caption{Auxiliary linkage data provide structural information that can be utilized to identify significant SNPs.}
	\label{LD.fig}
\end{figure}

Let $m$ denote the number of SNPs and $[m]=\{1,\ldots, m\}$. The primary data, as provided by GWAS, consists of a list of test statistics $\pmb T=\{T_i: i\in [m]\}$, which assess the association strength of individual SNPs with a phenotype of interest. Let $\pmb P =\{P_i: i\in [m]\}$ denote the corresponding $p$-values. The auxiliary data, as provided by linkage analysis, is a matrix comprising pairwise correlations $\pmb S=(r_{ij}: i,j\in[m])$, where $r_{ij}$ measures the LD correlation between SNP $i$ and SNP $j$. 
To illustrate, we construct an undirected LD graph based on the pairwise LD correlation
in Figure \ref{LD.fig}, where each node represents an SNP, and an edge is drawn to connect two SNPs if their correlation exceeds a pre-determined cutoff. 

Incorporating the network structured auxiliary data, such as the LD correlation matrix, is desirable as it may improve the power and interpretability of analysis. However, developing a principled approach that cross-utilizes data from different sources is a challenging task. Firstly, the primary data $\pmb T$ and the auxiliary data $\pmb S$ are usually collected from different populations. For instance, in our data analysis detailed in Section \ref{sec:application-gwas}, the target population in GWAS consists of 77,418 individuals with Type 2 diabetes, while the auxiliary data for linkage analysis are collected from the general population [1000 Genomes (1000G) Phase 3 Database]. Secondly, the dimensions of $\pmb T$ and $\pmb S$ may not match as in the conventional settings. Specifically, $\pmb{T} \in \mathbb{R}^m$ is a vector of statistics while $\pmb S \in \mathbb{R}^{m\times m}$ is a network matrix. 


Motivated by the observation that SNPs in the same sub-network can exhibit similar distributional characteristics and tend to work together in GWAS \citep{2012schaub}, this article develops a locally adaptive structure learning algorithm (LASLA)
which integrates the structural knowledge in auxiliary data by computing a set of data-driven weights $\cbrac{w_i: i \in [m]}$ to adjust the $p$-values $\cbrac{P_i: i \in [m]}$. In summary, LASLA follows a two-step strategy: 
\begin{description}
\item Step 1: Learn the relational or structural knowledge of the high-dimensional parameter through auxiliary data.
\item Step 2: Apply the structural knowledge by adaptively placing differential weights $w_i$ on $p$-values $P_i$, for $i\in [m]$. 
\end{description}

While LASLA is unique in its ability to directly leverage network side information, its versatility extends beyond networks. For instance, when identifying significant regions using functional Magnetic Resonance Imaging (fMRI) data, as discussed in Section \ref{sec:fmri}, LASLA can effectively incorporate 3D spatial locations as auxiliary information. Additionally, in high-dimensional regression models aimed at identifying disease-related genetic variants, LASLA can integrate data from related diseases to enhance detection power by prioritizing shared risk factors and genetic variants, as detailed in Section \ref{app:additional-applications} of the Supplementary Material. LASLA offers a flexible and powerful tool that adapts to various forms of auxiliary information, making it suitable for a wide range of applications beyond its original focus on network data integration.

\subsection{Related work and our contribution} \label{sec:lit-review}

Structured multiple testing has gained increasing attention in recent years \citep{Caietal19, LiBar19, Casetal20, RenCan20}. {The strategy is to augment the sequence of primary statistics $\pmb T=\{T_i: i\in [m]\}$, which may consist of $p$-values or $z$-values, with an auxiliary sequence $\pmb U=\{U_i: i \in [m]\}$ in order to enhance the statistical power and accuracy of inference. The auxiliary data can be collected through various methods, including: 
(a) external sources, such as prior studies and secondary datasets (\citealp{Ignetal16, Basetal18}); (b) internal sources within the same dataset, achieved by carefully constructing independent sequences (\citealp{Caietal19, Xiaetal19}); (c) intrinsic patterns associated with the data, such as the natural order of data streams (\citealp{FosSti08, Lynetal17}) or spatial locations (\citealp{Leietal17, Caietal20}).} 
The aforementioned studies primarily focus on an auxiliary sequence $\pmb U \in \mathbb{R}^m$ that has the same dimension as $\pmb T \in \mathbb{R}^m$. {However, there has been limited research conducted on exploiting a wider variety of side information, such as LD matrices or multiple auxiliary sequences.} 

Moreover, existing structure-adaptive methods (\citealp{Caietal19, LiBar19, Xiaetal19}) primarily focus on leveraging sparsity structures in auxiliary data. However, research has shown that other forms of structural information can also significantly enhance inference (\citealp{RoeWas09, Penetal11, LeiFit18, Fuetal19}). For instance, \citet{RoeWas09} illustrates the power gains achieved by constructing weights based on signal amplitudes, \citet{Fuetal19} demonstrates the advantages of adjusting for heteroscedasticity, and \citet{Penetal11} highlights the benefits of adapting to hierarchical structures. Despite these advancements, most methods typically concentrate on a specific type of structural information. Therefore, developing a unified framework that can systematically integrate various types of side information is highly desirable.

Our contributions are three-fold. {Firstly, LASLA employs a carefully designed distance matrix to characterize the relational knowledge of the $p$-values, providing a principled and generic strategy for integrative inference. Secondly, LASLA can leverage various types of structural information by deriving data-driven weights to emulate a hypothetical oracle (\citealp{SunCai07, HelRos19}), resulting in significant power improvement compared to existing methods \citep{RoqVan09, IgnHub20, Caietal20}. Lastly, we have developed theory to demonstrate that LASLA asymptotically controls the False Discovery Proportion (FDP) and the False Discovery Rate (FDR) for both independent and weakly dependent $p$-values, while also achieving proven power gain under mild conditions.}

The rest of this article is organized as follows. Section \ref{method:sec} presents the problem formulation and provides details on the development of the LASLA procedure. Section \ref{theory:sec} investigates the theoretical properties of LASLA in the independent case and establishes its superiority in ranking. Section \ref{sec:simulation} presents simulation results comparing LASLA with existing methods. In Section \ref{sec:application}, we illustrate LASLA through two real-world applications. Additional simulation results, theoretical results for the dependent case, and proofs of the theories can be found in the Supplementary Material.


\section{A Locally Adaptive Structure Learning Algorithm}
\label{method:sec}

We present the problem formulation and the basic framework in Section \ref{sec:framework}, followed by the derivation of the oracle rule under a suitable working model in Section \ref{sec:oracle}. Subsequently, we discuss the strategies for approximating the related unknown quantities and derive the oracle-assisted weights in Section~\ref{sec:data-driven-weight}. We address the selection of a threshold for the weighted p-values and present the LASLA procedure in Section \ref{sec:lasla-threshold}. Finally, we demonstrate the superiority of the LASLA weights over conventional sparsity-adaptive weights in Section~\ref{sec:weight-comparison}. 


\subsection{Problem formulation and basic framework}\label{sec:framework}

Suppose we are interested in testing $m$ hypotheses:
\begin{equation} \label{eq:hypothesis}
{H_{0, i}}: \theta_{i}=0 \quad \text{vs.} \quad H_{1, i}: \theta_{i} = 1, \quad i\in[m],
\end{equation}
where $\pmb\theta=(\theta_i: i\in[m])$ is a vector of binary random variables indicating the existence or absence of signals at the testing locations.
A multiple testing rule can be represented by a binary vector $\pmb\delta=(\delta_i: i\in[m])\in\{0,1\}^m$, where $\delta_i=1$ indicates that we reject {$H_{0, i}$} and $\delta_i=0$ otherwise. In this paper we focus on problem \eqref{eq:hypothesis} with the goal of controlling the FDP and FDR \citep{bh95} defined respectively by:
$$
\text{FDP}(\pmb{\delta}) = \frac{\sum_{i=1}^m(1-\theta_i)\delta_i}{\max(\sum_{i=1}^m\delta_i,1 )}, \quad \text{and FDR}= \EE\left\{\text{FDP}(\pmb{\delta})\right\}.
$$
An ideal procedure should maximize power, which is interpreted as the expected number of true positives:
$$
{\text{ETP} = \EE \left\{ \sum_{i=1}^{m} \theta_i \delta_i \right\}.}
$$

Our approach involves the calculation of a distance matrix $\pmb D=(1-r^2_{ij}: i, j\in [m])$ based on the LD matrix $\pmb{S} = (r^2_{ij}: i, j\in [m])$, where $ r_{ij} \in [0,1]$ represents the Pearson's correlation coefficient that measures the LD between SNPs $i$ and $j$. Note that the distance matrix $\pmb{D}$ contains the same information as $\pmb{S}$, as the transformation from $\pmb{S}$ to $\pmb{D}$ involves a simple linear operation: $1 - r^2_{ij}$, applied to each entry in the LD matrix. This transformation facilitates our definition of the \text{local neighborhood} and subsequent data-driven algorithms. Specifically, SNPs with higher correlation $r_{ij}$ are considered ``closer'' in distance. {The driving assumption behind our methodological development is that clusters of SNPs with small pairwise distances are part of the same local neighborhood and demonstrate similar distributional characteristics. Conversely, SNPs that are widely separated are more likely to belong to different neighborhoods and exhibit distinct patterns in terms of sparsity levels and distributions.}
Deriving a distance matrix $\pmb{D}$ from the LD matrix $\pmb{S}$ is a relatively straightforward process. In Section \ref{app:local-neighborhood} of the Supplementary Material, we discuss extensions that allow for the construction of distance matrices using other types of side information.
{Harnessing the heterogeneity across different neighborhoods plays a crucial role in constructing informative weights for the $p$-values. By incorporating weights that encode the network structure, the identification of SNP clusters is facilitated, enabling a more accurate representation of the underlying biological processes and genetic basis of complex diseases. These insights serve as the motivation for our proposed LASLA procedure, which aims to enhance both the interpretability and power of FDR analysis.

}

\subsection{The oracle rule under a working model}\label{sec:oracle}

One effective {strategy} to incorporate relevant domain knowledge in multiple testing is {through the use of} $p$-value weighting (\citealp{Genetal06, Sunetal15, Basetal18}). 
The proposed LASLA builds upon this strategy.  
This subsection introduces an oracle procedure under an ideal setting. In Sections~\ref{sec:data-driven-weight}, we {present details for the development of} a data-driven algorithm that emulates the oracle. {This algorithm utilizes a distance matrix learned from auxiliary data to construct informative weights for the $p$-values.} 
 

Assume the primary test statistics $\{T_i: i\in[m]\}$ follow a hierarchical model: 
\beq\label{model-hypo}
\theta_i\overset{\text{ind}}{\sim} \text{Bernoulli}(\pi_i^*),\quad \PP(T_i\leq t|\theta_i)=(1-\theta_i)F_0(t)+\theta_iF_{1i}^*(t),
\eeq
where $F_0(t)$ and {$F_{1i}^*(t)$} respectively represent the null and alternative cumulative distribution functions (CDF) of $T_i$. Then $T_i$ obeys the following mixture distribution
\beq\label{eq:oracle-distribution}
F_i^*(t) \coloneqq \PP(T_i\leq t) =(1-\pi_i^*)F_0(t)+\pi_i^*F_{1i}^*(t).
\eeq 

{We would like to provide further clarification on our notations. Firstly, Model \eqref{model-hypo} is utilized as a working model to provide an approximation of the true data-generating process. Its primary goal is to facilitate the derivation of our data-driven weights. Secondly, the incorporation of $\pi_i^*$ and $F_{1i}^*(t)$ aims to account for the potential heterogeneity across testing units, which arises due to the availability of auxiliary data.
 In contrast, the null distribution $F_0(t)$ is assumed to be known and remains the same across all testing units.}

The optimal testing rule, which has the largest ETP among all valid marginal FDR/FDR procedures (\citealp{SunCai07, Caietal19, HelRos19}), is a thresholding rule based on the local false discovery rate (lfdr):  
$$
L_i^*= \PP(\theta_i=0|T_i)={(1-\pi_i^*)f_0(T_i)}/{f_i^*(T_i)},
$$
where $f_0(\cdot)$ and $f_i^*(\cdot)$ respectively denote the density functions corresponding to the CDFs $F_0(\cdot)$ and $F_i^*(\cdot)$.
A step-wise algorithm (e.g. \citealp{SunCai07}) can be used to determine a data-driven threshold along the $L_i^*$ ranking. Let $L_{(1)}^*\leq\cdots\leq\ L_{(m)}^*$ be the ordered statistics of $L_i^*$, and  $H_{(1)},\ldots,H_{(m)}$ be the corresponding null hypotheses. 
At FDR level $\alpha$, the rejection threshold is determined by  
$
k_*=\max\left\{j:j^{-1}\sum^j_{i=1}{L_{(i)}^*\leq \alpha}\right\}
$, and the algorithm rejects $H_{(1)}, \ldots, H_{(k_*)}$.

\subsection{Data-driven weights} \label{sec:data-driven-weight}

It is essential to note that $\pi_i^*$ and $f_i^*(t)$ represent hypothetical and non-accessible oracle quantities that cannot be directly employed in data-driven algorithms. Therefore, we develop {data-driven quantities} to approximate the oracle quantities. 

{In our derivation of the data-driven quantity to approximate $\pi_i^*$, we utilize the screening strategy outlined in the work of \citet{Caietal20}. To ensure completeness, we reiterate the key steps below.}
Let $D_{ij}$ be the distance between entries $i$ and $j$. A key assumption is that $\pi_i^*$ is close to $\pi_j^*$ if $D_{ij}$ is small (and the distributional informations for the two hypotheses are identical if $D_{ij}$=0), which enables us to {estimate $\pi_i^*$ by borrowing strength from neighboring points}. 
Let $K$: $\mathbb{R}\rightarrow \mathbb{R}$ be a positive, bounded, and symmetric kernel function that satisfies the following conditions:
\begin{align}\label{kernel}
\int_{\mathbb{R}}K(x)dx=1, \quad \int_{\mathbb{R}}xK(x)dx=0,\quad \int_{\mathbb{R}}x^2K(x)dx=\sigma_K^2<\infty.
\end{align}
Define $K_h(x)=h^{-1}K(x/h)$, where $h$ is the bandwidth, and $V_h(i,j)=K_h(D_{ij}) / K_h(0)$. 
We construct a data-driven quantity that resembles $\pi_i^*$ by 
\begin{align}\label{eq:pi-est}
\pi_i=1-\frac{\sum_{j \neq i}\brac*{{V_h(i,j)\II\cbrac{P_j>\tau}}}}{(1-\tau)\sum_{j \neq i}{V_h(i,j)}}, \quad i\in[m],
\end{align} 
{where $\tau$ is a screening parameter that will be chosen in Section \ref{app:numeric_results} of the Supplementary Material.}
Similarly, we introduce a data-driven quantity that resembles the hypothetical density $f_i^*(t)$ for $i \in [m]$ by:
\begin{equation}\label{eq:density-est}
   f_i(t)=\frac{\sum_{j \neq i}\brac*{{V_h(i,j)K_h(T_j-t)}}}{\sum_{j \neq i}{V_h(i,j)}},
\end{equation}
which describes the likelihood of $T_i$ taking a value in the vicinity of $t$.
Consequently, the data-driven lfdr is given by
\begin{equation} \label{eq:clfdr-est}
L_i=\paren{1-\pi_i}f_0(T_i)/f_i(T_i).
\end{equation} 
Let $L_{(1)}\leq \ldots \leq  L_{(m)}$ denote the sorted values of $L_i$. Following the thresholding approach in Section \ref{sec:oracle}, we choose $L_{(k)}$ to be the rejection threshold, where $k=\max \{j: j^{-1}\sum^j_{i=1}{L_{(i)}\leq \alpha}\}$.

{Now we are ready to develop the oracle-assisted weights that mimic thresholding rules based on lfdr given in \eqref{eq:clfdr-est}.} Without loss of generality we assume that the function $f_0$ is symmetric about zero. In cases where this assumption does not hold, we can always transform the primary statistics into $z$-statistics or $t$-statistics. Intuitively, the thresholding rule $L_i<t$ is approximately equivalent to:
\beq
\label{approx-equiv}
T_i<t_i^- \text{ or } T_i>t_i^+.
\eeq
Here $t_i^- \leq 0$ and $t_i^+ \geq 0$ are coordinate-specific thresholds that lead to asymmetric rejection regions, which are useful for capturing structures of the alternative distribution, such as unequal proportions of positive and negative signals \citep{SunCai07, LiBar19, Fuetal19}. 

The next step is to derive weights that can emulate the rule \eqref{approx-equiv}. When $T_i \geq 0$, let $t^+_i=\infty$ if $(1-\pi_i)f_0(t)/f_i(t) > L_{(k)} \text{ for all } t \geq 0,$ else let
\begin{align}\label{eq:t+}
t^+_i=\inf \cbrac*{t\geq0: {(1-\pi_i)f_0(t)/f_i(t)}\leq  L_{(k)}}
\end{align}
 The rejection rule $T_i > t_i^+$ is equivalent to the $p$-value rule $P_i<1-F_{0}(t_i^+)$, where $P_i$ is the one-sided $p$-value. Define the weighted $p$-values as $\{P_i^w=P_i/w_i: i\in[m]\}$. Intuitively, a larger rejection region $T_i > t_i^+$ implies a proportionally larger weight $w_i$ to prioritize the rejection of $H_i$. Therefore, if we choose a universal threshold for all $P_i^w$, using weight $w_i=1-F_{0}(t^+_i)$ can effectively emulate the rule $T_i > t_i^+$.

Similarly, when $T_i < 0$, let $t^-_i=-\infty$ if $(1-\pi_i)f_0(t)/f_i(t) > L_{(k)} \text{ for all } t \leq 0,$ else let
\begin{align}\label{eq:t-}
t^-_i=
    \sup \cbrac*{t\leq0: (1-\pi_i)f_0(t)/f_i(t)\leq L_{(k)}}
\end{align}
The corresponding weight is given by $w_i=F_{0}(t^-_i)$. 

To ensure the robustness of the algorithm, we set $w_i = \max\{w_i, \xi\}$ and $w_i = \min\{w_i, 1-\xi\}$, where $0<\xi<1$ is a small constant. In all numerical studies, we set $\xi=10^{-5}$. To expedite the calculation {and facilitate our theoretical analysis}, we propose to use only a subset of $p$-values in the neighborhood to obtain $\pi_i$ and $f_i(t)$ for $i \in [m]$. Specifically, we define $\cN_i = \{j \neq i: D_{ij} \leq a_{\epsilon}\}$ as a neighborhood set that only contains indices with distance to $i$ smaller than $a_{\epsilon}$ and satisfies $|\cN_i| = m^{1-\epsilon}$ for some small constant $\epsilon>0$. Moreover, we require that $D_{ij_1} \leq D_{ij_2}$ for any $j_1\in \cN_i$, $j_2 \notin \cN_i$, and $j_2\neq i$. This approach has minimal impact on numerical performance, as $p$-values that are far apart contribute little. We summarize the computation of the data-driven weights in Algorithm~\ref{alg:simple_weights} below.

\begin{algorithm}[htbp]
    \caption{Data-driven weights}
    \label{alg:simple_weights}
    \begin{algorithmic}[1]
        \STATE \textbf{Input:} Nominal FDR level $\alpha$; $\epsilon$ specifying the size of the sub-neighborhood;
        \STATE \textcolor{white}{''Input:} kernel function $K(\cdot)$; primary statistics $\pmb T = \{T_i: i \in [m]\}$ and 
        \STATE \textcolor{white}{''''Input:}distance matrix $\pmb D = (D_{ij}:   i,j \in [m])$. 
        \STATE \textbf{For} $i \in [m]$ \textbf{do}:
        \STATE \qquad Estimate $\pi_i$ as given by (\ref{eq:pi-est}) with summation taken over $\cN_i$.
        \STATE \qquad Estimate $f_i(t)$ as given by (\ref{eq:density-est}) with summation taken over $\cN_i$. 
        \STATE \qquad Compute $L_i$ by (\ref{eq:clfdr-est}). Denote the sorted statistics by $L_{(1)}\leq \ldots \leq  L_{(m)}$.
        \STATE Let $L_{(k)}$ be the oracle threshold, where $k=\max \{j \in [m]: j^{-1}\sum^j_{i=1}{L_{(i)}\leq \alpha}\}$. 
        \STATE \textbf{For} $i \in [m]$ \textbf{do}:
        \STATE \qquad \textbf{If}{ $T_i \geq 0$}:
        Compute $t^+_i$ by (\ref{eq:t+}), and the weight by $w_i=1-F_{0}(t^+_i)$.
        \STATE \qquad \textbf{If}{ $T_i < 0$}:
        Compute $t^-_i$ by (\ref{eq:t-}), and the weight by $w_i=F_{0}(t^-_i)$.
        \STATE \textbf{Output:} Data-driven weights $\{w_i: i \in [m]\}$.
    \end{algorithmic}
\end{algorithm}

\subsection{The LASLA procedure}\label{sec:lasla-threshold}

{Consider the weighted $p$-values $\{P^w_i: i \in [m]\}$, where $P^w_i\coloneqq P_i/w_i$ and $\{w_i: i \in [m]\}$ represent the data-driven LASLA weights.}
Assume that $P_i$ is independent with $w_i$, for $i \in [m]$. This assumption only serves as a motivation for the thresholding rule, and is not required for the theoretical analysis. It follows that,  given the set of weights, the expected number of false positives (EFP) for a threshold $t^w$ can be calculated as
   $ t^w \sum_{i=1}^m w_i \paren*{1-\pi_i^*}
$
under the working model \eqref{model-hypo}.
Suppose we reject $j$ hypotheses along the ranking provided by $P_{(i)}^w$ for $i \in [m]$, namely, set the threshold as $t^w = P^w_{(j)}$. Then the FDP can be estimated as:
$
\widehat{\text{FDP}}=\paren*{P^{w}_{(j)}/j}\sum_{i=1}^m w_i \paren*{1-\pi_i}.
$ 
We choose the largest possible threshold such that $\widehat{\text{FDP}}$ is less than the nominal level $\alpha$. Specifically, we find
        \begin{equation}\label{eq:lasla-threshold}
        k^w=\max\bigg\{j:\paren*{P^{w}_{(j)}/j}\sum_{i=1}^m w_i \paren*{1-\pi_i} \leq \alpha \bigg\},
        \end{equation} 
and reject $H_{(1)}, \cdots, H_{(k^w)}$, which are the hypotheses with weighted $p$-values no larger than $P_{(k^w)}^w$.

\begin{remark}\label{rmk:wbh}
One could replace \eqref{eq:lasla-threshold} with other thresholding rules, such as the weighted BH (WBH) procedure \citep{Genetal06}. However, empirical evidence suggests that WBH tends to be overly conservative. The adaptive WBH method~\citep{2019ramadaspfilter} improves the power of WBH by adjusting to an estimated global sparsity level, but it is less effective when the sparsity level is heterogeneous. See Section \ref{sec:alt-thresholding} of the Supplementary Material for a detailed numerical comparison.
\end{remark}

\subsection{Oracle-assisted weights vs sparsity-adaptive weights}\label{sec:weight-comparison}

This section presents a comparison between the oracle-assisted weights and the sparsity-adaptive weights discussed in Section \ref{app:sparsity-review} of the Supplementary Material. In summary, the sparsity-adaptive weights only utilize the signal sparsity level $\pi_i^*$, whereas the oracle-assisted weights can leverage the structural information embedded in the mixture density $f_i^*(t)$. Consider the oracle sparsity-adaptive weights $w_i^{\text{laws}}=\pi_i^*/(1-\pi_i^*)$ for $i \in [m]$, as reviewed in Section \ref{app:sparsity-review}. In the case where neighborhood information is provided by spatial locations, the sparsity-adaptive weights are equivalent to the weights employed by the LAWS procedure (\citealp{Caietal20}). Intuitive examples are provided to illustrate potential information loss associated with the sparsity-adaptive weights (referred to as LAWS weights), and the advantages of the newly proposed weights (referred to as LASLA weights) are highlighted.
We focus on scenarios where $\pi_i^*$ is homogeneous, but it's important to note that the inadequacy of sparsity-adaptive weights extends to cases where $\pi_i^*$ is heterogeneous. This shortcoming arises from the fact that the sparsity-adaptive weights ignore structural information in alternative distributions.

In the following comparisons, the oracle LASLA weights are obtained via Algorithm \ref{alg:simple_weights}, with $\pi_i$ and $f_i$ replaced by their oracle counterparts $\pi_i^*$ and $f_i^*(t)$. We implement both weights with oracle quantities to emphasize the methodological differences. Consider two examples where $\{T_i:  i \in [m]\}$ follow the mixture distribution in (\ref{eq:oracle-distribution}) with $m=1000$ and $\pi_i^* = 0.1$.  

\begin{example}\label{example:asymmetry}\rm{
Set $F^*_{1i}(t) = \gamma N(3, 1)+(1-\gamma)N(-3, 1)$, where $\gamma$ controls the relative proportions of positive and negative signals. We vary $\gamma$ from 0.5 to 1; when $\gamma$ approaches 1, the level of asymmetry increases.
}
\end{example}

\begin{example} \label{example:shape-hetero}
\rm{Set $F^*_{1i}(t) = 0.5N(3, \sigma_i^2)+0.5N(-3, \sigma_i^2)$, where $\sigma_i$ controls the shape of the alternative distribution, and $\mathbb{P}(\sigma_i=1)=0.5$ and $\mathbb{P}(\sigma_i=\sigma)=0.5$. We vary $\sigma$ from 0.2 to 1; the heterogeneity is most pronounced when $ \sigma=0.2$. 
}
\end{example}

\begin{figure*}[ht]
  \centering
  \includegraphics[width=0.85\linewidth]{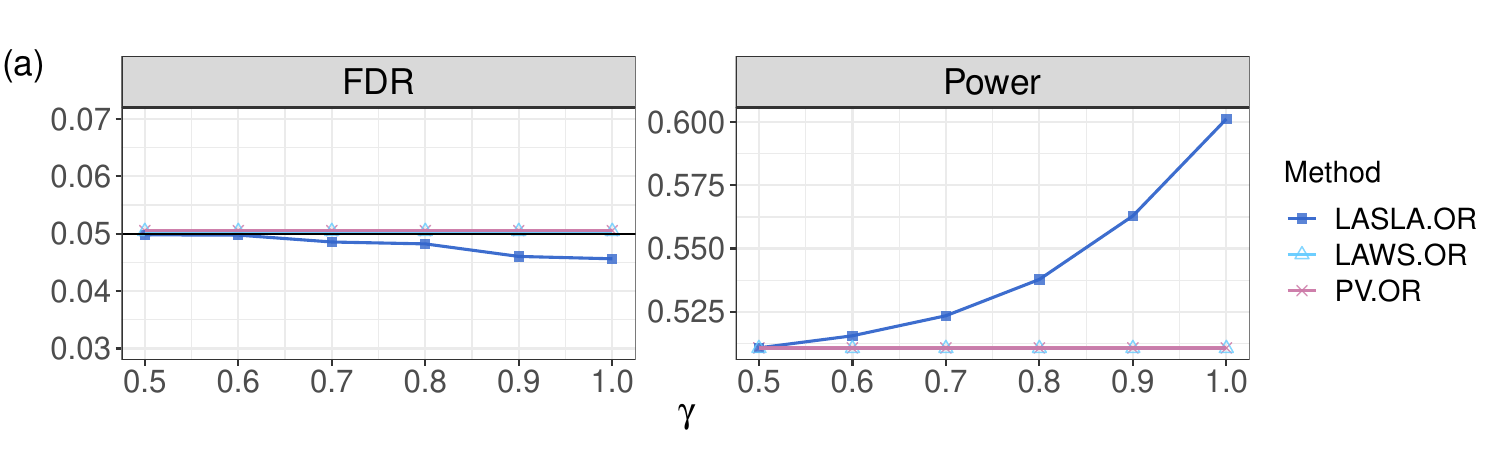}
  \includegraphics[width=0.85\linewidth]{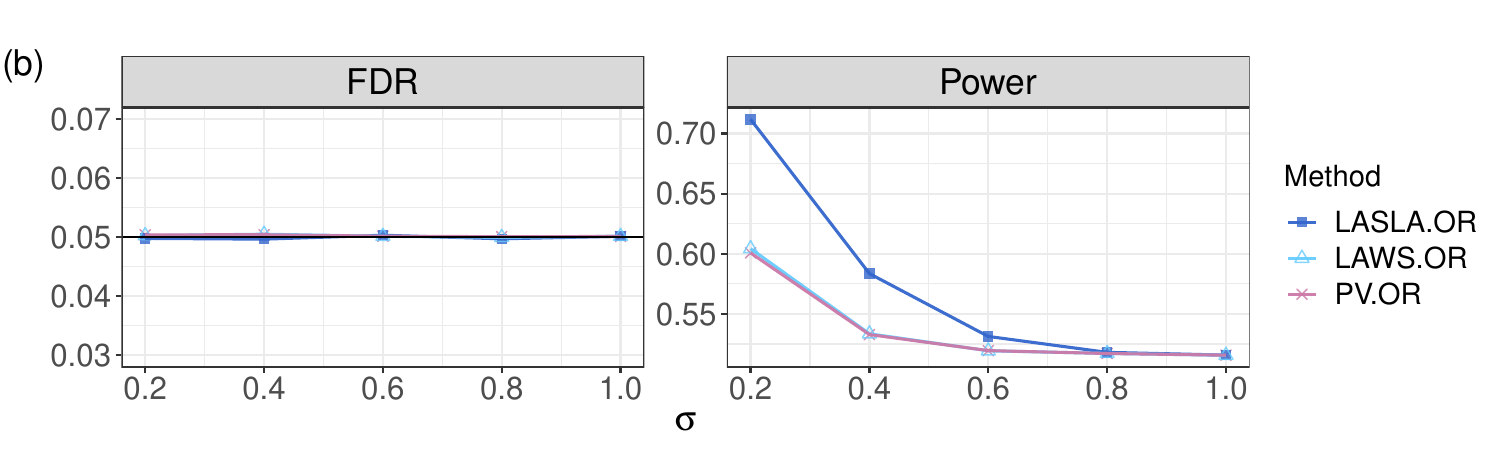}
  \caption{Empirical FDR and power comparison for oracle LASLA (LASLA.OR), oracle LAWS (LAWS.OR) and unweighted oracle $p$-value procedure (PV.OR) that leverages $\pi_i^*$ to determine the rejection thresholds, with nominal FDR level $\alpha=0.05$. Results for LASLA, LAWS and the unweighted oracle procedure are derived by applying the thresholding rule in Section~\ref{sec:lasla-threshold} respectively with LASLA weights, LAWS weights, and weights equal to $1$.  (a): Example \ref{example:asymmetry}: increasing asymmetry level $\gamma$; (b): Example \ref{example:shape-hetero}: decreasing shape heterogeneity $\sigma$.} 
  \label{fig:sparsity-comp}
\end{figure*}

In both examples, since $\pi_i^*$ and $w_i^{\text{laws}}$ remain constant for all $i \in [m]$, the oracle LAWS reduces to the unweighted $p$-value procedure \citep{GenWas02}. On the contrary, LASLA weights are able to capture the signal asymmetry in Example \ref{example:asymmetry} and the shape heterogeneity in the alternative distribution in Example \ref{example:shape-hetero},, resulting in notable power gain in both settings. Additional examples are provided in Section \ref{app:hetero-alt} to further demonstrate LASLA’s strength in leveraging hypothesis-specific information.

\section{Theoretical Analysis} 
\label{theory:sec}
{We first introduce in Section \ref{sec:alt-weights} a modified version of the data-driven weights to facilitate the theoretical analysis,
 and then study in Section~\ref{sec:indep_fdp} the FDP and FDR control of LASLA under a marginal independence assumption. Finally, Section~\ref{sec:power} demonstrates the theoretical power gain of the proposed weighting strategy.}

\subsection{Modified data-driven weights}\label{sec:alt-weights}
{To show the asymptotic validity of LASLA with minimal assumptions as shown in Section \ref{sec:indep_fdp}, we slightly modify the data-driven weights from Algorithm~\ref{alg:simple_weights}. It is worthy noting that, by imposing stronger conditions as presented in Section \ref{app:dependent-theory} of the Supplementary Material, such modification is no longer needed.

Specifically, to facilitate the theoretical analysis, we propose to compute an index-dependent threshold $L^i_{(k)}$ which only relies on $|\cN_i|$ statistics as opposed to the universal threshold $L_{(k)}$ defined in Section \ref{sec:data-driven-weight}. This modification arises from the subtle fact that, though $L_i$ in \eqref{eq:clfdr-est} is estimated without using $T_i$ itself {(the sign of $T_i$ is used, but it will not affect the validity of the algorithm)}, $w_i$ still correlates with $T_i$ since the determination of $k$ in the universal threshold relies on all primary statistics. This intricate dependence structure poses challenges for the theoretical analysis. Therefore, we compute the index-dependent threshold $L^i_{(k)}$ to ensure that given the sign of $T_i$, $w_i$ is independent of $T_i$  under the null hypothesis at the cost of computational efficiency. This modification has little impact on the numerical
performance, but greatly facilitates the establishment of the asymptotic validity of the corresponding weighted multiple testing strategy in Section \ref{sec:indep_fdp}. 
We summarize the modified method in Algorithm \ref{alg:complicated_weights}.
}

\begin{algorithm}
    \caption{Modified data-driven weights}
    \label{alg:complicated_weights}
    \begin{algorithmic}[1]
        \STATE \textbf{Input:} Nominal FDR level $\alpha$; $\epsilon$ specifying the size of the sub-neighborhood;
        \STATE {\color{white}\textbf{Input:}} kernel function $K(\cdot)$; primary statistics $\pmb T = \{T_i: i \in [m]\}$ and 
        \STATE {\color{white}\textbf{Input:}} distance matrix $\pmb D = (D_{ij}:  i,j \in [m])$. 
        \STATE \textbf{For} $i \in [m]$ \textbf{do}:
        {
        \STATE \quad Estimate $\pi_i$ by (\ref{eq:pi-est}) and $f_i(t)$ by (\ref{eq:density-est}) with summation taken over $\cN_i$.
        \STATE \quad \textbf{For} $j \in \cN_i$ \textbf{do}:
        \STATE \quad \quad Estimate $\pi^i_j$ by (\ref{eq:pi-est}) and $f^i_j(t)$ by (\ref{eq:density-est}) with the summation taken over $\cN_i$.}
        \STATE \quad \quad Compute $L^i_j$ as in (\ref{eq:clfdr-est}) with $\pi_j$ replaced by $\pi^i_j$; $f_j(t)$ by $f^i_j(t)$. 
        \STATE \quad Denote the sorted statistics by $L^i_{(1)}\leq \ldots \leq  L^i_{(|\cN_i|)}$.
        \STATE \quad Let $k=\max \{j\in [|\cN_i|]: j^{-1}\sum^j_{l=1}{L^i_{(l)}\leq \alpha}\}$.    
        \STATE \quad \textbf{If}{ $T_i \geq 0$} \textbf{then:} 
        \STATE
        \quad \quad Compute $t^+_i$ as given by (\ref{eq:t+}), with 
        $L_{(k)}$ replaced by $L^i_{(k)}$; 
        \STATE
        \quad \quad   Compute the weight by $w_i=1-F_{0}(t^+_i)$.
        \STATE \quad \textbf{If}{ $T_i < 0$} \textbf{then:} \STATE
        \quad \quad Compute $t^-_i$ as given by (\ref{eq:t-}), with 
        $L_{(k)}$ replaced by $L^i_{(k)}$. 
        \STATE
        \quad \quad 
        Compute the weight by $w_i=F_{0}(t^-_i)$.
        \STATE \textbf{Output:} Modified weights $\{w_i: i \in [m]\}$.
    \end{algorithmic}
\end{algorithm}

We remark on the distinctions between Algorithm~\ref{alg:simple_weights} and Algorithm~\ref{alg:complicated_weights}. Theoretically, the weighted multiple testing procedure based on Algorithm \ref{alg:complicated_weights} can achieve asymptotic FDR and FDP control with minimal assumptions as shown in Section \ref{sec:indep_fdp}, while the validity of the alternative procedure based on Algorithm \ref{alg:simple_weights} requires more technical assumptions, as discussed in Section \ref{app:dependent-theory} of the Supplementary Material. Practically, weights obtained through both methods are very similar, leading to nearly identical empirical FDR and power performance. Hence, we recommend using Algorithm \ref{alg:simple_weights} for practical applications due to its computation efficiency. Note that throughout the paper, the alternative weights from Algorithm \ref{alg:complicated_weights} are only employed for theoretical purposes in Theorems \ref{thm:FDR_idp} and \ref{power}.

\subsection{Asymptotic validity of LASLA}\label{sec:indep_fdp}

We establish the asymptotic validity of LASLA when each $p$-value is marginally independent of other $p$-values under the null. This assumption formally outlined in~\ref{cond:marg-indep} provides a comprehensible starting point for our analysis and methodology development, and the dependent cases will be studied in Section \ref{app:dependent-theory} of the Supplementary Material.
\begin{enumerate}[label=(A\arabic*), series=A]
   \item  \label{cond:marg-indep}
   $P_i$ is marginally independent of $\pmb P_{-i}=\{P_j, j\in[m], j\neq i\}$ under $H_{0,i}$,  for $i \in [m]$.
\end{enumerate}

By convention, we assume that the auxiliary variables only affect the alternative distribution but not the null distribution of the primary statistics. 
{Furthermore, we assume that $\theta_j$ has no impact on the null distribution of $T_i$ for $i\neq j$, i.e. $\pr(T_i \leq t \mid \theta_i =0, \theta_j) = \pr(T_i \leq t \mid \theta_i =0) = F_0(t)$. This assumption is mild and can be fulfilled by the class of structured probabilistic models (e.g. \citealp{Goodfellow-et-al-2016}). 
}

We start with the theoretical analysis on the data-driven estimator $\pi_i$, as its consistency is needed for valid FDR control.
{Let $\pmb{D}_i$ be the $i$th column of the observed distance matrix $\pmb{D}$, and the corresponding observed network of the indices $i \in [m]$ is considered as a \textit{partial} network. In contrast, we refer to the network as a \textit{full} network when the nodes encompass the entire population under the oracle setting. The associated distance matrix is denoted as $\mathcal{D}$.
Define $\cD_i$ to be the collection of distances from all nodes in the entire population to index $i$.
Adopting the fixed-domain asymptotics in \cite{stein1995fixed},
we assume that $\pmb D_i\rightarrow\cD_i$ uniformly for all $i \in [m]$ as $m\rightarrow \infty$, 
where $\cD_i$ is a continuous\footnote[1]{The assumption on the continuity of $\cD_i$ can be relaxed. For example, our theory can be easily extended to the case where $\cD_i$ is the union of disjoint continuous domains.} finite domain (with respect to coordinate $i$) in $\mathbb{R}$ with positive measure, and each $d\in \cD_i$ is a distance and $0\in \cD_i$.
Note that, $\pmb D_i$ and $\cD_i$ can be viewed as collections of distances respectively measured from {partial and full networks} with $\pmb D_i\subset\cD_i$.
The details on the theoretical analysis employing fixed-domain asymptotics, i.e., $\pmb D_i\rightarrow\cD_i$ as $m\rightarrow \infty$, are relegated to Section \ref{app:proof} in the Supplementary Material.}
The next assumption requires that,  in light of full network information, the distributional quantities vary smoothly in the vicinity of location $i$.

\begin{enumerate}[resume*=A]
    \item \label{A1} For all $i,j$, $\pr(P_j > \tau | \cD_j, D_{ij} = x)$ is continuous at $x$, and has bounded first and second derivatives.
\end{enumerate}

It is important to note that marginally independent $p$-values will become dependent conditional on auxiliary data. The next assumption generalizes the commonly used ``weak dependence'' notion in \cite{Sto03}. It requires that most of the neighborhood $p$-values (conditional on auxiliary data) do not exhibit strong pairwise dependence. 

\begin{enumerate}[resume*=A]
   { \item \label{A2} $\Var\left(\sum_{j \in \cN_i}K_h(D_{ij})\II \cbrac{P_j>\tau} | \pmb D\right) \leq C\sum_{j \in \cN_i}\Var\left(K_h(D_{ij})\II \cbrac{P_j>\tau} | \pmb D_j\right)$}
for some constant $C>1$, for all $i\in[m]$.
\end{enumerate}

\begin{remark}\label{rmk:bw-relaxation}
Though marginal independence is assumed, we allow $p$-values to be weakly dependent conditional on auxiliary data. Additionally, Condition~\ref{A2} can be further relaxed by selecting a larger bandwidth $h$ for the kernel $K_h(\cdot)$. For instance, if we choose $h \sim m^{-1/6}$, a common bandwidth selection, then by the proof of Proposition~\ref{prop_pi}, Assumption \ref{A2} can be relaxed to: for some constant $C' > 1$, and for all $i \in [m]$, $\Var\left(\sum_{j \in \cN_i} K_h(D_{ij}) \II \cbrac{P_j > \tau} | \pmb{D}\right) \leq m^{c} C' \sum_{j \in \cN_i} \Var\left(K_h(D_{ij}) \II \cbrac{P_j > \tau} | \pmb{D}_j\right)$ for $c < 5/6$. This relaxation permits the $p$-values to be highly correlated given the side information.
\end{remark}

 {The next proposition establishes the convergence of $\pi_i$ (estimated by \eqref{eq:pi-est} with summation taken over $\cN_i$) to an intermediate quantity $\pi_i^{\tau}$:
$$
    \pi_i^{\tau} = 1 - \frac{\pr(P_i > \tau  \mid \mathcal{D}_i)}{1-\tau}, \quad 0 < \tau < 1,
$$
which offers a good approximation of $\pi_i^*$ with a suitably chosen $\tau$. Intuitively, when $\tau$ is large, null $p$-values are dominant in the tail area $[\tau, 1]$. Hence $\pr(P_i > \tau \mid \mathcal{D}_i)/(1-\tau)$ approximates the overall null proportion.
}
\begin{proposition}\label{prop_pi}
   Recall that $|\cN_i| = m^{1-\epsilon}$.  Under Assumptions \ref{A1} and \ref{A2}, if { $m^{-1} \ll h \ll m^{-\epsilon}$}, we have, uniformly for all $i \in [m]$,
$\ep\left[(\pi_i-\pi_i^{\tau})^2 | \pmb D_i \right] \rightarrow 0, \hspace{3mm} \text{as } \pmb D_i \rightarrow \cD_i.$
\end{proposition}

Define $Z_i=\Phi^{-1}(1-P_i/2)$ and denote by $\Z = (Z_1,\ldots,Z_m)^\T$ .
We collect below several regularity conditions for proving the asymptotic validity of LASLA. 

\begin{enumerate}[resume*=A]
\item \label{A3}
Assume that $\sum_{i=1}^{m}\pr(\theta_i=0 | \cD_i)\geq cm$ for some constant $c>0$ and that $\Var\left[\sum_{i=1}^{m}I\{\theta_i=0\} | \cD\right] =  O(m^{1+\zeta})$ for some $0\leq \zeta<1$, where $\cD = \{\cD_i\}_{i \in [m]}$.

\item \label{A4}
Define $\mathcal{S}_\rho= \left\{i: 1\leq i\leq m, |\mu_i| \geq (\log m)^{(1+\rho)/2}\right\},$ where $\mu_i=\EE(Z_i)$. For some $\rho>0$ and $\delta>0$, $|\mathcal{S}_\rho| \geq [{1}/({\pi}^{1/2}\alpha)+\delta]({\log m})^{1/2}$, where $\pi\approx 3.14$ is a constant.
\end{enumerate}

\begin{remark}{\rm
 Condition \ref{A3} assumes that the model is sparse and that $\{\theta_i\}_{i=1}^m$ are not perfectly correlated (conditional on auxiliary data).
 Condition \ref{A4} requires that there are a slowly growing number of signals having magnitude of the order $\{(\log m)/n\}^{(1+\rho)/2}$ if $n$ samples are employed to obtain the $p$-value. 
}\end{remark}

Let $t^w = P^w_{(k^w)}$, where $k^w$ is calculated based on the step-wise algorithm \eqref{eq:lasla-threshold} with weights from Algorithm \ref{alg:complicated_weights}. {Denote by $\pmb{\delta}^w \equiv \pmb{\delta}^w(t^w)=\{\delta_i^w(t^w): i \in [m]\}$ the set of decision rules, where $\delta_i^w(t^w)=\II \cbrac{P^w_i \leq t^w}$.} Then the FDP and FDR of LASLA are respectively given by
\[
\text{FDP}(\pmb{\delta}^w) = \frac{\sum_{i=1}^m(1-\theta_i)\II\cbrac{P_i^w \leq t^w}}{\max\{\sum_{i=1}^m \II\cbrac{P_i^w \leq t^w},1\}}, \text{ and } {\text{FDR} }= \EE\left\{\text{FDP}(\pmb{\delta}^w)\right\}.
\]
The next theorem states that LASLA controls both the FDP and FDR at the nominal level asymptotically. 

\begin{theorem}\label{thm:FDR_idp}
Under Assumptions~\ref{cond:marg-indep}, \ref{A3}, \ref{A4} and the conditions in Proposition \ref{prop_pi}, we have for any $\varepsilon>0$:
\begin{align*}
    \limsup\limits_{\pmb D_i \rightarrow \cD_i, \forall i} {\text{FDR}} \leq \alpha, \hspace{2mm} and \hspace{2mm} \lim\limits_{\pmb D_i \rightarrow \cD_i, \forall i}\prob{\text{FDP}(\pmb{\delta}^w) \leq \alpha+\varepsilon} =1. 
\end{align*}
\end{theorem}


In Section~\ref{app:dependent-theory} of the Supplementary Material, we extend our study to examine the asymptotic control of FDP and FDR for dependent $p$-values. This analysis is crucial, as it more accurately reflects real-world applications, such as the motivating GWAS example, where complex dependency structures are common. We demonstrate that with additional sufficient regularity conditions on the density functions of the primary statistics, LASLA remains theoretically valid for weakly dependent $p$-values. Numerical simulations in Section~\ref{sec:simulation-dependent} further confirm that LASLA performs robustly across various types of dependencies.

\subsection{Asymptotic power analysis}\label{sec:power}

This section provides a theoretical analysis to demonstrate the benefit of the proposed weighting strategy. 
 To simplify the analysis, we assume that the distance matrix $\cD$ of the full network is known. 

Denote by $\pmb{\delta}^v(t)=\{\delta_i^v(t): i \in [m]\}$ a class of testing rules based on weighted $p$-values, where $\delta_i^v(t)=\II \cbrac{P^v_i \leq t}$, $P^v_i = P_i/v_i$ is the weighted $p$-value, and $v_i$ is the pre-specified weight. 
It can be shown that (e.g. Proposition 2 of \cite{Caietal20}) under mild conditions, the FDR of $\pmb{\delta}^v(t)$ can be written as $\text{FDR}\{\pmb{\delta}^v(t)\}=Q^v(t|\cD)+o(1)$, where
\begin{align*}
Q^v(t|\cD)=\frac{\sum_{i=1}^m(1 - \pi_i^*)v_it}{\sum_{i=1}^m(1 - \pi_i^*)v_it+\sum_{i=1}^m\pi_i^*F_{1i}^*(v_it |\cD_i)}
\end{align*}
corresponds to the limiting value of the FDR. In what follows, we omit the conditioning on $\cD$ for the simplicity of notation. 
The power of $\pmb{\delta}^v(t)$ can be evaluated using the expected number of true positives
{with threshold $t$: $\Psi^v(t) = \sum_{i=1}^m\pi_i^*F_{1i}^*(v_it).$} 

Let $\tilde w_i  = w_i \brac{\sum_{j=1}^m (1 - \pi_j^*)}/\brac{\sum_{j=1}^m (1 - \pi_j^*) w_j}$, where $w_i$'s are the modified weights from Algorithm \ref{alg:complicated_weights}. It is easy to see that LASLA and $\pmb{\delta}^{\tilde w}(t)$ share the same ranking of hypotheses. The goal is to compare the LASLA weights $(v_i=\tilde w_i: i \in [m])$ with the naive weights $\{v_i=1: i \in [m]\}$.  
Denote by $t_o^v=\sup\{t:Q^v(t) \leq \alpha\}$ the oracle threshold for the $p$-values with generic weights $\{v_i: i \in [m]\}$. The oracle procedure with LASLA weights and the (unweighted) oracle $p$-value procedure \citep{GenWas02} are denoted by $\delta^{\tilde w}(t_o^{\tilde w})$ and $\delta^{1}(t_o^1)$, respectively.

Next, we discuss some assumptions needed in our power analysis. The first condition states that weights should be ``informative'' in the sense that on average, small/large $\pi_i^*$ correspond to small/large $w_i$. A similar assumption has been used in \cite{Genetal06}. 

\begin{enumerate}[resume*=A]
\item \label{cond:informative} 
The oracle-assisted weights satisfy
$$\frac{\sum_{i=1}^m(1 - \pi_i^*)}{\sum_{i=1}^m(1 - \pi_i^*)w_i} \cdot \frac{\sum_{i=1}^m\pi_i^*}{\sum_{i=1}^m\pi_i^*w_i^{-1}} \geq 1.$$
\end{enumerate}

The second condition is concerned with the shape of the alternative $p$-value distributions. When the densities are homogeneous, i.e. $F_{1i}^*(t)\equiv F_1^*(t)$, it reduces to the condition that $x \rightarrow F_{1}^*(t/x)$ is a convex function, which is satisfied by commonly used density functions \citep{Huetal10, Caietal20}.

\begin{enumerate}[resume*=A]
\item \label{cond:shape}
For any $0\leq a_i\leq 1$, $\min_{i \in [m]}\tilde w_i^{-1} \leq x_i \leq \max_{i \in [m]}\tilde w_i^{-1}$ and $t_o^{1}/\min_{i \in [m]}\tilde w_i^{-1}\leq 1$, 
\begin{align*}
   \sum_{i=1}^m a_i F_{1i}^*\paren*{\frac{t}{x_i}} \geq \sum_{i=1}^m a_i F_{1i}^*\paren*{ \frac{\sum_{j=1}^m a_j t}{\sum_{j=1}^m a_j x_j}}. 
\end{align*}
\end{enumerate}

The above two conditions are mild in many practical settings. For example, we checked that both are easily fulfilled in all our simulation studies with the proposed LASLA weights. The next theorem provides insights into why the weighting strategy used in LASLA provides power gain, as we shall see in our numerical studies.

\begin{theorem}\label{power}
Assume Conditions \ref{cond:marg-indep}, \ref{cond:informative} and {\ref{cond:shape}} hold. Then
    (a) $Q^{\tilde w}(t_o^1) \leq Q^1(t_o^1) \leq \alpha$; and 
    (b) $\Psi^{\tilde w}(t_o^{\tilde w}) \geq \Psi^{\tilde w}(t_o^1) \geq \Psi^1(t_o^1)$.
\end{theorem}

The theorem implies that (a) if the same threshold $t_o^1$ is used, then $\delta^{\tilde w}(t_o^1)$ has smaller FDR and larger power than $\delta^{1}(t_o^1)$; (b) the thresholds satisfy $t_o^{\tilde w}\geq t_o^1$. Since $\Psi^v(t)$ is non-decreasing in $t$, we conclude that $\delta^{\tilde w}(t_o^{\tilde w})$ (oracle procedure with LASLA weights) dominates $\delta^{\tilde w}(t_o^1)$ and hence $\delta^{1}(t_o^1)$ (unweighted oracle $p$-value procedure) in power.

\section{Simulation} 
\label{sec:simulation}

This section considers a model that mimics the scenario with network structured side information. Additional simulation results for high-dimensional regression, latent variable model, multiple auxiliary samples, and the implementation details including parameter tuning are relegated to Section \ref{app:numeric_results} of the Supplementary Material. Simulations with dependent data can be found in Section~\ref{sec:simulation-dependent}. Software implementing the algorithms and replicating all data experiments are available online at \url{https://github.com/ZiyiLiang/r-lasla}.

For $i \in [m]$, let $\theta_i \overset{\text{ind}}{\sim} \text{Bernoulli}(0.1)$ denote the existence or absence of the signal at index $i$. 
The primary data $\pmb T = (T_i: i\in [m])$ are independently generated as $T_i \sim (1-\theta_i)N(0,1)+\theta_i N(\mu_1,1)$ with $\mu_1$ controlling the signal strength.
The distance matrix $\pmb D =(D_{ij})_{1 \leq i,j \leq m}$ follows
$D_{ij} \sim I_{\{\theta_i=\theta_j\}}|N(\mu_2,0.7)|+I_{\{\theta_i \neq \theta_j\}}|N(1,0.7)|,$
where $0\leq \mu_2 \leq 1$ controls the informativeness of the distance matrix. Intuitively, if $\theta_i = \theta_j$, then $D_{ij}$ should be relatively small. We investigate two settings. 
\begin{itemize}
\item
\text{Setting 1}: Fix $\mu_2=0$, $m=1200$, vary $\mu_1$ from 2.5 to 3 by 0.1; 
\item
\text{Setting 2}: Fix $\mu_1=3$, $m=1200$, vary $\mu_2$ from 0 to 1 by 0.2. $\pmb D$ becomes less informative as $\mu_2$ gets closer to 1.  
\end{itemize}

\begin{figure*}[ht]
  \centering
  \includegraphics[width=0.85\linewidth]{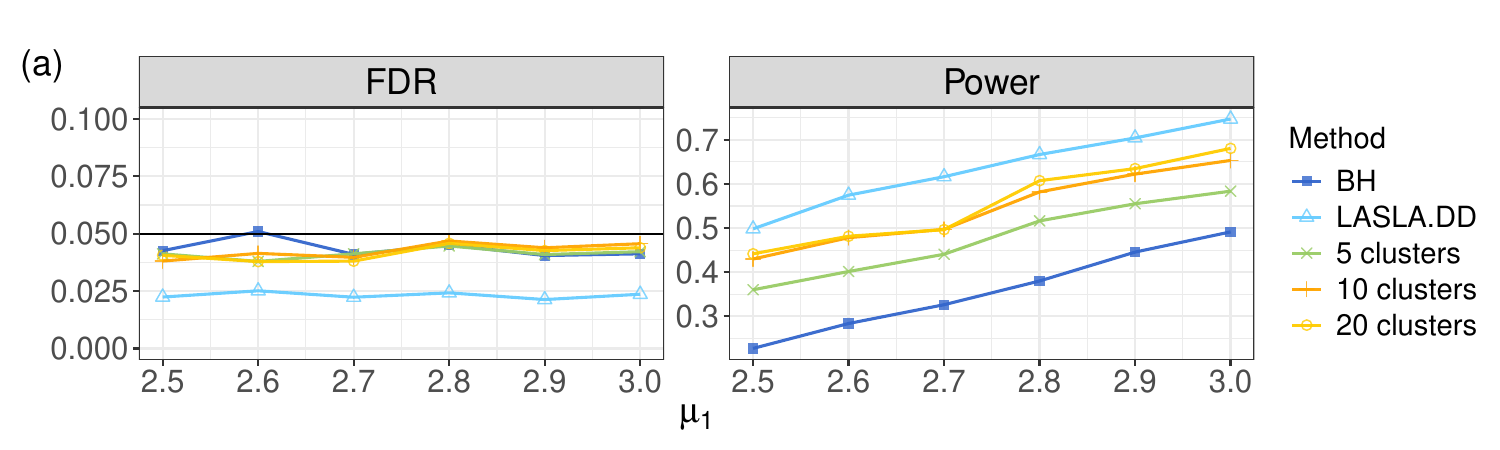}
  \includegraphics[width=0.85\linewidth]{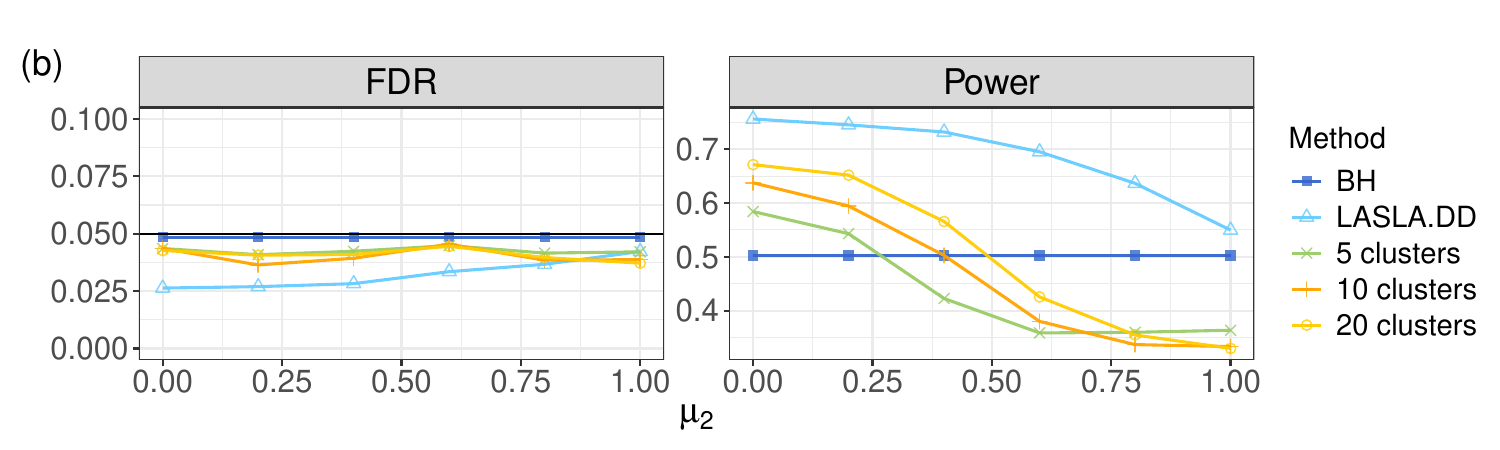}
  \caption{Empirical FDR and power comparison for data-driven LASLA, AdaPT and BH. (a): Setting 1: increasing signal strength $\mu_1$; (b): Setting 2: decreasing informativeness of network side information (controlled by $\mu_2$).}
  \label{fig:network}
\end{figure*}

Although existing structured multiple testing methods cannot directly incorporate network side information, \citet{Yurko2020PNAS} proposes a gradient-boosted trees implementation of the AdaPT procedure \citep{LeiFit18}, which indirectly utilizes network information by preprocessing it through hierarchical clustering. For each index $i$, they generate a vector of indicators denoting cluster memberships, which is then used as side information. In our simulation, we adopt this strategy by clustering the distance matrix $\pmb{D}$ into 5, 10, and 20 clusters, respectively. We then compare the data-driven LASLA (LASLA.DD) with the clustering-based AdaPT approach and the vanilla BH method, which disregards auxiliary information.  The simulation results, averaged over 100 randomized datasets, are summarized in Figure~\ref{fig:network}.

All methods successfully control the FDR at the nominal level. However, in both settings, LASLA demonstrates a substantial power advantage over both the BH and AdaPT methods. In addition, an incorrect choice of the number of clusters for the AdaPT method can lead to either a substantial computational burden or a significant loss of power.

In Setting 2, we evaluate the performance of LASLA as the informativeness of $\pmb D$ varies. As $\mu_2$ approaches 1, $\pmb D$ becomes less informative. Notably, even when $\pmb D$ becomes completely non-informative ($\mu_2=1$), LASLA still outperforms BH. This is because LASLA effectively captures the asymmetry within the alternative distribution of the primary statistics. This finding is consistent with \cite{SunCai07}, which shows that the lfdr (\citealp{Efr01}) procedure dominates BH in power. In contrast, the performance of the AdaPT procedure heavily depends on the quality of the side information. This phenomenon is consistently observed in data-sharing high-dimensional regression and latent variable settings, as shown in Figures~\ref{fig:reg}-\ref{fig:latent} in the Supplementary Material. 

Furthermore, we extend this network simulation to dependent scenarios in Sections~\ref{sec:block-dependency}-\ref{sec:random-dependency}, mimicking the potential dependency structure of GWAS dataset. The results demonstrate that LASLA's performance remains robust across various types of dependencies.

\section{Real Data Applications}
\label{sec:application}
\subsection{Detecting T2D-associated SNPs with auxiliary data from linkage analysis}\label{sec:application-gwas}
This section focuses on conducting association studies of Type 2 diabetes (T2D), a prevalent metabolic disease with strong genetic links. Our primary goal is to identify SNPs associated with T2D in diverse populations. We construct the distance matrix from LD information to gain valuable insights into the genetic basis of complex diseases as previously described in Section \ref{sec:framework}.

\citet{SNPs} performs a meta-analysis to combine 23 studies on a total of 77,418 individuals with T2D and 356,122 controls. For illustration purposes, we randomly choose $m=5000$ SNPs from Chromosome 6 to be the target of inference. Primary statistics are the $z$-values provided in \cite{SNPs} and the auxiliary LD matrix is constructed by the genetic analysis tool \texttt{Plink} from the 1000 Genomes (1000G) Phase 3 Database. It is important to note that the primary and auxiliary data are collected from different populations and are not matched in dimension.

We apply BH and LASLA at different FDR levels and compare them with the Bonferroni correction which is commonly used in GWAS to control Family-wise Error Rate (FWER). The numbers of rejections by different methods are summarized in Tables \ref{snps_table} and \ref{bonferroni}.  Both BH and LASLA are more powerful than the Bonferroni method. Moreover, at the same FDR level, LASLA makes notably more rejections than BH, and the discrepancy becomes larger as the nominal FDR level increases.

\begin{table}[ht]
\caption{\label{snps_table}
Number of rejections by different methods}
\centering
\setlength\tabcolsep{30pt}
\begin{tabular}{@{}lllll@{}}
\toprule
FDR           & 0.001 & 0.01 & 0.05 & 0.1 \\ \midrule
\text{BH}   & 35    & 61   & 128  & 179 \\
\text{LASLA} & 36    & 101   & 184  & 271 \\ \bottomrule
\end{tabular}
\end{table}

\begin{table}[ht]
\caption{\label{bonferroni}
Number of rejections by Bonferroni Correction}
\centering
\setlength\tabcolsep{28.5pt}
\begin{tabular}{@{}lllll@{}}
\toprule
FWER       & 0.001 & 0.01 & 0.05 & 0.1 \\ \midrule
\text{Bonferroni} & 21   & 29   & 35   & 43 \\ \bottomrule
\end{tabular}
\end{table}

\begin{figure}[ht]
    \centering
    \includegraphics[scale=0.7]{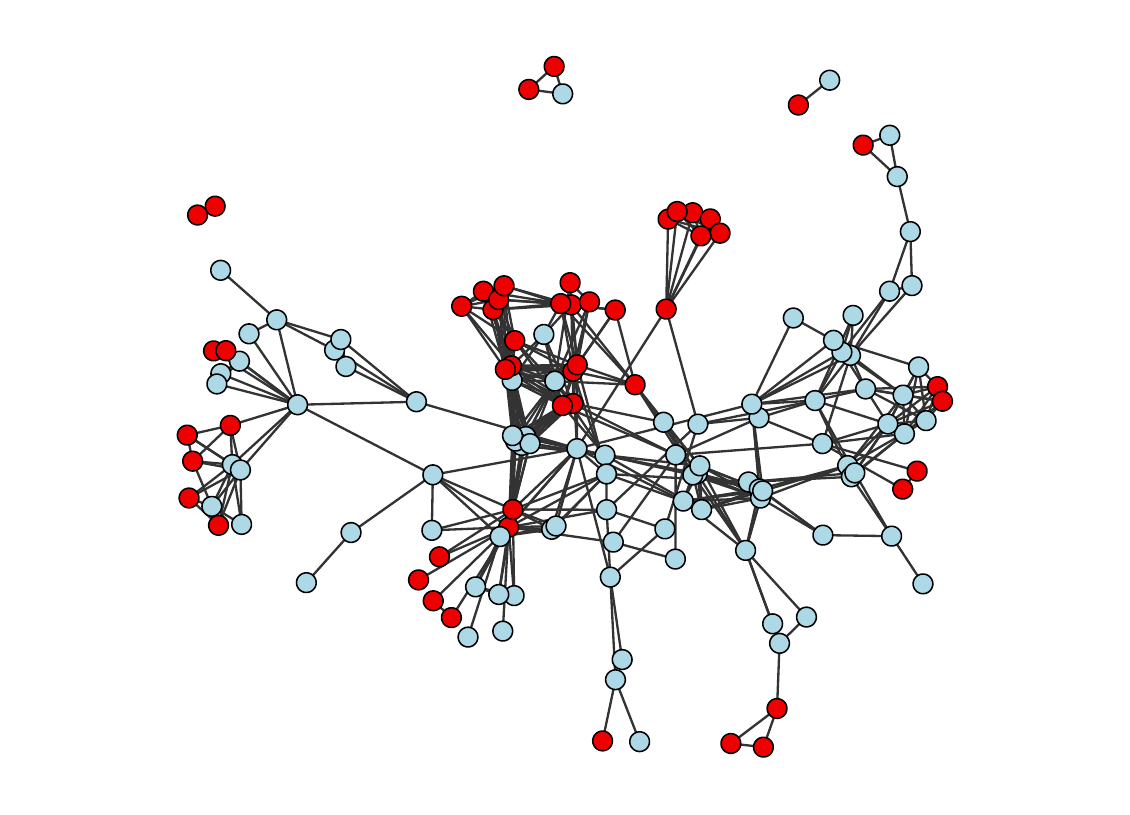}
    \caption{Sub-network identified by LASLA (all nodes). BH only detects the blue nodes.}
    \label{fig:snps}
\end{figure}

To illustrate the power gain of LASLA over BH, we visualize the rejected hypotheses in Figure \ref{fig:snps}. 
Red nodes in the figure present SNPs detected by LASLA but not by BH at FDR level 0.05. Nodes connected by an edge are in linkage disequilibrium. The graph highlights LASLA's ability to leverage the LD matrix's network structure for inference, leading to the identification of clusters of SNPs in LD. In contrast, BH could potentially miss important variants. Notably, LASLA detects T2D-risk variants within the gene CCHCR1, a candidate gene for T2D reported by \cite{brenner2020}.

\subsection{Detecting significant brain regions in ADHD patients using fMRI data}\label{sec:fmri}

In this section, we focus on identifying significant brain regions in patients with Attention Deficit Hyperactivity Disorder (ADHD) using functional Magnetic Resonance Imaging (fMRI) data, incorporating the spatial distance network as side information. The dataset, obtained from the ADHD-200 Sample Initiative, was preprocessed by the Neuro Bureau and is publicly available at \url{http://neurobureau.projects.nitrc.org/ADHD200/Data.html}.

The dataset includes 931 individuals, comprising 356 diagnosed with ADHD and 575 neurotypical controls. Following \citet{Li2017parsimonious}, we downsize the original MRI images, with a resolution of $256\times198\times256$, to a lower resolution of $30\times36\times30$ by aggregating neighboring pixels into larger blocks. This additional preprocessing step provides two key advantages. First, it enhances the signal-to-noise ratio and reduces potential misalignment issues during measurement, thereby improving the accuracy of the analysis. Second, it effectively reduces dependency by capturing localized correlations within each block.

\begin{figure}[ht]
    \centering
    \begin{minipage}[b]{0.49\textwidth}  
        \centering
        \includegraphics[width=\textwidth]{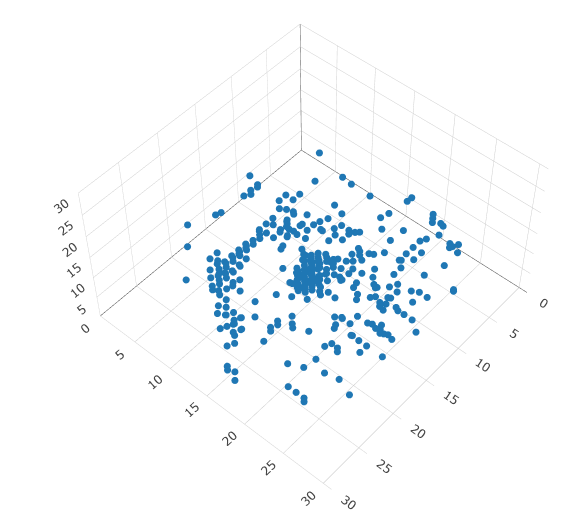}  
        \text{(a) BH rejection regions}  
    \end{minipage}
    \hfill
    \begin{minipage}[b]{0.49\textwidth}  
        \centering
        \includegraphics[width=\textwidth]{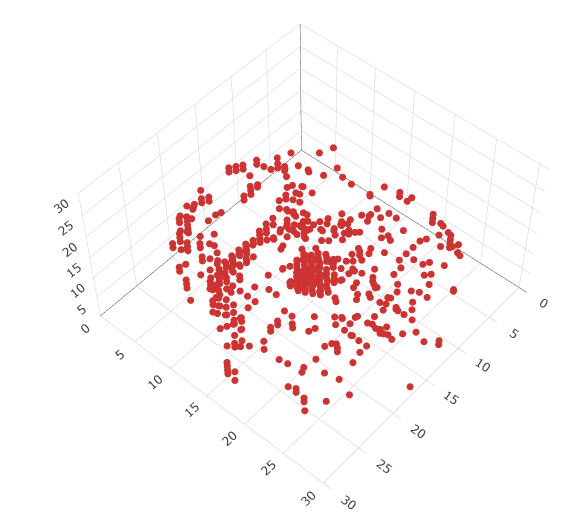}  
        \text{(b) LASLA rejection regions}  
    \end{minipage}
    
    \caption{Differential brain regions between ADHD patients and controls identified by BH and LASLA at an FDR level of 0.05.}
    \label{fig:fmri}
\end{figure}

Next, we perform two-sample $t$-tests on the aggregated blocks to compare the two groups, using a normal approximation to calculate the $p$-values. The results are summarized in Figure~\ref{fig:fmri}.  LASLA reveals a clearer pattern of brain regions that exhibit significant differences between individuals with ADHD and those without. Specifically, the BH procedure identified 349 regions, while LASLA detected 540 at an FDR level of 0.05.

\section*{Acknowledgments}
The research of Yin Xia was supported in part by the National Natural Science Foundation of China (Grant No. 12331009, 12022103).
The research of Tony Cai was supported in part by the National Science Foundation (Grant DMS-2413106) and the National Institutes of Health (Grants R01-GM129781 and R01-GM123056).

\printbibliography

@book{Goodfellow-et-al-2016,
    title={Deep Learning},
    author={Ian Goodfellow and Yoshua Bengio and Aaron Courville},
    publisher={MIT Press},
    note={\url{http://www.deeplearningbook.org}},
    year={2016}
}

@article{Casetal20,
  title={On spike and slab empirical Bayes multiple testing},
  author={Castillo, Isma{\"e}l and Roquain, {\'E}tienne},
  journal={Ann. Statist.},
  volume={48},
  number={5},
  pages={2548--2574},
  year={2020},
  publisher={Institute of Mathematical Statistics}
}

@article{Penetal11,
  title={Power-enhanced multiple decision functions controlling family-wise error and false discovery rates},
  author={Pe{\~n}a, Edsel A and Habiger, Joshua D and Wu, Wensong},
  journal={Ann. Statist.},
  volume={39},
  number={1},
  pages={556},
  year={2011},
  publisher={NIH Public Access}
}

@article{HelRos19,
  title={Optimal control of false discovery criteria in the two-group model},
  author={Heller, Ruth and Rosset, Saharon},
  journal={J. Roy. Statist. Soc. B},
  volume={83},
  number={1},
  pages={133--155},
  year={2021},
  publisher={Oxford University Press}
}

@article{Caietal20,
	author   = {Cai, T. Tony and Sun, Wenguang and Xia, Yin},
	doi      = {10.1080/01621459.2020.1859379},
	issn     = {0162-1459},
	journal  = {J. Am. Statist. Assoc.},
	pages    = {1370--1383},
	title    = {{LAWS: A Locally Adaptive Weighting and Screening Approach to Spatial Multiple Testing}},
	url      = {https://www.tandfonline.com/doi/full/10.1080/01621459.2020.1859379},
	volume  = {117},
	year     = {2022}
}

@article{Lynetal17,
  title={The control of the false discovery rate in fixed sequence multiple testing},
  author={Lynch, Gavin and Guo, Wenge and Sarkar, Sanat K and Finner, Helmut and others},
  journal={Electron. J. Stat.},
  volume={11},
  number={2},
  pages={4649--4673},
  year={2017},
  publisher={The Institute of Mathematical Statistics and the Bernoulli Society}
}

@article{RenCan20,
  title={Knockoffs with side information},
  author={Ren, Zhimei and Cand{\`e}s, Emmanuel},
  journal={Ann. Appl. Stat.},
  volume={17},
  number={2},
  pages={1152--1174},
  year={2023},
  publisher={Institute of Mathematical Statistics}
}

@article{Ignetal16,
  title={Data-driven hypothesis weighting increases detection power in genome-scale multiple testing},
  author={Ignatiadis, Nikolaos and Klaus, Bernd and Zaugg, Judith B and Huber, Wolfgang},
  journal={Nat. Methods},
  volume={13},
  number={7},
  pages={577},
  year={2016},
  publisher={Nature Publishing Group}
}

@article{Basetal18,
  title={Weighted False Discovery Rate Control in Large-Scale Multiple Testing},
  author={Basu, Pallavi and Cai, T Tony and Das, Kiranmoy and Sun, Wenguang},
  journal={J. Am. Statist. Assoc.},
  volume={113},
  number={523},
  pages={1172--1183},
  year={2018},
  publisher={Taylor \& Francis}
}

@article{IgnHub20,
  title={Covariate powered cross-weighted multiple testing},
  author={Ignatiadis, Nikolaos and Huber, Wolfgang},
  journal={J. Roy. Statist. Soc. B},
  volume={83},
  number={4},
  pages={720--751},
  year={2021},
  publisher={Oxford University Press}
}

@article{Fuetal19,
  title={Heteroscedasticity-adjusted ranking and thresholding for large-scale multiple testing},
  author={Fu, Luella and Gang, Bowen and James, Gareth M and Sun, Wenguang},
  journal={J. Am. Statist. Assoc.},
  volume={117},
  number={538},
  pages={1028--1040},
  year={2022},
  publisher={Taylor \& Francis}
}

@article{Kru87,
  title={A valuation of state of object based on weighted Mahalanobis distance},
  author={Krusi{\'n}ska, E},
  journal={Pattern Recognit.},
  volume={20},
  number={4},
  pages={413--418},
  year={1987},
  publisher={Elsevier}
}

@article{Caietal19,
  title={{CARS}: {C}ovariate assisted ranking and screening for large-scale two-sample inference (with discussion)},
  author={Cai, T. Tony and Sun, Wenguang and Wang, Weinan},
  journal={J. Roy. Statist. Soc. B},
  year={2019},
  volume={81},
  pages={187--234}
}

@article{LiBar19,
  title={Multiple testing with the structure-adaptive Benjamini--Hochberg algorithm},
  author={Li, Ang and Barber, Rina Foygel},
 JOURNAL = {J. R. Stat. Soc. B},
  FJOURNAL = {Journal of the Royal Statistical Society. Series B.
              Statistical Methodology},
  volume={81},
  number={1},
  pages={45--74},
  year={2019},
  publisher={Wiley Online Library}
}

@article{Leietal17,
   title={A general interactive framework for false discovery rate control under structural constraints},
   volume={108},
   ISSN={1464-3510},
   url={http://dx.doi.org/10.1093/biomet/asaa064},
   DOI={10.1093/biomet/asaa064},
   number={2},
   journal={Biometrika},
   publisher={Oxford University Press (OUP)},
   author={Lei, Lihua and Ramdas, Aaditya and Fithian, William},
   year={2020},
   pages={253–267} }

@article{LeiFit18,
  title={AdaPT: an interactive procedure for multiple testing with side information},
  author={Lei, Lihua and Fithian, William},
  JOURNAL = {J. R. Stat. Soc. B},
  FJOURNAL = {Journal of the Royal Statistical Society. Series B.
              Statistical Methodology},
  volume={80},
  number={4},
  pages={649--679},
  year={2018},
  publisher={Wiley Online Library}
}

@article{Xiaetal19,
	Author = {Xia, Yin and Cai, T Tony and Sun, Wenguang},
	Title = {{GAP: A General Framework for Information Pooling in Two-Sample Sparse Inference}},
  journal={J. Am. Statist. Assoc.},
  year={2020},
  volume={115},
  pages={1236--1250},
  publisher={Taylor \& Francis}
}

@article{Huetal10,
  title={False discovery rate control with groups},
  author={Hu, James X and Zhao, Hongyu and Zhou, Harrison H},
  journal={J. Am. Statist. Assoc.},
  year={2010},
  volume={105},
  pages={1215--1227},
  publisher={Taylor \& Francis}
}

@article{Caietal16-bio,
  title={Structured matrix completion with applications to genomic data integration},
  author={Cai, Tianxi and Cai, T Tony and Zhang, Anru},
  journal={J. Am. Statist. Assoc.},
  volume={111},
  number={514},
  pages={621--633},
  year={2016},
  publisher={Taylor \& Francis}
}

@article{Medetal10,
author = {Medina, Ignacio and Carbonell, José and Pulido, Luis and Madeira, Sara and Götz, Stefan and Conesa, Ana and Tárraga, Joaquín and Pascual-Montano, Alberto and Nogales-Cadenas, Ruben and Santoyo-Lopez, Javier and García-García, Francisco and Marba, Martina and Montaner, David and Dopazo, Joaquin},
year = {2010},
month = {07},
pages = {W210-3},
title = {Babelomics: An integrative platform for the analysis of transcriptomics, proteomics and genomic data with advanced functional profiling},
volume = {38},
journal = {Nucleic acids research},
doi = {10.1093/nar/gkq388}
}

@article{Zhoetal18,
  title={{Integrative DNA copy number detection and genotyping from sequencing and array-based platforms}},
  author={Zhou, Zilu and Wang, Weixin and Wang, Li-San and Zhang, Nancy Ruonan},
  journal={Bioinformatics},
  volume={34},
  number={14},
  pages={2349--2355},
  year={2018},
  publisher={Oxford University Press}
}

@article{FosSti08,
  title={$\alpha$-investing: a procedure for sequential control of expected false discoveries},
  author={Foster, Dean P and Stine, Robert A},
 JOURNAL = {J. R. Stat. Soc. B},
  FJOURNAL = {Journal of the Royal Statistical Society. Series B.
              Statistical Methodology},
  volume={70},
  number={2},
  pages={429--444},
  year={2008},
  publisher={Wiley Online Library}
}

@article{RoeWas09,
  title={Genome-wide significance levels and weighted hypothesis testing},
  author={Roeder, Kathryn and Wasserman, Larry},
  journal={Statistical science: a review journal of the Institute of Mathematical Statistics},
  volume={24},
  number={4},
  pages={398},
  year={2009},
  publisher={NIH Public Access}
}

@article{RoqVan09,
  title={Optimal weighting for false discovery rate control},
  author={Roquain, Etienne and Van De Wiel, Mark A},
  journal={Electron. J. Stat.},
  volume={3},
  pages={678--711},
  year={2009},
  publisher={The Institute of Mathematical Statistics and the Bernoulli Society}
}

@article{Sunetal15,
  title={False discovery control in large-scale spatial multiple testing},
  author={Sun, Wenguang and Reich, Brian J and Cai, T. T. and Guindani, Michele and Schwartzman, Armin},
  JOURNAL = {J. R. Stat. Soc. B},
  FJOURNAL = {Journal of the Royal Statistical Society. Series B.
              Statistical Methodology},
  volume={77},
  number={1},
  pages={59--83},
  year={2015},
  publisher={Wiley Online Library}
}

@article {Efr01,
    AUTHOR = {Efron, Bradley and Tibshirani, Robert and Storey, John D. and
              Tusher, Virginia},
     TITLE = {Empirical {B}ayes analysis of a microarray experiment},
   JOURNAL = {J. Amer. Statist. Assoc.},
  FJOURNAL = {J. Am. Statist. Assoc.},
    VOLUME = {96},
      YEAR = {2001},
 
     PAGES = {1151--1160},
      ISSN = {0162-1459},
     CODEN = {JSTNAL},
   MRCLASS = {62C12 (62P10)},
  MRNUMBER = {MR1946571},
}

@article {GenWas02,
    AUTHOR = {Genovese, Christopher and Wasserman, Larry},
     TITLE = {Operating characteristics and extensions of the false
              discovery rate procedure},
   JOURNAL = {J. R. Stat. Soc. B},
  FJOURNAL = {Journal of the Royal Statistical Society. Series B.
              Statistical Methodology},
    VOLUME = {\textbf{64}},
      YEAR = {2002},
     PAGES = {499--517},
      ISSN = {1369-7412},
   MRCLASS = {62F03},
  MRNUMBER = {MR1924303 (2003h:62027)},
}

@article{Genetal06,
  title={False discovery control with p-value weighting},
  author={Genovese, Christopher R and Roeder, Kathryn and Wasserman, Larry},
  journal={Biometrika},
  volume={93},
  number={3},
  pages={509--524},
  year={2006},
  publisher={Biometrika Trust},
}

@article {Sto03,
    AUTHOR = {Storey, John D.},
     TITLE = {The positive false discovery rate: a {B}ayesian interpretation
              and the {$q$}-value},
   JOURNAL = {Ann. Statist.},
  FJOURNAL = {Ann. Statist.},
    VOLUME = {31},
      YEAR = {2003},
 
     PAGES = {2013--2035},
      ISSN = {0090-5364},
     CODEN = {ASTSC7},
   MRCLASS = {62F03 (62F15 62H30)},
  MRNUMBER = {MR2036398 (2004k:62055)},
MRREVIEWER = {Mohan Delampady},
}

@article {SunCai07,
    AUTHOR = {Sun, Wenguang and Cai, T. Tony},
     TITLE = {Oracle and adaptive compound decision rules for false
              discovery rate control},
   JOURNAL = {J. Amer. Statist. Assoc.},
  FJOURNAL = {J. Am. Statist. Assoc.},
    VOLUME = {\textbf{102}},
      YEAR = {2007},
     PAGES = {901--912},
      ISSN = {0162-1459},
     CODEN = {JSTNAL},
   MRCLASS = {Database Expansion Item},
  MRNUMBER = {MR2411657},
}

@article {SNPs,
	author = {Spracklen, C.N. and Horikoshi, M. and Kim, Y.J. et al.},
	title = {Identification of type 2 diabetes loci in 433,540 East Asian individuals.},
	volume = {582},
	pages = {240-245},
	year = {2020},
	journal = {Nature},
}

@article{stein1995fixed,
author = { Michael L.   Stein },
title = {Fixed-Domain Asymptotics for Spatial Periodograms},
journal = {J. Am. Statist. Assoc.},
volume = {90},
number = {432},
pages = {1277-1288},
year  = {1995},
publisher = {Taylor & Francis},
doi = {10.1080/01621459.1995.10476632},
}

@article{brenner2020,
author = {Brenner, Laura N et al},
title = {Analysis of Glucocorticoid-Related Genes Reveal CCHCR1 as a New Candidate Gene for Type 2 Diabetes},
journal = {J. Endocr. Soc.},
volume={4},
  number={11},
  pages={bvaa121},
  year={2020},
  publisher={Oxford University Press US}
}

@article{2012schaub,
    title={Linking disease associations with regulatory information in the human genome},
  author={Schaub, Marc A and Boyle, Alan P and Kundaje, Anshul and Batzoglou, Serafim and Snyder, Michael},
  journal={Genome Res.},
  volume={22},
  number={9},
  pages={1748--1759},
  year={2012},
  publisher={Cold Spring Harbor Lab}
}

@article{bh95,
 ISSN = {00359246},
 author = {Yoav Benjamini and Yosef Hochberg},
 journal = {J. Roy. Statist. Soc. B},
 number = {1},
 pages = {289--300},
 publisher = {[Royal Statistical Society, Wiley]},
 title = {Controlling the False Discovery Rate: A Practical and Powerful Approach to Multiple Testing},
 volume = {57},
 year = {1995}
}

@ARTICLE{Joiret2019-sl,
  title     = {Confounding of linkage disequilibrium patterns in large scale
               DNA based gene-gene interaction studies},
  author    = {Joiret, Marc and Mahachie John, Jestinah M and Gusareva, Elena S
               and Van Steen, Kristel},
  journal   = {BioData Min.},
  publisher = {Springer Science and Business Media LLC},
  volume    =  {12},
  number    =  {1},
  pages     = {11},
  year      =  {2019},
  copyright = {https://creativecommons.org/licenses/by/4.0},
}

@article{Yurko2020PNAS,
author = {Ronald Yurko  and Max G’Sell  and Kathryn Roeder  and Bernie Devlin },
title = {A selective inference approach for false discovery rate control using multiomics covariates yields insights into disease risk},
journal = {Proceedings of the National Academy of Sciences},
volume = {117},
number = {26},
pages = {15028-15035},
year = {2020},
doi = {10.1073/pnas.1918862117},
URL = {https://www.pnas.org/doi/abs/10.1073/pnas.1918862117},
eprint = {https://www.pnas.org/doi/pdf/10.1073/pnas.1918862117}
}

@article{Li2017parsimonious,
author = {Lexin Li and Xin Zhang},
title = {Parsimonious Tensor Response Regression},
journal = {Journal of the American Statistical Association},
volume = {112},
number = {519},
pages = {1131--1146},
year = {2017},
publisher = {ASA Website},
doi = {10.1080/01621459.2016.1193022},
}

@article{2019ramadaspfilter,
author = {Aaditya K. Ramdas and Rina F. Barber and Martin J. Wainwright and Michael I. Jordan},
title = {{A unified treatment of multiple testing with prior knowledge using the p-filter}},
volume = {47},
journal = {The Annals of Statistics},
number = {5},
publisher = {Institute of Mathematical Statistics},
pages = {2790 -- 2821},
year = {2019},
doi = {10.1214/18-AOS1765},
URL = {https://doi.org/10.1214/18-AOS1765}
}

@book{Scott1992,
  title = {Multivariate Density Estimation: Theory,  Practice,  and Visualization},
  ISBN = {9780470316849},
  ISSN = {1940-6347},
  url = {http://dx.doi.org/10.1002/9780470316849},
  DOI = {10.1002/9780470316849},
  journal = {Wiley Series in Probability and Statistics},
  publisher = {Wiley},
  author = {Scott,  David W.},
  year = {1992},
  month = {8}
}
\newpage

\appendix
\renewcommand{\thesection}{A\arabic{section}}
\renewcommand{\theequation}{A\arabic{equation}}
\renewcommand{\thetheorem}{A\arabic{theorem}}
\renewcommand{\thecorollary}{A\arabic{corollary}}
\renewcommand{\theproposition}{A\arabic{proposition}}
\renewcommand{\thelemma}{A\arabic{lemma}}
\renewcommand{\thetable}{A\arabic{table}}
\renewcommand{\thefigure}{A\arabic{figure}}
\renewcommand{\thealgorithm}{A\arabic{algorithm}}







\section{Additional applications}\label{app:additional-applications}

LASLA has a wide range of applications aside from the network-structured data like the GWAS example discussed in the main article. In this section, we introduce two additional challenging settings:  data-sharing regression and integrative inference with multiple auxiliary data sets. In both scenarios, traditional frameworks are not applicable since the auxiliary data $\pmb U$ and the primary data $\pmb T$ do not match in dimension.

\noindent\textbf{Example 1. Data-sharing high-dimensional regression.} Suppose we are interested in identifying genetic variants associated with type II diabetes (T2D). Consider a high-dimensional regression model:
\begin{equation}\label{eq:reg_main} 
\pmb Y = \pmb \mu + \pmb X \pmb \beta +\pmb\epsilon,
\end{equation}  
where $\pmb Y=(Y_{1},\ldots, Y_{n})^{\T}$ are measurements of phenotypes, $\pmb \mu=\mu\pmb 1^{\T}$ is the intercept, with $\pmb 1^{\T}$ being a vector of ones, $\pmb\beta=(\beta_{1},\dots,\beta_{m})^{\T}$ is the vector of regression coefficients, {$\X \in \mathbb{R}^{n\times m}$} is the matrix of measurements of genomic markers, and $\eps=(\epsilon_{1}, \dots,\epsilon_{n})^\T$ are random errors. 

Both genomics and epidemiological studies have provided evidence that complex diseases may have shared genetic contributions. The power for identifying T2D-associated genes can be enhanced by incorporating data from studies of related diseases such as cardiovascular disease (CVD) and ischaemic stroke. Consider models for other studies:
\begin{equation}\label{eq:reg_side} 
{\pmb Y}^{k} = {\pmb \mu}^{k} + {\pmb X}^{k} {\pmb \beta}^{k} +{\pmb\epsilon}^{k},
\end{equation}  
where the superscript $k$ indicates that the auxiliary data are collected from disease type $k\in[K]$. The notations ${\pmb Y}^k$, $\pmb\mu^k$, $\pmb\be^k$, $\X^k$ and $\eps^k$ have similar explanations as above. 
The identification of genetic variants associated with T2D can be formulated as a multiple testing problem \eqref{eq:hypothesis}, where $\pmb\theta=(\theta_i:  i\in [m])=\{\II(\beta_{i}\neq 0):  i\in [m]\}$ is the primary parameter of interest. The primary and auxiliary data sets are $\pmb T=(\Y, \X)$ and $\pmb U=\{(\Y^k, \X^k): k \in [K]\}$, respectively. The auxiliary data $\pmb U$ can provide useful guidance by prioritizing the shared risk factors and genetic variants. 
\medskip

\noindent\textbf{Example 2. Integrative ``omics'' analysis with multiple auxiliary data sets.} The rapidly growing field of {integrative genomics} calls for new frameworks for combining various data types to identify novel patterns and gain new insights. Related examples include (a) the analysis of multiple genomic platform (MGP) data, which consist of several data types, such as DNA copy number, gene expression, and DNA methylation, in the same set of the specimen (\citealp{Caietal16-bio}); (b) the integrated copy number variation (iCNV) caller that aims to boost statistical accuracy by integrating data from multiple platforms such as whole exome sequencing (WES), whole genome sequencing (WGS) and SNP arrays (\citealp{Zhoetal18}); (c) the integrative analysis of transcriptomics, proteomics and genomic data (\citealp{Medetal10}). The identification of significant genetic factors can be formulated as \eqref{eq:hypothesis} with mixed types of auxiliary data.

\section{Forming local neighborhoods: illustrations}\label{app:local-neighborhood}

Recall that, in Section \ref{sec:intro}, LASLA first summarizes the structural knowledge in a distance matrix $\pmb D \in \mathbb{R} ^{m \times m}$ where $m$ is the number of hypotheses. The distance matrix describes the relation between each pair of hypotheses in the light of the auxiliary data. For the GWAS example detailed in Section \ref{sec:intro}, $\pmb D = (1-r^2_{ij} : i, j\in [m])$ where $r_{ij}$ measures the linkage disequilibrium between the two SNPs $i \text{ and } j$. 

In Example 1 (data-sharing regression) from Section \ref{app:additional-applications}, we can extract the structural knowledge provided by the related regression problems via Mahalanobis distance \citep{Kru87}. Specifically, let $\{\hat{\pmb\beta}^k = (\hat{\beta}^k_1, \dots, \hat{\beta}^k_m)^{\T}:  k\in [K]\}$ denote the estimation of $\{\pmb\beta^k = (\beta^k_1, \dots, \beta^k_m)^{\T}:  k\in [K]\}$. Denote by $\hat{\pmb\beta_i}=(\hat{\beta}^k_i:  k\in [K])$ the vector of estimated coefficients for the $i$th genomic marker across $K$ different studies. The distance matrix $\pmb D=(D_{ij})_{i,j\in[m]}$ is then constructed via Mahalanobis distance with $D_{ij}=({\hat{\pmb\beta}_i}- {\hat{\pmb\beta}_j}) {\hat\Sigma}_{\pmb\beta}^{-1}({\hat{\pmb\beta}_i}- {\hat{\pmb\beta}_j})^{\T}$ , where ${\hat\Sigma}_{\pmb\beta}$ is the estimated covariance matrix based on $\{\hat{\pmb\beta}_i:  i\in [m]\}$. Similarly, in Example 2 (analysis with multiple auxiliary data sets), suppose we collect a multivariate variable $\pmb U_i$ from different platforms as the side information for gene $i$, then the Mahalanobis distance can be used to construct a distance matrix $\pmb D=(D_{ij})_{i,j\in[m]}$ with $D_{ij}=({\pmb U_i}- {\pmb U_j}) {\hat\Sigma}_U^{-1}({\pmb U_i}- {\pmb U_j})^{\T}$, where $\hat\Sigma_{U}$ is the estimated covariance matrix based on the auxiliary sample $\{\pmb U_i:  i\in [m]\}$.

We emphasize that LASLA is not limited to the aforementioned examples. Most of the traditional covariate-assisted methods focus on the array-like auxiliary data $\pmb U=\{U_i: i=1, 2, \ldots\}$ that matches primary data coordinate by coordinate. LASLA can also handle this dimension-matching side information as the latter can be represented by a distance matrix $\pmb D$ through simple manipulations. Below, we provide a list of practical types of side information and their corresponding methods for constructing the distance matrix.

\begin{itemize}

\item [(a)] A vector of categorical covariates. The elements in $\pmb U$ take discrete values and the local neighborhoods can be defined as groups. With suitably chosen weights LASLA reduces to the methods considered in \cite{Huetal10}, \cite{LiBar19}, and \cite{Xiaetal19} that are developed for multiple testing with groups.  

\item [(b)] A vector of continuous covariates. We can define distance as either the absolute difference or the standardized difference in rank $D_{ij}=|\hat F_m(U_i)-\hat F_m(U_j)|$, where $\hat F_m(t)$ is the empirical CDF.

\item [(c)] Spatial locations. Such structures have been considered in, for example, \cite{Lynetal17}, \cite{LeiFit18} and \cite{Caietal20}. The locations are viewed as covariates and $D_{ij}$ is the Euclidean distance between locations $i$ and $j$.  

\item [(d)] The correlations in a network or partial correlations in graphical models. See the GWAS example discussed in Section \ref{sec:intro} of the main article.

\item [(e)] Multiple auxiliary samples. The Mahalanobis distance or its generalizations  \citep{Kru87} can be used to calculate the distance matrix $\pmb D$.     

\end{itemize}
Note that in practical applications, it could be beneficial to ``standardize" the distance matrix $\pmb{D}$; this step ensures algorithm robustness. A more comprehensive discussion on the implementation details is relegated to Section \ref{app:implementation}.

\section{Details on sparsity-adaptive weights}\label{app:sparsity-review}

Recall the definition from Section \ref{sec:oracle} that the primary statistics $T_i$ has the hypothetical mixture distribution:
\begin{align*}
    F^*_i(t) = (1-\pi_i^*)F_0(t) + \pi_i^*F^*_{1i}(t)
\end{align*}
for $i \in [m]$. The quantity $\pi_i^*$ indicates the sparsity level of signals at location $i$, and $\pi_i^*$ is allowed to be heterogeneous across $m$ testing locations.  

The key idea in existing weighted FDR procedures such as GBH \citep{Huetal10}, SABHA (\citealp{LiBar19}) and LAWS (\citealp{Caietal20}) is to construct weights that leverage $\pi_i^*$ by prioritizing the rejection of the null hypotheses in groups or at locations where signals appear to be more frequent. Specifically, SABHA defines the weight as $w^{\text{sabha}}_i = 1/(1-\pi_i^*)$, and LAWS as $w^{\text{laws}}_i = \pi_i^*/(1-\pi_i^*)$. The sparsity adaptive-weights have an intuitive interpretation. Consider the LAWS weight $w^{\text{laws}}_i$, if $\pi_i^*$ is large, indicating a higher occurrence of signals at location $i$, the weighted $p$-value $P^w_i := P_i/w^{\text{laws}}_i = (1-\pi_i^*)P_i/\pi_i^*$ will be smaller, up-weighting the significance level of hypothesis $i$. However, compared to the proposed weights,  such weighting scheme  ignores structural information in alternative distributions as discussed in Section \ref{sec:weight-comparison}.

\section{Additional numerical results with marginally independent data}\label{app:numeric_results}
In this section, we provide the numerical implementation details and collect additional simulation results for data-sharing high-dimensional regression, latent variable model and multiple auxiliary samples under the marginal independence assumption \ref{cond:marg-indep}.

\subsection{Implementation Details}
\label{app:implementation}
In all of our numerical results, the bandwidth $h$ for the kernel estimations in \eqref{eq:pi-est} and \eqref{eq:density-est} is chosen automatically by applying the \texttt{density} function with the option ``SJ-ste'' in R package \texttt{stats}. For the size of neighborhoods $m^{1-\epsilon}$, the default choice for $\epsilon$ is 0.1 for marginally independent $p$-values, while for dependent $p$-values, we set $\epsilon=0$ to comply with our FDR control theory under weak dependence in Section \ref{app:dependent-theory}. For the screening parameter $\tau$, we choose the threshold through the BH algorithm with FDR level $\alpha=0.8$. This choice ensures that the screening set $\{i\in [m]: P_i > \tau\}$ is predominantly composed of null indices. See \citet{Caietal20} for a more comprehensive discussion on the choice of $\tau$.

To enhance algorithmic robustness and numerical performance, we perform a data-driven scaling of the distance matrix $\pmb{D}$ by a constant factor $a$. A practical guideline is to ensure that the spread of entries in the scaled distance matrix $\pmb{D}/a$ is similar to that of the entries in $\pmb{T}$.  We use the interquartile range (IQR) to measure data spread, a strategy similarly employed in \citet{Scott1992}. All additional details can be found in the public repository containing all experiments implementation at \url{https://github.com/ZiyiLiang/r-lasla}. 


\subsection{Heterogeneous alternative distributions}\label{app:hetero-alt}
As highlighted in Section~\ref{sec:weight-comparison}, LASLA can handle hypothesis-specific alternative densities, unlike many popular methods that only accommodate heterogeneous sparsity levels \citep{LiBar19, Caietal20}. Building on the discussion in Section~\ref{sec:weight-comparison}, we present two additional examples involving heterogeneous data to further demonstrate LASLA's strength in leveraging hypothesis-specific information. Consistent with the analysis in Section~\ref{sec:weight-comparison}, we compare the oracle LASLA procedure with the oracle LAWS to illustrate the methodological distinctions without the influence of practical implementation.

Extending the analysis of Example \ref{example:asymmetry} in Section \ref{sec:weight-comparison}, which considers asymmetry in the sign of the signal, we explore a more challenging setting that allows for varying levels of asymmetry across hypotheses.

\begin{example}\label{example:hetero-asymmetry}\rm{
Set $F^*_{1i}(t) = \gamma_i N(3, 1)+(1-\gamma_i)N(-3, 1)$, where $\gamma_i$ controls the relative proportions of positive and negative signals for the $i$th hypothesis. Each $\gamma_i$ follows a uniform distribution over interval $[0.5-r, 0.5+r]$, i.e., $\gamma_i \sim U(0.5-r, 0.5+r)$. We vary $r$ from 0 to 0.5 by 0.1.
}
\end{example}

In practice, the heterogeneity of alternative densities can be complex, potentially involving a mixture of factors such as signal strength, signal sign, and the shape of the alternative density. In the following setting, we introduce substantial heterogeneity across each of these components:

\begin{example} \label{example:hetero-alt}
\rm{Set $F^*_{1i}(t) = \gamma_iN(\mu_i, \sigma_i^2)+(1-\gamma_i)N(-\mu_i, \sigma_i^2)$, where each parameter follows a uniform distribution: $\gamma_i \sim U(0,1); \mu_i \sim U(2.5,3.5) \text{ and } \sigma_i \sim \text U(0.1,1)$. 
}
\end{example}

\begin{figure}[ht]
  \centering
  \includegraphics[width=0.8\linewidth]{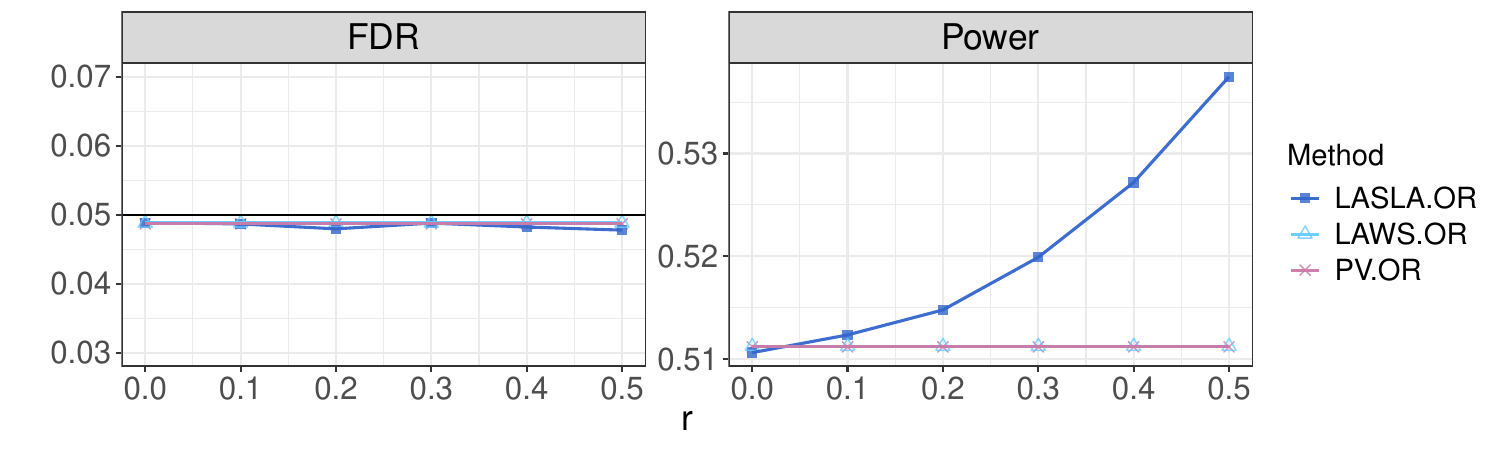}
  \caption{Comparison of oracle LASLA and benchmarks with nominal FDR level $\alpha=0.05$ under the heterogeneous asymmetry setting in Example~\ref{example:hetero-asymmetry}. Parameter $r$ controls the level of asymmetry in each individual hypothesis. All other details remain consistent with Figure~\ref{fig:sparsity-comp}.} 
  \label{fig:hetero-asym}
\end{figure}

\begin{figure}[ht]
  \centering
  \includegraphics[width=0.75\linewidth]{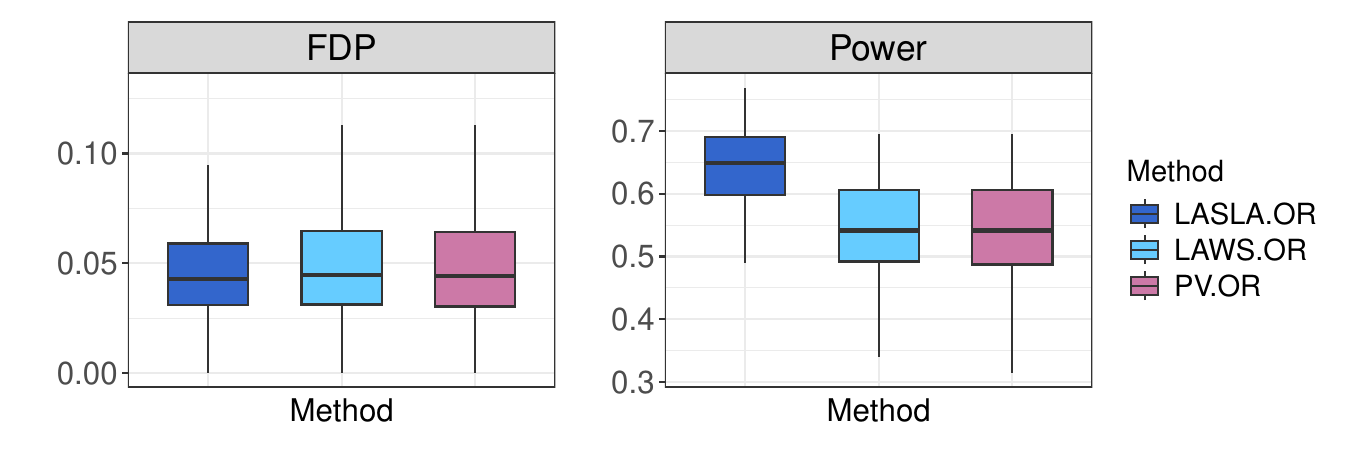}
  \caption{Comparison of oracle LASLA and benchmarks with nominal FDR level $\alpha=0.05$ under the setting in Example~\ref{example:hetero-alt}, which introduces substantial heterogeneity across hypotheses. } 
  \label{fig:hetero-alt}
\end{figure}

The results for the two examples above are summarized in Figure~\ref{fig:hetero-asym} and Figure~\ref{fig:hetero-alt}, respectively. We again observe LASLA's ability in capturing the heterogeneities, thereby improving the power across the settings.

\subsection{Data-sharing high-dimensional regression}
\label{sec:regression}
Example 2 in Section \ref{app:additional-applications} discussed how the knowledge in regression models from related studies can be transferred to improve the inference on regression coefficients from the primary model. This section designs simulation studies to illustrate the point.

Consider the regression model (\ref{eq:reg_main}) defined in Section \ref{app:additional-applications} with $\X_{ij} \sim N(0,1)$ for $i\in [n], j\in [m]$ where $\X_{ij}$ denotes the entry of $\X$ at coordinate $(i,j)$; $\epsilon_i \sim N(0,1)$ for $i \in [n]$. 
Let $\prob{\beta_i=0}=0.9$. For the non-null locations, $\beta_i \sim (-1)^u |N(\mu,0.1)|$; $u \sim \text{Bernoulli}(0.2)$. Note that signals will be more likely to take positive signs, hence asymmetric rejection rules are desired.

Models from $K$ related studies are generated by (\ref{eq:reg_side}). If the auxiliary model is closely related to the primary model, they tend to share similar coefficients. Therefore, we generate the coefficients for study $k\in [K]$ as 
$
    {\pmb \beta}^k = \pmb \beta + \pmb \sigma,
$
where each coordinate of $\pmb \sigma$ is drawn from normal distribution $N(0,\sigma^2)$. Other quantities are defined similarly as the primary model.

We compute the distance matrix $\pmb D$ using the Mahalanobis distance on the estimated coefficients as specified in Section \ref{app:local-neighborhood}. Fix $K=3, n=1000, m=800$, consider the following settings:
\begin{itemize}
    \item \text{Setting 1}: Fix $\mu=0.25$, vary the noise level $\sigma$ from 0.1 to 0.2 by 0.02.
    \item \text{Setting 2}: Fix $\sigma = 0.15$, vary the signal strength $\mu$ from 0.25 to 0.3 by 0.01.
\end{itemize}

\begin{figure*}[!htb]
  \centering
  \includegraphics[width=0.8\linewidth]{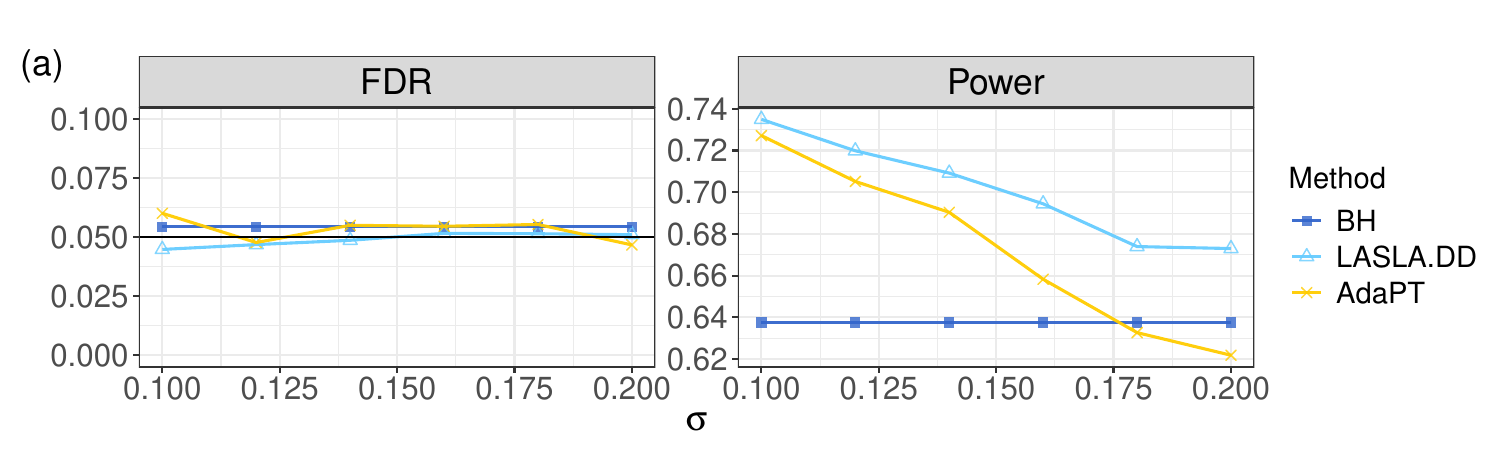}
  \includegraphics[width=0.8\linewidth]{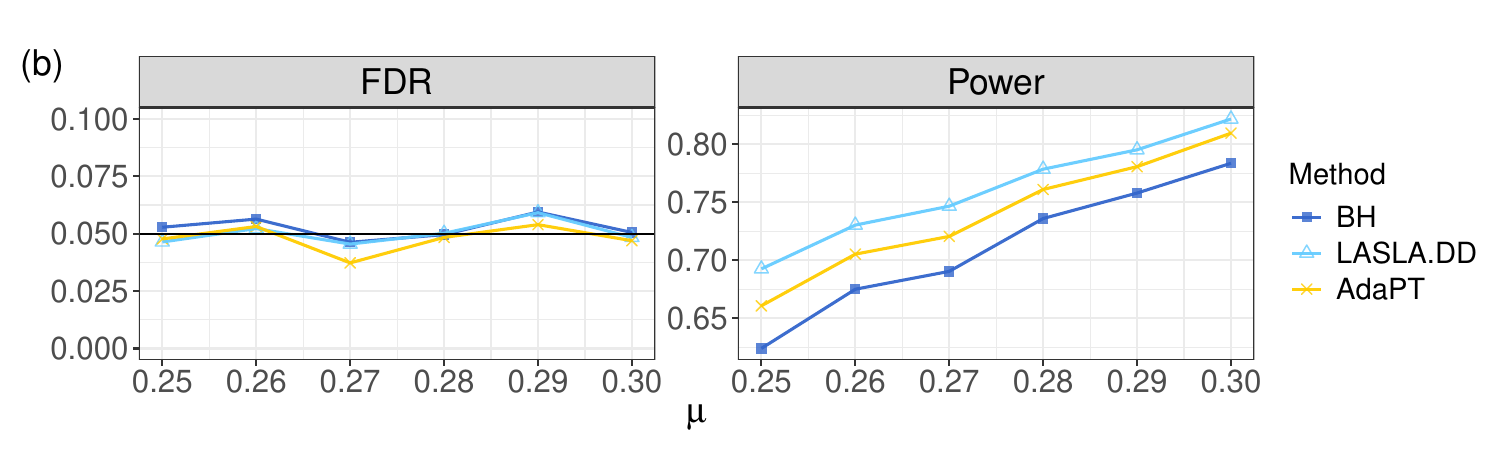}
  \caption{
 Empirical FDR and power comparison of data-driven LASLA and benchmarks at nominal FDR level $\alpha=0.05$ under the data-sharing regression setting in Section~\ref{sec:regression}. (a): Regression setting 1: increasing the noise level $\sigma$ in the auxiliary data; (b): Regression setting 2: increasing signal strength $\mu$. }
  \label{fig:reg}
\end{figure*}

We compare data-driven LASLA with BH and AdaPT method~\citep{Yurko2020PNAS}. To apply LASLA, it's essential to have knowledge of the null distribution for the test statistics. In this simulation we use the ordinary least square estimators and $T_i$ follows a $t$-distribution. Alternatively, one can explore the approach outlined in \cite{Xiaetal19}, where the test statistics follow the $N(0,1)$ distribution asymptotically. Figure \ref{fig:reg} shows that LASLA can effectively leverage the side information from related studies and outperforms both BH and AdaPT.

\subsection{Latent variable setting}
\label{latent_simu}

Suppose the primary and auxiliary data are associated with a common latent variable $\boldsymbol{\xi}=(\xi_{i}:i \in [m])$ where
$ \xi_i \sim (1-\theta_i)\Delta_0 + \theta_i N(\mu,1)$ and $\Delta_0$ is the Dirac delta function, namely, $\xi_i=0$ if $\theta_i=0$. The primary data $\pmb T=(T_i:i \in [m])$ and auxiliary data $\pmb U=(U_i: i \in [m])$ respectively follow:
\beq\label{eq:latent-model}
 T_i \sim N(\xi_i,1), \quad  U_i \sim N(\xi_i,\sigma_s^2),
\eeq
where  $\sigma_s$ controls the informativeness the auxiliary data. Our goal is to test $m$ hypotheses on $\theta_i$ as stated in (\ref{eq:hypothesis}). Fix $m=1200$ and let $\theta_i \overset{\text{ind}}{\sim} \text{Bernoulli}(0.1)$, for $i \in [m]$. We consider two settings:
\begin{itemize}
    \item \text{Setting 1}: Fix $\mu=2.5$, vary $\sigma_s$ from 0.5 to 2 by 0.25.
    \item \text{Setting 2}: Fix $\sigma_s=1$ , vary $\mu$ from 3 to 4 by 0.2.
\end{itemize}

We compute the distance matrix $\pmb D$ from the auxiliary data $\pmb U$ using the Euclidean distance, i.e. $D_{ij} = |U_i - U_j|$.
We then compare LASLA with the BH procedure, data-driven SABHA (SABHA.DD) as reviewed in Section \ref{app:sparsity-review} and AdaPT method~\citep{Yurko2020PNAS}.

\begin{figure*}[!htb]
  \centering
  \includegraphics[width=0.8\linewidth]{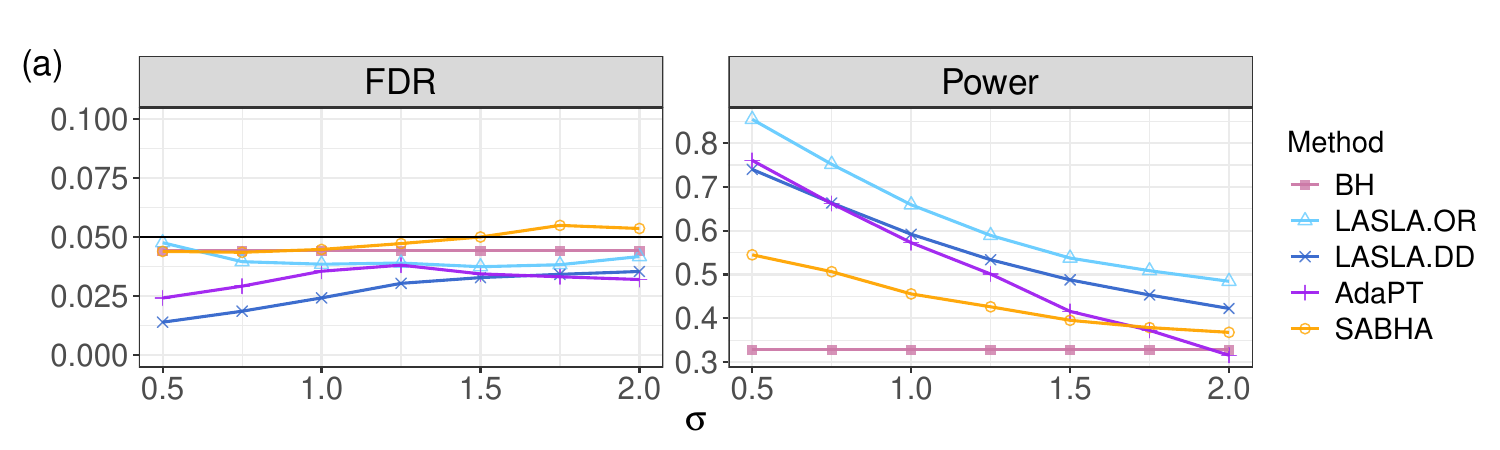}
  \includegraphics[width=0.8\linewidth]{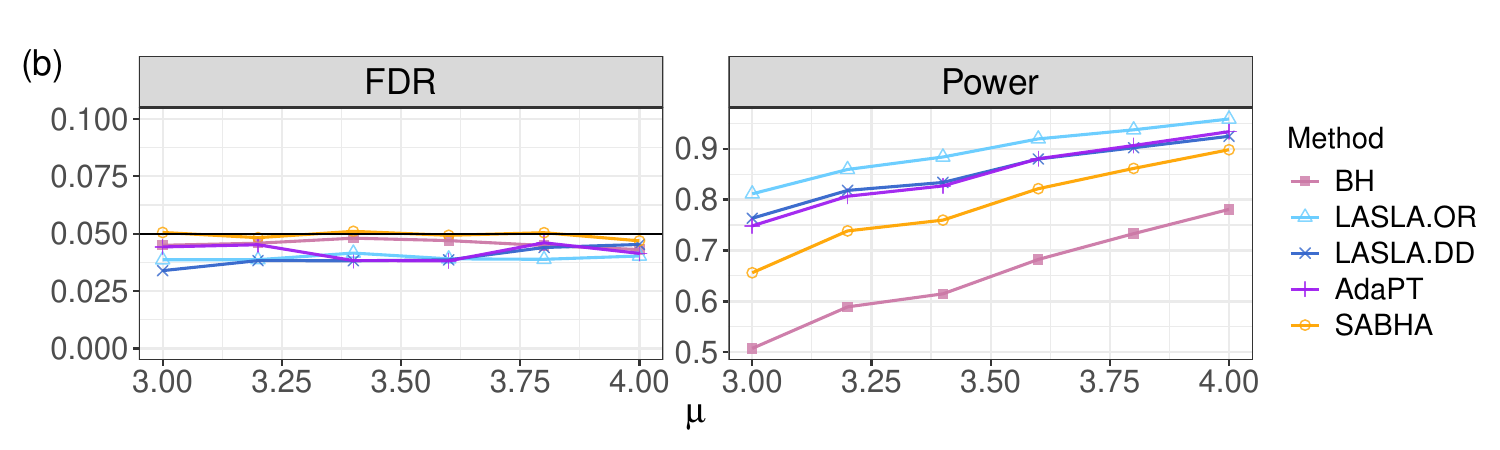}
  \caption{Empirical FDR and power comparison of LASLA and benchmarks at nominal FDR level $\alpha=0.05$ under the latent variable setting in Section~\ref{latent_simu}. (a): Latent setting 1: increasing the noise level $\sigma$ in the auxiliary data; (b): Latent setting 2: increasing signal strength $\mu$.}
  \label{fig:latent}
\end{figure*}

The results are summarized in Figure \ref{fig:latent}. In both settings, LASLA achieves a lower FDR than SABHA while still outperforming it in power. This is because SABHA relies solely on $p$-values and uses weights that only account for sparsity. The AdaPT method performs comparably to data-driven LASLA when the noise level in the auxiliary data is low, but its performance rapidly deteriorates as the auxiliary information becomes noisier.
  
\subsection{Multiple auxiliary samples}\label{sec:latent_mult}
We explore two scenarios with multiple auxiliary samples: (1) all samples are informative; (2) some samples are non-informative. Similar to the previous section, consider a latent variable $\boldsymbol{\xi}=(\xi_{i}:i \in [m])$ where
$ \xi_i \sim (1-\theta_i)\Delta_0 + \theta_i N(\mu,1)$ and $\theta_i \sim \text{Bernoulli}(0.1)$. The primary statistics $T_i \sim N(\xi_i,1)$ for $i \in [m]$. The goal is to make inference on the unknown $\theta_i$. Let $\pmb{U}^k = (U^k_i: i \in [m])$ denote the $k$th auxiliary sequence for $k \in [K]$. If $\pmb{U}$ is informative, it should carry knowledge on the underlying signal $\theta_i$. Hence we introduce the first setting where all auxiliary samples are associated with the latent variable $\boldsymbol{\xi}$:
\begin{itemize}
    \item \text{Setting 1:} $U^k_i \sim N(\xi_i, \sigma^2_s)$ for $i \in [m], k = 1, ..., 4$.
\end{itemize}
Let $\gamma_i \sim \text{Bernoulli}(0.1)$ for $i \in [m]$ independently of everything else, and $\psi_i \sim (1-\gamma_i)\Delta_0 + \gamma_i N(\mu,1)$. Consider:
\begin{itemize}
    \item \text{Setting 2:} $ U^k_i \sim N(\xi_i, \sigma^2_s)$ , for $k = 1,2$; $U^k_i \sim N(\psi_i, \sigma^2_s)$ , for $k = 3,4$.
\end{itemize}

Note that $\gamma_i$ being independent of $\theta_i$ can lead to significant divergence between the latent variables $\psi_i$ and $\xi_i$, potentially making $\pmb{U}^3$ and $\pmb{U}^4$ anti-informative. The construction of $\pmb{D}$ from $\cbrac{\pmb{U}^k}_{k \in [K]}$ is not unique, we explore two different methods: using Mahalanobis distance vs using Euclidean distance with the averaged data $U^{\text{avg}}_i = \frac{1}{4}(U_i^1+U_i^2+U_i^3+U_i^4)$ for $i \in [m]$.  We assess their effectiveness under varying degrees of informativeness exhibited by the auxiliary samples. 

\begin{figure*}[!htb]
  \centering
  \includegraphics[width=0.8\linewidth]{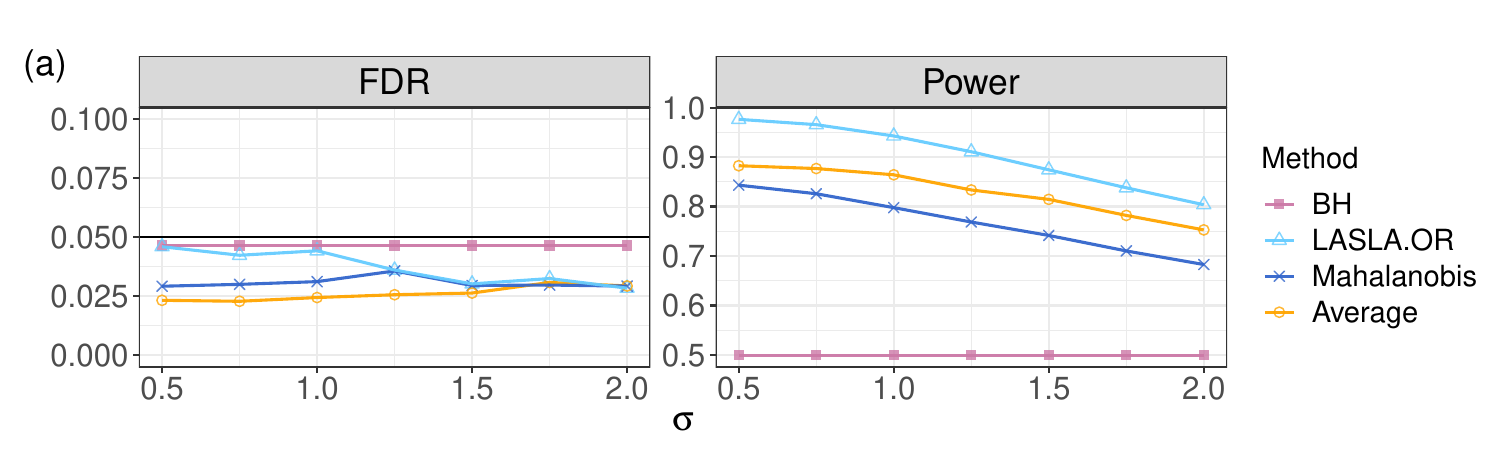}
  \includegraphics[width=0.8\linewidth]{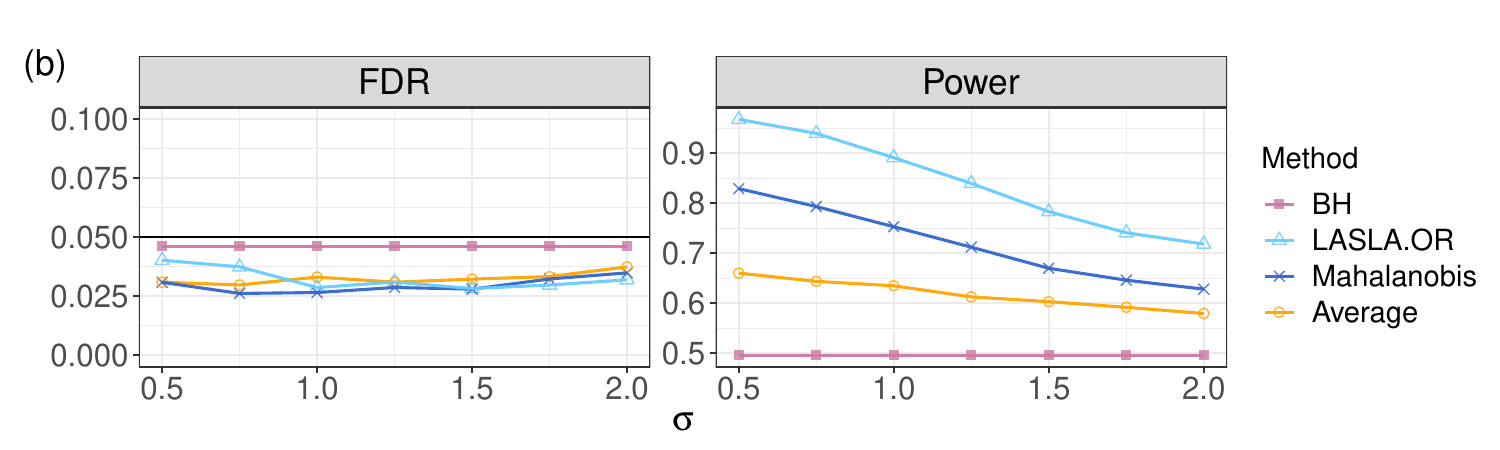}
  \caption{Empirical FDR and power comparison of LASLA with different distance computations at a nominal FDR level of $\alpha=0.05$ under the multiple auxiliary sample setting in Section~\ref{sec:latent_mult}. (a): Multiple auxiliary samples, setting 1: all auxiliary samples are informative; (b): Multiple auxiliary samples, setting 2: half of the auxiliary samples are uninformative.}
  \label{fig:mult}
\end{figure*}

In both settings, we fix $m=1200, \mu=3$, and change $\sigma_s$ from 0.5 to 2 by 0.25. The results are summarized in Figure \ref{fig:mult}. Intuitively, the averaging method reduces variance when all auxiliary samples are informative and leads to power gain over the Mahalanobis approach. However, the latter appears to be more robust when some samples are anti-informative.

\subsection{Comparison of alternative thresholding rules}\label{sec:alt-thresholding}
As mentioned in Remark~\ref{rmk:wbh} of Section~\ref{sec:lasla-threshold}, it is possible to substitute the LASLA thresholding rule in \eqref{eq:lasla-threshold} with other types of weighted thresholding rules such as the weighted BH procedure (WBH)~\citep{Genetal06} and the adaptively adjusted WBH procedure (Adj-WBH)~\citep{2019ramadaspfilter}. In this section, we first describe both approaches and then present numerical comparisons.

\begin{figure}[t!]
  \centering
  \includegraphics[width=0.75\linewidth]{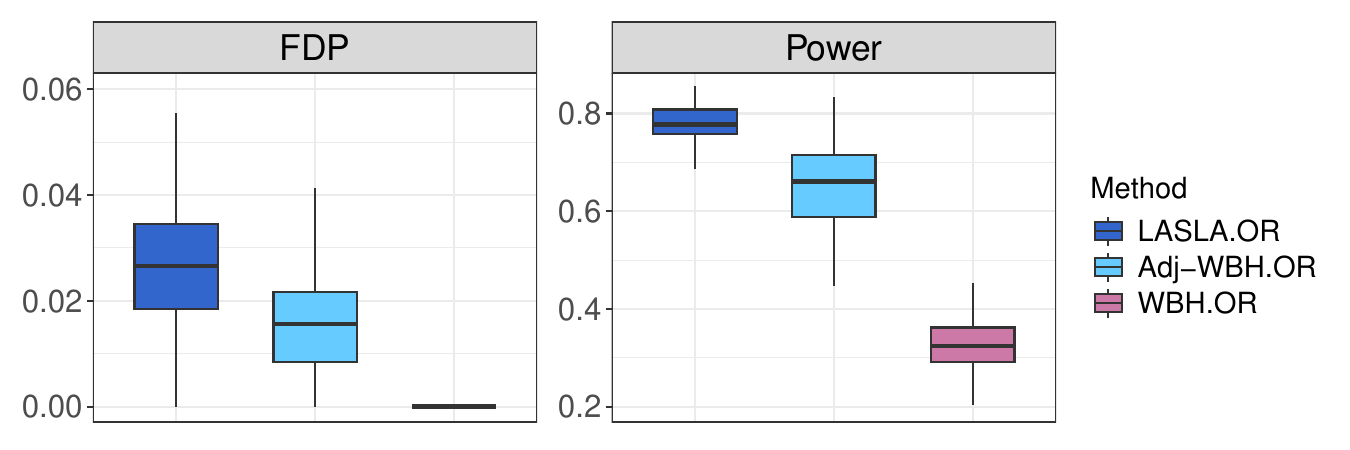}
  \caption{Empirical FDR and power comparison for the oracle LASLA, WBH and adjusted WBH for the heterogeneous sparsity level setting. All methods utilize the oracle LASLA weights.}\label{fig:hetereo-sparsity-wbh}
\end{figure}

The WBH method applies the BH procedure to the weighted p-values $\{P_i^w=P_i/w_i: i\in[m]\}$ at level $\alpha$, 
while the Adjusted WBH method~\citep{2019ramadaspfilter} improves on WBH by applying the BH procedure at an adaptively adjusted FDR level. Specifically, it computes a ``weighted sparsity estimator"
\begin{align}\label{eq:weighted-nonnull-est}
    \hat{\pi} = 1- \frac{|\mathbf{w}|_{\infty}+\sum_{i=1}^m w_i I\{P_i >\tau\}}{m(1-\tau)}
\end{align}
where $\mathbf{w}=(w_1, \dots, w_m)$ are the weights, $|\cdot|_{\infty}$ is the infinity norm and $\tau >0$ is a screening parameter similar to the one used in \eqref{eq:pi-est}. Note that the weighted sparsity estimator does not account for potential heterogeneity in the sparsity level across the hypotheses. When such heterogeneity is significant, the adaptive adjustment may not be optimal, as we shall see later in Figure~\ref{fig:hetereo-sparsity-wbh}.
The rejection threshold of the adjusted WBH method is then computed by applying the BH procedure to the weighted p-values $\{P_i^w=P_i/w_i: i\in[m]\}$ at level $\alpha/(1-\hat{\pi})$. 

\begin{figure}[!htb]
  \centering
  \includegraphics[width=0.8\linewidth]{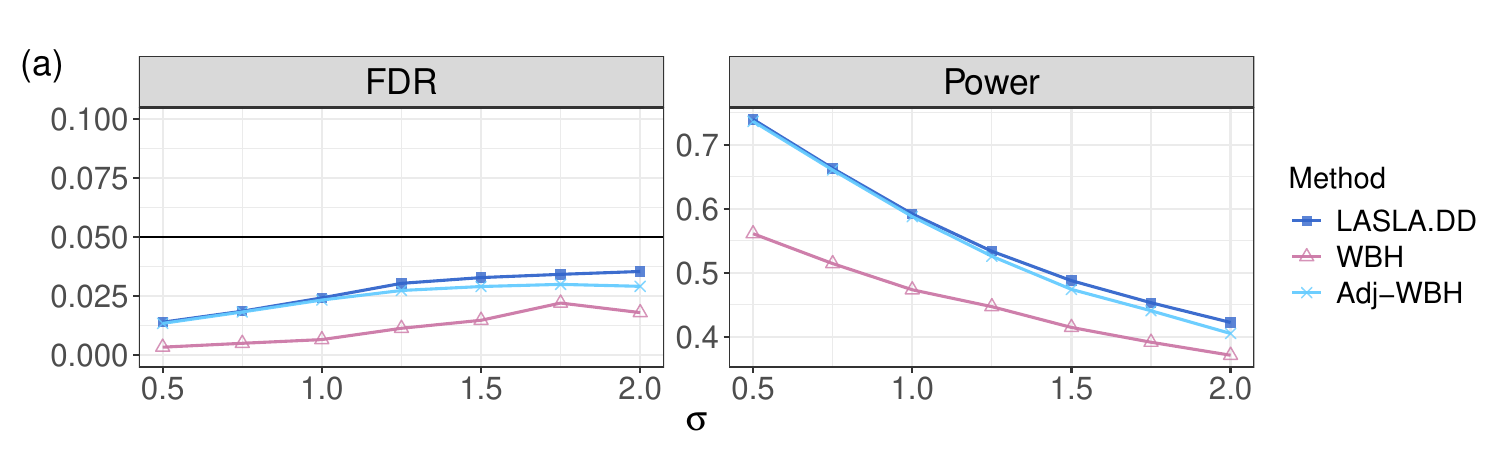}
  \includegraphics[width=0.8\linewidth]{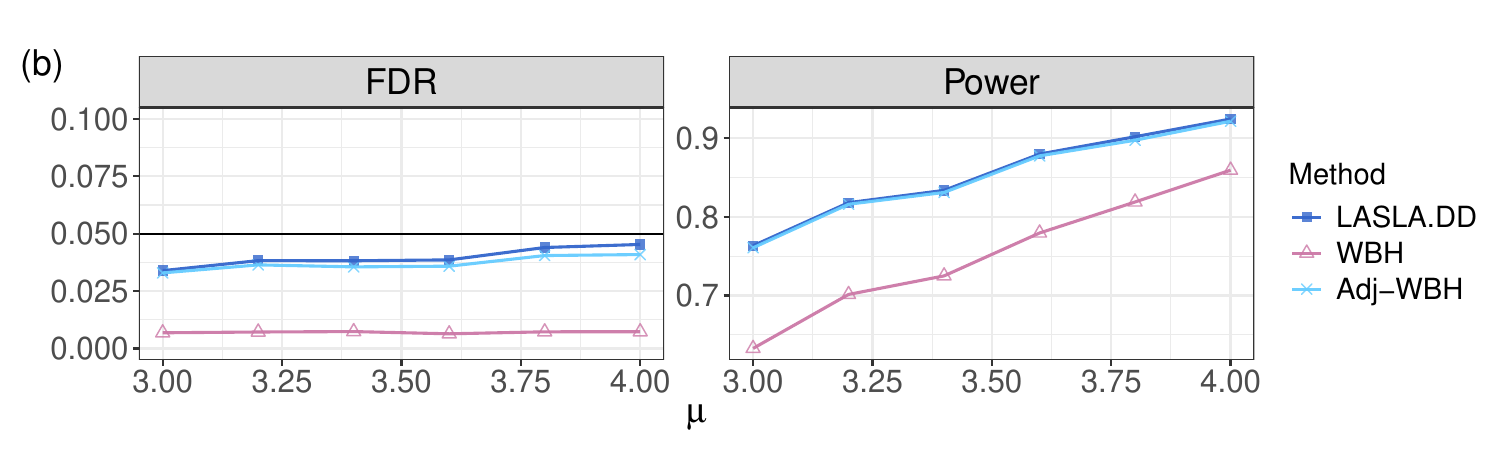}
  \caption{Empirical FDR and power comparison of different thresholding rules at nominal FDR level $\alpha=0.05$ under the latent variable setting in Section~\ref{latent_simu}. Other details are the same as in Figure~\ref{fig:latent}.}
  \label{fig:latent-wbh}
\end{figure}

Next, we provide some numerical comparisons of different thresholding approaches. Under the latent variable setting detailed in Section~\ref{latent_simu}, we compare LASLA with WBH and adjusted WBH, both benchmarks are implemented with data-driven LASLA weights. The purpose of this comparison is to isolate the effects of different thresholding rules, as the weights remain consistent across methods. Figure~\ref{fig:latent-wbh} shows that WBH is overly conservative, whereas both LASLA and adjusted WBH achieve higher power by adjusting to the sparsity level in the dataset. We emphasize that in this latent variable setting, the sparsity level is constant across all locations $i \in [m]$, hence LASLA and adjusted WBH have nearly identical performance. In the next experiment, we examine a synthetic setting with heterogeneous sparsity levels.

Consider a simple scenario where the primary data $\pmb T = (T_i: i\in [m])$ are generated as $T_i \sim (1-\theta_i)N(0,1)+\theta_i N(2,1)$, where $\theta_i \sim \text{Bernoulli}(\pi_i)$ with
\begin{align}\label{eq:hetero-sparsity}
    \pi_i \sim U(0.8, 0.9), \text{ for }i=1, \dots,200; \quad \pi_i = 0.01, \text{ for }i=201,\dots,1000. 
\end{align}
Here, the sparsity levels are heterogeneous, with elevated levels in the first 200 indices. For simplicity and direct methodological comparison, we use oracle quantities across all methods in this heterogeneous setting. Both adjusted and unadjusted WBH methods utilize the oracle LASLA weights, and the weighted non-null proportion in \eqref{eq:weighted-nonnull-est} is also computed using the oracle LASLA weights.  The results in Figure~\ref{fig:hetereo-sparsity-wbh} demonstrate that LASLA effectively utilizes the heterogeneity in sparsity levels and outperforms the adjusted WBH method. The adjusted WBH falls short because the weighted sparsity level in~\eqref{eq:weighted-nonnull-est} only captures the global sparsity level, neglecting potential heterogeneity.

\section{Numerical experiments for dependent data}\label{sec:simulation-dependent}
In this section we conduct more numerical studies under data dependency. Following a similar setup as in Section~\ref{sec:simulation}, in all the subsequent experiments, $\theta_i \sim \text{Bernoulli}(0.1)$ indicates the presence or absence of a signal at index $i$. The primary statistics $T_i$ are marginally distributed as $N(0,1)$ when $\theta_i=0$ and are distributed as $N(3,1)$ when $\theta_i=1$. The distance matrix $\pmb D =(D_{ij})_{1 \leq i,j \leq m}$ is defined by
$D_{ij} \sim I_{\{\theta_i=\theta_j\}}|N(0,0.7)|+I_{\{\theta_i \neq \theta_j\}}|N(1,0.7)|$. We fix $m=1000$ and the FDR level at $\alpha=0.05$.
The correlation structure will be specified in each setting below. Section~\ref{sec:block-dependency} examines LASLA's performance in a weakly dependent setting, while Section~\ref{sec:random-dependency} considers stronger dependency scenarios.

\subsection{Block dependency}\label{sec:block-dependency}
In this setting, we consider a ``block'' dependency type where variables within the same block are equally correlated with each other, while variables in different blocks are uncorrelated. This structure offers a simplified model to mimic the scenarios where there are distinct clusters or groups of highly correlated variables that have little or no correlation with variables in other groups, similar to GWAS data where certain clusters of SNPs may work together to influence specific phenotypes.

\begin{figure}[t!]
  \centering
  \includegraphics[width=0.8\linewidth]{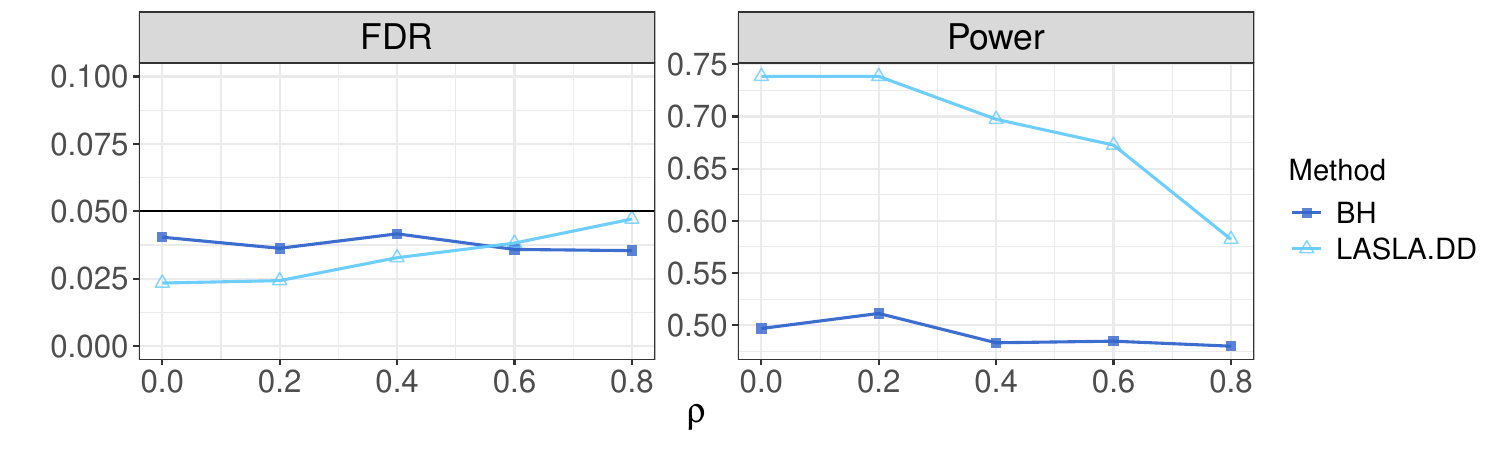}
  \caption{Empirical FDR and power comparison for data-driven LASLA and BH under the block dependency setting. The correlation strength increases as $\rho$ increases.}\label{fig:dependent1}
\end{figure}

We divide the $m=1000$ indices into $10$ blocks, each has a size of $100$. For $k \in [10]$, let $b_{k} \subset [m]$ denote the collection of indices in block $k$. For $i, j \in [m]$, define the correlation matrix $\pmb{\Sigma}$ as
\begin{align}\label{eq:block-correlation}
    \pmb{\Sigma}_{ij} = 
    \begin{cases}
        1, & \text{if } i=j;\\
        \rho, & \text{if } i\neq j, \text{ and }i,j \in b_k, \text{ for some }k \in [10];\\
        0, & \text{otherwise},
    \end{cases}
\end{align}
where $\rho$ controls the correlation strength. We vary $\rho$ from $0$ to $0.8$ by step of $0.2$, and summarize the result in Figure~\ref{fig:dependent1}. We observe that LASLA's performance remains robust under block dependency, consistently controlling the FDR within the nominal level.

\subsection{Random dependency}\label{sec:random-dependency}
\begin{figure}[t!]
  \centering
  \includegraphics[width=0.8\linewidth]{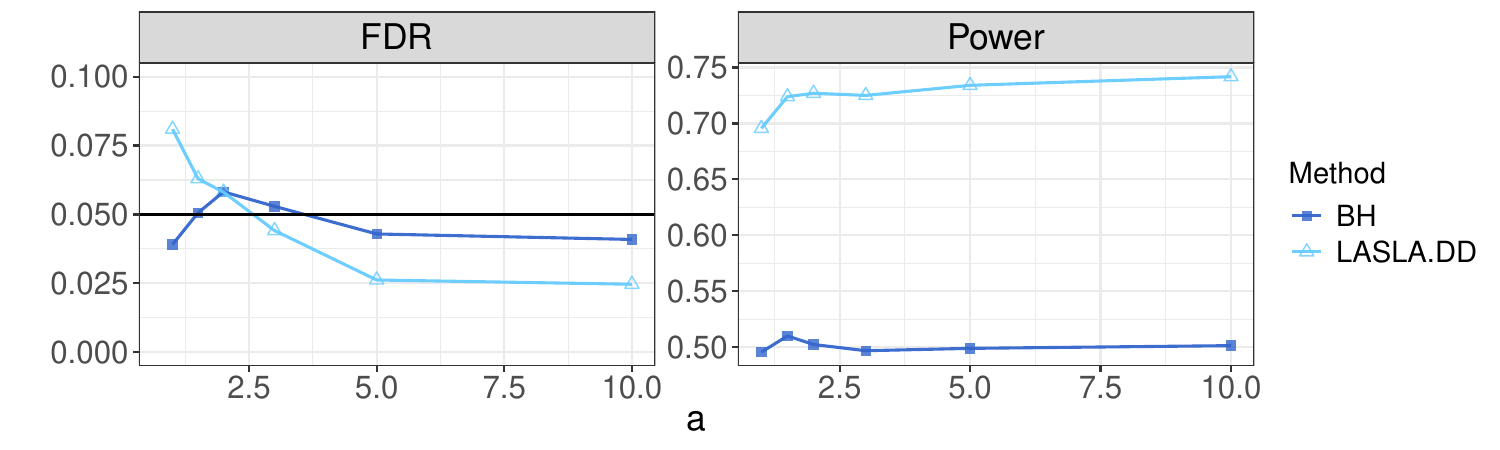}
  \caption{Empirical FDR and power comparison for data-driven LASLA and BH under the random dependency setting. The correlation strength decreases as $a$ increases.}\label{fig:dependent2}
\end{figure}

\begin{figure}[t!]
  \centering
  \includegraphics[width=0.8\linewidth]{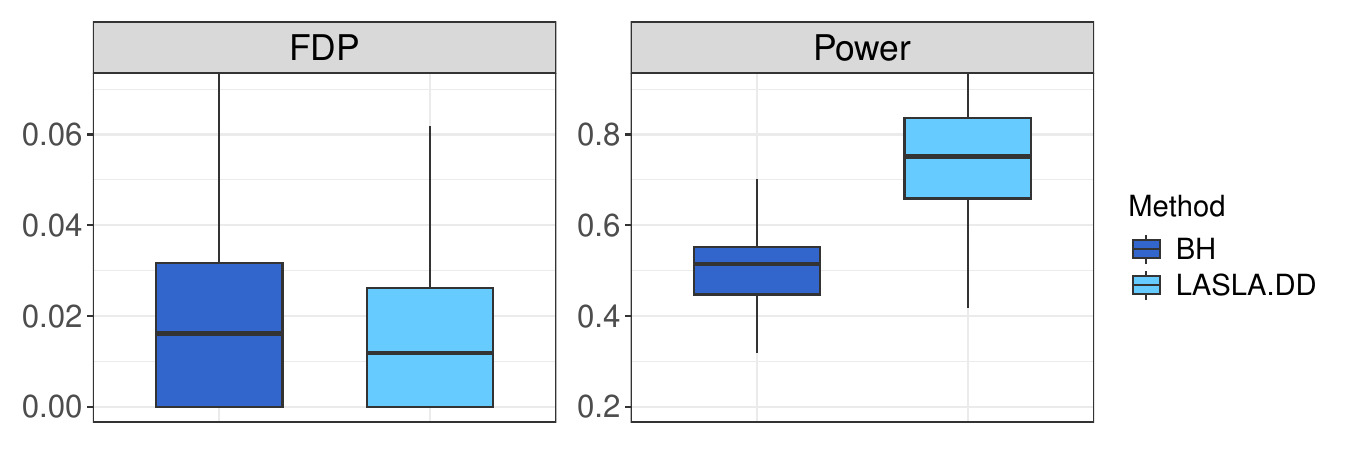}
  \caption{Empirical FDR and power comparison for data-driven LASLA and BH under the random dependency setting with $a=1$ and enlarged bandwidth $h=m^{-1/6}$.}\label{fig:dependent3}
\end{figure}

In this section we consider a ``random'' dependency structure where the correlation matrix is generated randomly, with no specific pattern or clustering. We first generate a random factor vector $\pmb{v} \in \mathbb{R}^m$, where each entry follows a standard normal distribution. Define the correlation matrix as
\begin{align}\label{eq:random-correlation}
    \pmb{\Sigma}_{ij} = s(\pmb{v}\pmb{v}^{\top}+\pmb{\eps}),
\end{align}
where we add a diagonal matrix $\pmb{\eps}$, with the diagonal elements uniformly distributed from $[0,1]$, ensuring the positive definiteness of the correlation matrix. The function $s(\cdot)$ is then applied to standardize the matrix, ensuring that all diagonal elements are equal to 1. The off-diagonal elements approximately follow a uniform distribution from $[-1,1]$. Unlike the block correlation in~\eqref{eq:block-correlation}  where indices are weakly and positively correlated, the random correlation in \eqref{eq:random-correlation} allows for both negative and positive correlations. Moreover, the correlation is strong and may even violate the weak dependency assumption in ~\ref{A2}. To adjust the dependency strength, we introduce a parameter $a$  that scales the off-diagonal elements of $\pmb{\Sigma}$ by dividing them by $a$. That is, for $i, j \in [m]$, the adjusted correlation matrix is defined as
\begin{align*}
    \tilde{\pmb{\Sigma}}_{ij} = 
    \begin{cases}
        1, & \text{if } i=j;\\
        \frac{\pmb{\Sigma}_{ij}}{a}, & \text{otherwise}.
    \end{cases}
\end{align*}
We examine LASLA's performance with $a$ taking values in $(1,1.5,2,3,5,10)$. Note that smaller values of $a$ signify stronger correlations. Figure~\ref{fig:dependent2} shows that the FDR of LASLA tends to rise above the nominal level under the strongest dependency setting. As discussed in Remark~\ref{rmk:bw-relaxation}, the dependency assumption can be further 
relaxed by choosing a larger bandwidth, for instance $m^{-1/6}$. Hence, we rerun the experiment with $a=1$ using the enlarged bandwidth, and Figure~\ref{fig:dependent3} demonstrates that LASLA effectively reduces the FDR level under this adjustment.

\section{Proof of Main Results}\label{app:proof}
Recall that $\pmb{D}_i$ is the $i$th column of $\pmb{D}$, and $\cD_i$ is a continuous finite domain (w.r.t. coordinate $i$) in $\mathbb{R}$ with positive measure by adopting the fixed-domain asymptotics in \cite{stein1995fixed}. Each $d\in \cD_i$ is a distance and $0\in \cD_i$. The two sets $\pmb D_i$ and $\cD_i$ can be viewed as collections of distances measured from the partial and full network respectively, and it follows that $\pmb D_i\subset\cD_i$. 

Throughout the proofs, we assume that $\pmb D_i \rightarrow \cD_i$ as $m\rightarrow\infty$ 
in the sense that, for any $d_0\in \cD_i$, there exists at least an index $j$ such that $|D_{ij} - d_0| = O(m^{-1})$ as $m\rightarrow \infty$.
\subsection{Proof of Proposition \ref{prop_pi}}

\begin{proof}
For simplicity of notation, throughout we omit the conditioning on $\cD$, and use $\pr(P_j > \tau |D_{ij} = x)$ and $\pr(P_i > \tau )$ to denote $\pr(P_j > \tau \mid \cD_j, D_{ij} = x)$ and $\pr(P_i > \tau \mid \cD_i)$ respectively. 
Recall that
\begin{align*}
    1-\pi_i &=\frac{\sum_{j \in \cN_i}[{K_h(D_{ij})\II\cbrac{P_j>\tau}}]}{(1-\tau)\sum_{j \in \cN_i}{K_h(D_{ij})}}
\end{align*}
Also note that, for all $i \in [m]$,
\begin{align*}
    \expect[\bigg]{\sum_{j \in \cN_i}[{K_h(D_{ij})\II\cbrac{{P_j>\tau}}} \mid \pmb D}=
    \sum_{j \in \cN_i}[{K_h(D_{ij})\prob{P_j>\tau \mid \pmb D_j}}]
\end{align*}
Then by $m^{-1}\ll h\ll m^{-\epsilon}$,  as $\pmb D_i\rightarrow \cD_i$, we have
\begin{align*}
    \frac{\expect{1-\pi_i \mid \pmb D}}{{\int_{\tilde\cD_i}{K_h(x)\prob{P_{j_x}>\tau \mid D_{ij_x} = x}} \,dx}/{(1-\tau)\int_{\tilde\cD_i}{K_h(x)} \,dx}} \rightarrow 1,
\end{align*}
where $j_x$ represents the index such that $D_{ij_x} = x$ and $\tilde\cD_i$ is the limit of $\{D_{ij}, j\in \cN_i\}$ in the asymptotic framework described at the beginning of Section~\ref{app:proof}.
Using Taylor expansion at $x=0$, combined with Assumption \ref{A1}, we have
\begin{align*}
    &\int_{\tilde\cD_i}{K_h(x)\prob{P_{j_x}>\tau \mid D_{ij_x} = x}} \,dx \\
    &= \prob{P_i>\tau}\int_{\tilde\cD_i} K_h(x) \,dx + \mathbb{P}'(P_i>\tau)\int_{\tilde\cD_i} xK_h(x) \,dx\\
    &+\frac{\mathbb{P}''(P_i>\tau)}{2}\int_{\tilde\cD_i} x^2K_h(x) \,dx +\textit{O}(h^2).
\end{align*}
Thus, by the assumptions of  $K(\cdot)$ in \eqref{kernel}, uniformly for all index $i$, there exists some constant $c>0$ such that
\begin{align*}
    &\brac*{\expect{\pi_i \mid \pmb D}-\pi_i^{\tau}}^2 \\
    &\leq \bigg(c\int_{\tilde\cD_i} xK_h(x) \,dx \bigg/ \int_{\tilde\cD_i} K_h(x) \,dx + c\int_{\tilde\cD_i} x^2K_h(x) \bigg/ \int_{\tilde\cD_i} K_h(x) \,dx\bigg)^2 + o(1)\\
    &\rightarrow 0, \hspace{1cm}\text{as } h \rightarrow 0.
\end{align*}
Now we inspect the variance term. By Condition \ref{A2}, there exists a constant $c'>1$,
\begin{align*}
    &\var[\bigg]{\sum_{j \in \cN_i}[K_h(D_{ij})\II\cbrac{P_j>\tau}] \mid \pmb D}\\ 
    &\leq c'\sum_{j \in \cN_i}[{K^2_h(D_{ij})\var[\big]{\II\cbrac{P_j>\tau} \mid \pmb D_j}}]\\
    &= c'\sum_{j \in \cN_i}\brac*{{K^2_h(D_{ij})\prob{P_j>\tau \mid \pmb D_j}\cbrac*{1-\prob{P_j>\tau \mid \pmb D_j}}}}.
\end{align*}
Hence, as $h \gg m^{-1}$, by the assumptions of  $K(\cdot)$ in \eqref{kernel} and that it is positive and bounded, we have
\beas
    \var{1-\pi_i}
    & \leq& c''m^{-1}\frac{\int_{\tilde\cD_i}K^2_h(x) \,dx}{[(1-\tau)\int_{\tilde\cD_i}K_h(x) \,dx]^2}\\
    & \leq& c''(mh)^{-1}\frac{\int_{\mathbb{R}} K^2(y) \,dy}{[(1-\tau)\int_{\tilde\cD_i}K_h(x) \,dx]^2}\\
    & \leq& c'''(mh)^{-1} = o(1),
\eeas
for some constant $c'',c'''>0$. Hence, as $\pmb D_i \rightarrow \cD_i$, by combining the bias term and variance term, the consistency result is proved.
\end{proof}

\subsection{Proof of Theorem \ref{thm:FDR_idp}}

\begin{proof}
{
For simplicity of notation, throughout we omit the explicit conditioning on $\cD$, and use $\pr(\theta_i=0)$ to denote $\pr(\theta_i=0 \mid \cD_i)$ and $\pr(P_i^{w} \leq t \mid \theta_i=0 ,w_i)$ to denote $\pr(P_i^{w} \leq t \mid \theta_i=0 ,w_i, \cD_i)$. 
Note that, by Algorithm \ref{alg:simple_weights}, 
the FDP of LASLA at the thresholding level $t$ can be calculated by

\begin{align*}
    FDP(t)&=\frac{\sum_{i=1}^{m} \II\{P_i^{w} \leq t, \theta_i=0\}}{\max[\sum_{i=1}^{m} \II\{P_i^{w} \leq t\},1]}\\
    &=\frac{\sum_{i=1}^{m} \pr(P_i^{w} \leq t \mid \theta_i=0 ,w_i)\pr(\theta_i=0)}{\max[\sum_{i=1}^{m} \II\{P_i^{w} \leq t\},1]} \cdot \frac{\sum_{i=1}^{m} \II\{P_i^{w} \leq t, \theta_i=0\}}{\sum_{i=1}^{m} \pr(P_i^{w} \leq t \mid \theta_i=0 ,w_i)\pr(\theta_i=0)},
\end{align*}

{\bf Step 1: } 
We first show that, uniformly for all $i \in [m]$, we have
\beq\label{show7}
\sum_{i=1}^{m} \pr(P_i^{w} \leq t \mid \theta_i=0 ,w_i)\pr(\theta_i=0)
\leq { [1+o_\pr(1 )]\sum_{i=1}^{m}w_i(1-\pi_i)t.}
\eeq

Note that, in Algorithm \ref{alg:complicated_weights}, $T_i$ is not used in the computation of $w_i$ given the sign of $T_i$. Then by the independence assumption~\ref{cond:marg-indep}, $T_i$ is independent of  $w_i$ conditioning on the sign, and it follows that:
\begin{align*}
    &\sum_{i=1}^{m} \pr(P_i^{w} \leq t \mid \theta_i=0 ,w_i)\pr(\theta_i=0) \\
    &= \brac*{\sum_{i=1}^{m} \pr(P_i^{w} \leq t \mid T_i >0, \theta_i=0 ,w_i)\pr(T_i>0 \mid \theta_i=0 ,w_i)(1-\pi_i^*)} \\
    &\textcolor{white}{=} +\brac*{\sum_{i=1}^{m} \pr(P_i^{w} \leq t \mid T_i <0, \theta_i=0 ,w_i)\pr(T_i<0 \mid \theta_i=0 ,w_i)(1-\pi_i^*)}\\
    &=\brac*{\sum_{i=1}^{m}\pr(T_i>0 \mid \theta_i=0)w_i (1-\pi_i^*)t}+\brac*{\sum_{i=1}^{m}\pr(T_i<0 \mid \theta_i=0)w_i (1-\pi_i^*)t}\\
    &= \sum_{i=1}^{m}w_i(1-\pi_i^*)t \leq \sum_{i=1}^{m}w_i(1-\pi_i^{\tau})t,
\end{align*}
where the last inequality follows from the fact that $\pi_i^{\tau}$ is a conservative approximation of $\pi_i^*$ as showed in \cite{Caietal20}.
By the result of Proposition \ref{prop_pi} and  Assumption \ref{A3}, together with  the fact that $\xi\leq w_i\leq 1$ for $i\in [m]$, we have
\begin{align*}
\sum_{i=1}^{m} \pr(P_i^{w} \leq t \mid \theta_i=0 ,w_i)\pr(\theta_i=0) 
\leq \sum_{i=1}^{m}w_i[1-\pi_i + o_\pr(1 )]t = { [1+o_\pr(1 )]\sum_{i=1}^{m}w_i(1-\pi_i)t.}
\end{align*}
Hence, \eqref{show7} is proved.

{\bf Step 2: } 
We next show that
\beq\label{show6}
\abs*{ \frac{\sum_{\theta_i=0} \pr(P_i^{w} \leq t \mid \theta_i=0, w_i) - \sum_{i=1}^{m} \pr(P_i^{w} \leq t \mid \theta_i=0 ,w_i)\pr(\theta_i=0)}{\sum_{i=1}^{m} \pr(P_i^{w} \leq t \mid \theta_i=0 ,w_i)\pr(\theta_i=0)} } \rightarrow 0,
\eeq
in probability. Define the event 
\[
B  = \brac*{ \cbrac*{\theta_i}_{i=1}^m, \sum_{i=1}^{m}\II\{\theta_i=0\} \geq cm \text{ for some constant }c>0 }.
\]
It follows from Condition \ref{A3} that $\pr(B) \rightarrow 1$. Then by the fact that $\xi \leq w_i \leq 1$, we have
\begin{align*}
    &\ep \brac*{ \ep \bigg(\bigg|\frac{\sum_{i=1}^{m} \left[ \pr({P_i^{w} \leq t \mid \theta_i=0, w_i})\II\{\theta_i=0\}-\pr(P_i^{w} \leq t \mid \theta_i=0 ,w_i)\pr(\theta_i=0)\right]}{\sum_{i=1}^{m} \pr(P_i^{w} \leq t \mid \theta_i=0 ,w_i)\pr(\theta_i=0)} \bigg|^2 \,\middle\vert\, w_i\bigg)}\\
    &=\ep \brac*{ \ep \bigg(\bigg| \frac{\sum_{i=1}^{m}\pr({P_i^{w} \leq t \mid \theta_i=0,w_i})\brac*{
    \II\{\theta_i=0\}-\pr(\theta_i=0)}}{\sum_{i=1}^{m}\pr({P_i^{w} \leq t \mid \theta_i=0, w_i})\pr(\theta_i = 0)} \bigg|^2 \,\middle\vert\, w_i \bigg)}\\
    &=\ep \brac*{ \Var \left(\sum_{i=1}^{m}\left[\pr({P_i^{w} \leq t|\theta_i=0})\II\{\theta_i=0\}\right] \,\middle\vert\, w_i\right)\bigg/\bigg(\sum_{i=1}^{m}w_i t\pr(\theta_i = 0)\bigg)^2 } \\
    &= O(m^{\zeta-1}),
\end{align*}
the last equality follows from the law of total variance and condition~\ref{A2} for some $0\leq \zeta < 1$. Hence \eqref{show6} is proved.
}

{\bf Step 3: }
Finally, we analyze the following quantity:
$$\frac{\sum_{i=1}^{m} \II\{P_i^{w} \leq t, \theta_i=0\}}{\sum_{i=1}^{m} \pr(P_i^{w} \leq t \mid \theta_i=0 ,w_i)\pr(\theta_i=0)}.$$
We first check the range of the cutoff $t$, or equivalently the threshold for the weighted $z$-values, i.e., $z_i^w=\Phi^{-1}(1-P_i^{w}/2)$, for $i \in [m]$.
Then as shown in \cite{Caietal20} and replace their weights $\frac{\pi_i}{1-\pi_i}$ by $w_i$, it is easy to see that, by applying  BH procedure at level $\alpha$ to the adjusted $p$-values with weights $w_i$, the corresponding threshold is no larger than the threshold of LASLA for the adjusted $p$-values with the same weights $w_i$. Hence it suffices to obtain the threshold for the weighted $z$-values $z_i^{w}=\Phi^{-1}(1-P_i^{w}/2)$ of such BH procedure with weights $w_i$.

Let $t_m=(2\log m-2\log\log m)^{1/2}$.
By Condition \ref{A4}, we have
\[
\sum_{\theta_i=1}\II\{|z_i|\geq (c\log m)^{1/2+\rho/4}\}\geq \{{1}/({\pi}^{1/2}\alpha)+\delta\}({\log m})^{1/2},
\]
with probability going to one.
Recall that we have $\xi\leq w_i\leq 1$ for some constant $\xi>0$. 
Thus, for those indices $i\in \cH_1$ (equivalently $\theta_i=1$) such that $|z_i|\geq (c\log m)^{1/2+\rho/4}$, we have 
\[
P_i^{w} \leq (1-\Phi((c\log m)^{1/2+\rho/4}))/w_i=o(m^{-M}),
\]
for any constant $M>0$.
Thus we have
\[
\sum_{ i\in [m]}\II\{z_i^{w}\geq (2\log m)^{1/2}\}\geq \{{1}/({\pi}^{1/2}\alpha)+\delta\}({\log m})^{1/2},
\]
with probability going to one. Hence, with probability tending to one, 
\[
\frac{2m}{\sum_{ i\in [m]}\II\{z_i^{w}\geq (2\log m)^{1/2}\}}\leq 2m\{{1}/({\pi}^{1/2}\alpha)+\delta\}^{-1}({\log m})^{-1/2}.
\]

{
Because $1-\Phi(t_m)\sim 1/\{(2\pi)^{1/2}t_m\}\exp(-t_m^2/2)$, it suffices to show that, 
\beq\label{show1}
\sup_{0\leq t\leq t_m}\Bigg{|} \frac{\sum_{i=1}^{m} \II\{z_i^{w} \geq t, \theta_i=0\} - \sum_{i=1}^{m} \pr(z_i^{w} \geq t \mid \theta_i=0, w_i)\pr(\theta_i=0)}{\sum_{i=1}^{m} \pr(z_i^{w} \geq t \mid \theta_i=0, w_i)\pr(\theta_i=0)}\Bigg{|}\rightarrow 0,
\eeq
in probability.
Let the event $A = \left[\{\theta_i\}_{i=1}^m: \eqref{show6}\text{ holds} \right]$. By the proofs in Step 2, we have $\pr(A) \rightarrow 1$. Hence, it is enough to show that, for $\{\theta_i\}_{i=1}^m \in A$, we have
\beq\label{show2}
\sup_{0\leq t\leq t_m}\Bigg{|} \frac{\sum_{\theta_i=0} \left[ \II\{z_i^{w} \geq t\} - \pr(z_i^{w} \geq t \mid \theta_i=0, w_i)\right]}{\sum_{\theta_i=0} \pr(z_i^{w} \geq t \mid \theta_i=0, w_i)}\Bigg{|}\rightarrow 0,
\eeq
in probability.
Let $0\leq t_0<t_1<\cdots<t_b=t_{m}$ such that $t_{\iota}-t_{{\iota}-1}=v_{m}$  for $1\leq {\iota}\leq b-1$ and $t_b-t_{b-1}\leq v_{m}$, where $v_{m}=1/\sqrt{\log m(\log_4 m)}$. Thus we have $b\sim t_{m}/v_{m}$. For any $t$ such that $t_{{\iota}-1}\leq t\leq t_{\iota}$, due to the fact that $G(t+o((\log m)^{-1/2}))/G(t)=1+o(1)$ with $G(t)=2(1-\Phi(t))$ uniformly in $0\leq t\leq c(\log m)^{1/2}$ for any constant $c$,
by \cite{Xiaetal19}, it suffices to prove that 
\beq\label{show3}
\max_{0\leq \iota \leq b}\Bigg{|} \frac{\sum_{\theta_i=0} \left[ \II\{z_i^{w} \geq t_\iota\} - \pr(z_i^{w} \geq t_\iota \mid \theta_i=0, w_i)\right]}{\sum_{\theta_i=0} \pr(z_i^{w} \geq t_\iota \mid \theta_i=0, w_i)}\Bigg{|}\rightarrow 0,
\eeq
in probability.
Thus, it suffices to show that, for any $\epsilon>0$,
\bea
\label{show4}
\int_0^{t_m}\pr\Bigg{\{}\Bigg{|}\frac{\sum_{\theta_i=0} \left[ \II\{z_i^{w} \geq t\} - \pr(z_i^{w} \geq t \mid \theta_i=0, w_i)\right]}{\sum_{\theta_i=0} \pr(z_i^{w} \geq t \mid \theta_i=0, w_i)}\Bigg{|}\geq \epsilon\Bigg{\}}dt
=o(v_m).
\eea
By the fact that $\xi\leq w_i\leq 1$ for some constant $\xi>0$, we have
\begin{align*}
    \pr(z_i^{w} \geq t \mid \theta_i=0, w_i) &= \pr(\Phi^{-1}(1-P_i^{w}/2) \geq t \mid \theta_i=0, w_i) \\
    &= \pr(P_i \leq 2w_i(1-\Phi(t)) \mid \theta_i=0, w_i)\\
    &= 2w_i(1-\Phi(t)) \geq  \xi G(t).
\end{align*}
It follows that
\begin{align*}
& \ep\left| \frac{\sum_{\theta_i=0} \left[ \II\{z_i^{w} \geq t\} - \pr(z_i^{w} \geq t \mid \theta_i=0, w_i)\right]}{\sum_{\theta_i=0} \pr(z_i^{w} \geq t \mid \theta_i=0, w_i)} \right|^2\\
&  \leq \frac{\ep\left|\sum_{\theta_i=0} \left[ \II\{z_i^{w} \geq t\} - \pr(z_i^{w} \geq t \mid \theta_i=0, w_i)\right]\right|^2}{ \cbrac*{\sum_{\theta_i=0} \xi G(t)}^2} \\
&= \frac{\ep\brac*{\sum_{\theta_i=0,\theta_j=0} \pr\paren{z_i^{w} \geq t, z_j^{w} \geq t \mid \theta_i=0,\theta_j=0, w_i, w_j} - \left\{\sum_{\theta_i=0} \pr(z_i^{w} \geq t \mid \theta_i=0, w_i)\right\}^2}}{\cbrac*{\sum_{\theta_i=0} \xi G(t)}^2}.
\end{align*}
Recall that, by Algorithm \ref{alg:complicated_weights} we only use $O(m^{1-\epsilon})$ neighbors to construct $w_i$ for any small enough constant $\epsilon>0$, 
Hence, we can divide the indices pairs $\tilde{\cH}_{0}=\{(i,j): \theta_i=0,\theta_j=0\}$ into two subsets:
\beas
\tilde{\cH}_{01}&=&\{(i,j)\in \tilde{\cH}_{0}, \mbox{ either $P_i^w$ is correlated with $P_j$ or $P_j^w$ is correlated with $P_i$}\},\cr
\tilde{\cH}_{02}&=&\tilde{\cH}_{0}\setminus \tilde{\cH}_{01},
\eeas
where 
$|\tilde{\cH}_{01}| = O(m^{2-\epsilon})$ while among them $m$ pairs with $i=j$ are perfectly correlated.

Note that, for $(i,j) \in \tilde{\cH}_{01}$,
\begin{align*}
&\frac{ \ep \brac*{ \sum_{(i,j)\in \tilde{\cH}_{01}} \cbrac*{ \pr\paren{z_i^{w} \geq t, z_j^{w} \geq t \mid \theta_i=0,\theta_j=0, w_i, w_j} - \prod_{h = i,j} \pr(z_h^{w} \geq t \mid \theta_h=0, w_h)}}}{\cbrac*{\sum_{\theta_i=0} \xi G(t)}^2}\\
& \leq \frac{\ep \brac*{\sum_{(i,j)\in \tilde{\cH}_{01}}  \pr\paren{z_i^{w} \geq t, z_j^{w} \geq t \mid \theta_i=0,\theta_j=0, w_i, w_j} }}{\cbrac*{\sum_{\theta_i=0} \xi G(t)}^2}.
\end{align*}
Recall that event $B  = \left[\{\theta_i\}_{i=1}^m, \sum_{i=1}^{m}\II\{\theta_i=0\} \geq cm \text{ for some constant }c>0 \right]$ and $\pr(B) \rightarrow 1$. 
For $\{\theta_i\}_{i=1}^m \in A \cap B$, we have
\begin{align*}
&\frac{ \ep \brac*{ \sum_{(i,j)\in \tilde{\cH}_{01}} \cbrac*{ \pr\paren{z_i^{w} \geq t, z_j^{w} \geq t \mid \theta_i=0,\theta_j=0, w_i, w_j} - \prod_{h = i,j} \pr(z_h^{w} \geq t \mid \theta_h=0, w_h)}}}{\cbrac*{\sum_{\theta_i=0} \xi G(t)}^2}\\
&\leq \frac{\sum_{\theta_i=0} \pr(z_i^{w} \geq t \mid \theta_i=0, w_i)}{\cbrac*{\sum_{\theta_i=0} \xi G(t)}^2} +  O\left(\frac{m^{2-\epsilon} }{m^2} \right)\\
&\leq O\left(\frac{1}{m G(t)} \right) + O\left(m^{-\epsilon}  \right), 
\end{align*}

where the first term reflects the pairs with $i=j$.
On the other hand,
\begin{align*}
    \ep \brac*{\sum_{(i,j)\in \tilde{\cH}_{02}} \cbrac*{ \pr\left\{z_i^{w} \geq t, z_j^{w} \geq t \mid \theta_i=0,\theta_j=0, w_i, w_j\right\} - \prod_{h = i,j} \pr(z_h^{w} \geq t \mid \theta_h=0, w_h)}} = 0.
\end{align*}
}
Then by the fact that
\[
\int_0^{t_m}\left\{ \frac{1}{m G(t)} + m^{-\epsilon} \right\} dt = o(v_m),
\]
and that $\pr(A\cap B) \rightarrow 1$, \eqref{show4} is proved and \eqref{show2} is thus proved. 
Combining \eqref{show2} and \eqref{show6}, we obtain \eqref{show1}.
This together with \eqref{show7} prove the result of Theorem \ref{thm:FDR_idp}.
\end{proof}

\subsection{Proof of Theorem 2}

\begin{proof} 
Note that
\begin{align*}
    Q^1(t)=\frac{\sum_{i=1}^m(1 - \pi_i^*)t}{\sum_{i=1}^m(1 - \pi_i^*)t+
    \sum_{i=1}^m{\pi_i^*F^*_{1i}(t \mid \cD)}}.
\end{align*}
Recall that $\tilde w_i  = w_i \brac{\sum_{j=1}^m (1 - \pi_j^*)}/\brac{\sum_{j=1}^m (1 - \pi_j^*) w_j}$, we have
\begin{align*}
    Q^{\tilde w}(t) & = \frac{\sum_{i=1}^m(1 - \pi_i^*) \tilde w_i t}{\sum_{i=1}^m(1 - \pi_i^*) \tilde w_i t+
    \sum_{i=1}^m{\pi_i^*F^*_{1i}(\tilde w_it \mid \cD)}}\\
    &=\frac{\sum_{i=1}^m(1 - \pi_i^*)t}{\sum_{i=1}^m(1 - \pi_i^*)t+
    \sum_{i=1}^m{\pi_i^*F^*_{1i}(\tilde w_it \mid \cD)}}.
\end{align*}
Under the Assumption \ref{cond:shape} we have,
\begin{align*}
    \sum_{i=1}^m{\pi_i^*F^*_{1i}(\tilde w_it \mid \cD)} &= \sum_{i=1}^m{\pi_i^*F^*_{1i}(t/\tilde w_i^{-1} \mid \cD)}\\
    &\geq 
    \sum_{i=1}^m\pi_i^*F^*_{1i}\left(\frac{\sum_{i=1}^m \pi_j^* t}{\sum_{j=1}^m\pi_j^*\tilde w_j^{-1}} \mid \cD\right).
\end{align*}
By Assumption \ref{cond:informative} and the construction that $\tilde w_i  = w_i \brac{\sum_{j=1}^m (1 - \pi_j^*)}/\brac{\sum_{j=1}^m (1 - \pi_j^*) w_j}$, we have $\brac*{\sum_{i=1}^m \pi_i^* }/\brac*{\sum_{i=1}^m\pi_i^*\tilde w_i^{-1}} \geq 1$. Therefore,
\begin{eqnarray*}
 \sum_{i=1}^m{\pi_i^*F^*_{1i}(\tilde w_it \mid \cD)}\geq\sum_{i=1}^m \pi_i^*F^*_{1i}(t \mid \cD).
\end{eqnarray*}
Hence, by the definition of $t_o^1$, it is easy to see that $Q^{\tilde w}(t_o^1) \leq Q^1(t_o^1) \leq \alpha$.
 It yields that $t_o^{\tilde w}\geq t_o^1$ and thus $\Psi^{\tilde w}(t_o^{\tilde w}) \geq \Psi^{\tilde w}(t_o^1) \geq \Psi^1(t_o^1)$.
\end{proof}

\section{Asymptotic theories under weak dependence} 
\label{app:dependent-theory}
In this section, we study the asymptotic control of FDP and FDR for dependent $p$-values. We collect some additional regularity conditions to develop the theories under weak dependence. We first introduce in Section \ref{oracle:sec} the benchmark oracle weight. Then the proofs are developed in two stages: Section \ref{weight_consistency} shows the consistency of the weight estimators; Section \ref{dependent_fdp} illustrates that the oracle-assisted LASLA controls FDP and FDR asymptotically. 

\subsection{Oracle weight} 
\label{oracle:sec}
With slight abuse of notation, we let $L_i^*=(1-\pi_i^{\tau})f_0(t_i)/f_i^*(t_i | \cD)$, where $f_i^*(\cdot | \cD)$ can be interpreted as the  density function of the primary statistic in light of the full network. Again we omit the conditioning on $\cD$ throughout for notation simplicity. 
Since $f_i(t)$ is calculated in light of the partial network $\pmb D_i$, it should become close to $f_i^*(t)$ as $\pmb D_i \rightarrow \cD_i$, which will be shown rigorously  later in Section \ref{weight_consistency}.

Similarly as the oracle-assisted weights defined in Section \ref{sec:data-driven-weight}, denote the sorted statistics by $L_{(1)}^*\leq \ldots \leq  L_{(m)}^*$. Let $L_{(k^*)}^*$ be the threshold, where $k^*=\max \{j: j^{-1}\sum^j_{i=1}{L_{(i)}^*\leq \alpha}\}$. 
Then for $T_i>0$, let $t^{*,+}_i=\infty$ if $(1-\pi^*_i)f_0(t)/f^*_i(t)\}\geq  \EE\left\{L_{(k^*)}^*\right\}$ for all $t \geq 0$, else:
\[
t^{*,+}_i=\inf\left[t\geq0: \{(1-\pi_i^{\tau})f_0(t)/f_i^*(t)\}\leq \EE\left\{L_{(k^*)}^*\right\}\right],
\] 
and define $w_i^*=1-F_{0}(t^{*,+}_i)$.
For $T_i < 0$, we let $t^{*,-}_i=-\infty$ if $(1-\pi^*_i)f_0(t)/f^*_i(t)\}\geq  \EE\left\{L_{(k^*)}^*\right\}$ for all $t \leq 0$, else:
\[
t^{*,-}_i=\sup\left[t\leq0: \{(1-\pi_i^{\tau})f_0(t)/f_i^*(t)\}\leq \EE\left\{L_{(k^*)}^*\right\}\right],
\] 
and the corresponding weight is given by $w_i^*=F_{0}(t^{*,-}_i)$.  Again, we let $w_i^* = \max\{w_i^*, \xi\}$ and $w_i^* = \min\{w_i^*, 1-\xi\}$ for any sufficiently small constant $0<\xi<1$. 
Then the oracle thresholding rule is provided by
\begin{align}\label{oracle_threshold}
    k^{*,w}=\max\left\{j:(1/j)\sum^m_{i=1}w_i^*(1-\pi_i^{\tau})P^{w^*}_{(j)} \leq \alpha \right\}.
\end{align}
We show next that the oracle-assisted weight $w_i$ in Algorithm \ref{alg:simple_weights} estimates $w_i^*$ consistently under some regularity conditions in the following section.

\subsection{Consistency of the weight estimator} 
\label{weight_consistency}

The weight consistency result  is built upon the consistency of sparsity estimator (\ref{eq:pi-est}) and density estimator (\ref{eq:density-est}). The theoretical properties of the former can be similarly proved as Proposition \ref{prop_pi} under conditions \ref{A1} and \ref{A2}, while letting $\cN_i = \{j \in [m], j\neq i\}$ and $h\gg m^{-1}$. We shall focus on the consistency of the density estimator below. Recall that 
\begin{align*}
    f_i(t)=\frac{\sum_{j \neq i}[{V_h(i,j)K_h(t_j-t)}]}{\sum_{j \neq i}{V_h(i,j)}}.
\end{align*}
We will focus on the cases when the support of the primary statistics $\pmb T=\{T_i:  i\in [m]\}$ is $\mathbb{R}$, e.g. $z$-statistics and $t$-statistics.  
\begin{enumerate}[label=(A\arabic*), series=A]
\setcounter{enumi}{7}
    \item \label{A7}Assume that for all $i,j$, {$f_j^*(t \mid D_{ij} = x)$ has bounded first and second partial derivatives at $t$ and $x$. }
    \item \label{A8}{Assume that, for all $i \in [m]$, 
    \[
    \Var\left\{\sum_{j=1}^m{K_h(D_{ij})K_h(t_j-t)} \mid  \pmb D\right\} \leq  C\sum_{j=1}^m\Var\left\{{K_h(D_{ij})K_h(t_j-t) \mid \pmb D_j}\right\}
    \] 
    for some constant $C>1$, for all $t$.}
\end{enumerate}

\begin{remark}
Assumption \ref{A7} is a mild regularity condition on the densities of the primary statistics. Condition \ref{A8} assumes that most of the primary statistics are weakly correlated. 
\end{remark}
\begin{lemma}\label{lem1}
Let $K(\cdot)$ be a kernel function that satisfies (\ref{kernel}) and let $T$ be a random variable with support $\mathbb{R}$. Assume that its conditional density $f(\cdot \mid \pmb D)$ has bounded first and second derivatives. Then for any fixed $t$, as the bandwidth $h \rightarrow 0$, we have
\begin{align*}
    \expect{K_h(T-t) \mid \pmb D} &=f(t \mid \pmb D)+ O(h^2)\sigma_K^2\\
    \expect{K^2_h(T-t) \mid \pmb D}&=f(t \mid \pmb D)\frac{R(K)}{h}+O(h)G(K),
\end{align*}
where
$
    R(K)=\int_{\mathbb{R}}K^2(x) \,dx$ and $G(K)=\int_{\mathbb{R}}x^2K^2(x) \,dx.
$
\end{lemma}
Once Lemma \ref{lem1} is developed, we can obtain the following proposition on density estimation consistency.
\begin{proposition}\label{prop_density}
    Under Assumptions \ref{A7} and \ref{A8}, if $h \gg m^{-1/2}$, we have for any $t$, uniformly for all $i \in [m]$,
    \begin{align*}
        \expect{\{f_i(t)-f_i^*(t)\} \mid  \pmb D}^2 \rightarrow 0, \hspace{3mm} \text{as } \pmb D_i \rightarrow \cD_i.
    \end{align*}
\end{proposition}

Next, we develop the consistency result of the oracle-assisted weight in Algorithm \ref{alg:simple_weights}. 
Without loss of generality, we assume that $-\infty < t_i^{*,-} \leq t_i^{*,+} < +\infty$ for all $i \in [m]$.
Let $g_i(t) = (1-\pi_i^{\tau})f_0(t)/f_i^*(t)$ and define functions $g_{i,+}^{-1}: x \rightarrow t$ and $g_{i,-}^{-1}: x \rightarrow t$ as
\[
g_{i,+}^{-1}(x) = \inf\{t\geq 0: g_i(t) \leq x\}, 
\]
and
\[
g_{i,-}^{-1}(x) = \sup\{t\leq 0: g_i(t) \leq x\}.
\]
We let $g_{i,+}^{-1}(x) = +\infty$ if $g_i(t) \geq x$ for all $t\geq 0$ and let $g_{i,-}^{-1}(x) = -\infty$ if $g_i(t) \geq x$ for all $t\leq 0$.
We also assume that $-\infty < t_i^{-} \leq t_i^{+} < +\infty$ for all $i \in [m]$ for simplicity. If not, the data-driven testing procedure will be more conservative than the oracle one and hence the asymptotic FDR control can again be guaranteed.
Then based on Proposition \ref{prop_density}, we obtain the following corollary.
\begin{corollary}\label{cor_weight}
Assume that $g_{i,+}^{-1}(x)$ and $g_{i,-}^{-1}(x)$ have bounded first derivative for all $0<x<1$ such that $-\infty < g_{i,-}^{-1}(x) \leq g_{i,+}^{-1}(x) < +\infty$ and there exists some constants $\alpha_1$ and $\alpha_2$ such that $\frac{1}{k^*}\sum_{i=1}^{k^*}L_{(i)}^*\leq \alpha_1 < \alpha <\alpha_2\leq \frac{1}{k^*+1}\sum_{i=1}^{k^*+1}L_{(i)}^*$ with probability tending to 1. 
Assume that $1/f_i^*(t)$ are bounded with probability tending to 1 uniformly for all $i \in [m]$. 
Further assume that {$\pi_i^{\tau} \leq 1-\xi$ for sufficiently small constant $\xi>0$ and 
$\Var \left(L_{(k^*)}^*\right) = o(1)$}. Then under the conditions of Propositions \ref{prop_pi} and \ref{prop_density}, we have, as $m\rightarrow \infty$,
$w_i = w_i^* +o_{\pr}(1)$, uniformly for all $i \in [m]$.
\end{corollary}

\begin{remark}\label{rem_weight_con}
The conditions on $g_{i,+}^{-1}$, $g_{i,-}^{-1}$ and $L_{(i)}^*$ are mild and can be easily satisfied by the commonly used distributions such as normal distribution, $t$-distribution, etc. The condition on $1/f_i^*(t)$ can be further relaxed by a more sophisticated calculation on the convergence rate of $f_i(t)$ in the proof of Proposition \ref{prop_density}.
The condition $\Var \left(L_{(k^*)}^*\right) = o(1)$ is mild and can be satisfied by most of the settings in the scope of this paper. For example, in Setting 1 of Section \ref{latent_simu}, $\Var \left(L_{(k^*)}^*\right)$ is of the order $10^{-2}$.
\end{remark}

\subsection{FDP and FDR control under weak dependence} 
\label{dependent_fdp}

Recall that we define the $z$-values by $Z_i=\Phi^{-1}(1-P_i/2)$, and let $\Z = (Z_1,\ldots,Z_m)^\T$.
We collect below one additional regularity condition for the asymptotic error rates control. We allow dependency to come from two sources: Dependence of the $\theta_i$'s and dependency of the $p$-values given $\theta_i$'s.  Our conditions on these two types of correlations are respectively specified in \ref{A3} and \ref{A9}.

\begin{enumerate}[resume*=A]

\item  \label{A9} Define $(r_{i,j})_{m\times m}= \R=\Corr(\Z)$. Assume $\max_{1\leq i<j\leq m}|r_{i,j}|\leq r< 1$ for some constant $r>0$. Moreover, there exists $\gamma>0$ such that $\max_{\{i: \theta_i = 0\}}|\Gamma_i(\gamma)|=o(m^\kappa)$ for some constant $0<\kappa<\frac{1-r}{1+r}$, where $\Gamma_i(\gamma)=\{j:1\leq j\leq m, |r_{i,j}|\geq (\log m)^{-2-\gamma}\}$. 
\end{enumerate}

We first consider the oracle case. 
Recall that
\begin{align*}
    k^{*,w}=\max\left\{j:\paren*{P^{w^*}_{(j)}/j}\sum^m_{i=1}w_i^*(1-\pi_i^{\tau}) \leq \alpha \right\}.
\end{align*}
{Denote the corresponding threshold for the weighted $p$-values as $t^{w^*}$ and the set of decision rules as $\pmb{\delta}^{w^*}$.}  The next theorem shows that both FDP and FDR are controlled at the nominal level asymptotically under dependency.

\begin{theorem}\label{FDR1}
Under \ref{A3}, \ref{A4} and \ref{A9}, 
we have for any $\varepsilon>0$,
\begin{align*}
    \limsup\limits_{\pmb D_i \rightarrow \cD_i, \forall i} {\text{FDR}(\pmb{\delta}^{w^*})} \leq \alpha, \hspace{2mm} and \hspace{2mm} \lim\limits_{\pmb D_i \rightarrow \cD_i, \forall i}\prob{\text{FDP}(\pmb{\delta}^{w^*}) \leq \alpha+\varepsilon} =1.
\end{align*}
\end{theorem}

The next theorem establishes the theoretical properties of the data-driven LASLA procedure. {Recall that $\pmb{\delta}^w \equiv \pmb{\delta}^w(t^w)=\{\delta_i^w(t^w): i \in [m]\}$ is the set of data-driven decision rules, where the LASLA weights are computed by Algorithm~\ref{alg:simple_weights} with $\cN_i = \{j \in [m], j\neq i\}$.} Based on the weight consistency result, the FDP and FDR of data-driven LASLA can be asymptotically controlled under dependency.
\begin{theorem}\label{FDR2}
Under the conditions in Corollary \ref{cor_weight} and Theorem \ref{FDR1}, we have for any $\varepsilon>0$,
\begin{align*}
   \limsup\limits_{\pmb D_i \rightarrow \cD_i, \forall i} {\text{FDR}(\pmb{\delta}^w)} \leq \alpha, \hspace{2mm} and \hspace{2mm} \lim\limits_{\pmb D_i \rightarrow \cD_i, \forall i}\prob{\text{FDP}(\pmb{\delta}^w) \leq \alpha+\varepsilon} =1.\\
\end{align*}
\end{theorem}

\section{Proof of the theoretical results under dependency}\label{sec:dep-proofs}
 
\subsection{Proof of Lemma \ref{lem1}}
\begin{proof}
By Taylor expansion of $f(y \mid \pmb D)$ at $y=t$, we have
\begin{align*}
    \expect{K_h(T-t) \mid \pmb D}&=\int K_h(y-t)f(y \mid \pmb D) \,dy\\
    &=\int K_h(y-t)\brac*{f(t \mid \pmb D)+f'(t \mid \pmb D)(y-t)+\frac{f''(t \mid \pmb D)}{2}(y-t)^2} \,dy + O(h^2) \\
    &=f(t \mid \pmb D)+O(h^2)\sigma_K^2.
\end{align*}
Similarly,
\begin{align*}
    \expect{K^2_h(T-t) \mid \pmb D}&=\int K^2_h(y-t)f(y \mid \pmb D) \,dy\\
    &=\int K^2_h(y-t)\brac*{f(t \mid \pmb D)+f'(t \mid \pmb D)(y-t)+\frac{f''(t \mid \pmb D)}{2}(y-t)^2} \,dy + O(h)\\
    &=\frac{f(t \mid \pmb D)R(K)}{h}+O(h)G(K).
\end{align*}
\end{proof}

\subsection{Proof of Proposition \ref{prop_density}}
\begin{proof}
By Lemma \ref{lem1}, we have
\begin{align*}
    \expect{f_i(t) \mid \pmb D} = \frac{\sum_{j \neq i}{K_h(D_{ij})\expect{K_h(t_j-t) \mid \pmb D_j}}}{\sum_{j \neq i}{K_h(D_{ij})}}=\frac{\sum_{j \neq i}{K_h(D_{ij}){ f_j^*(t \mid \pmb D_j)}}}{\sum_{j \neq i}{K_h(D_{ij})}} + \textit{O}(h^2).
\end{align*}
By $h\gg m^{-1/2}$, as $\pmb D_i \rightarrow \cD_i$, we have
\[
\frac{{\sum_{j \neq i}{K_h(D_{ij}){ f_j^*(t \mid \pmb D_j)}}}/{\sum_{j \neq i}{K_h(D_{ij})}} } {{\int_{\cD_i}{K_h(x)f_{j_x}^*(t \mid D_{ij_x}=x)}\,dx}/{\int_{\cD_i}{K_h(x)}\,dx}} \rightarrow 1,
\]
where $j_x$ represents the index such that $D_{ij_x} = x$.
 By Taylor expansion of $f_{j_x}^*(t \mid D_{ij_x}=x)$ at $x=0$, we have,
\begin{align*}
    \frac{\int_{\cD_i}{K_h(x)f_{j_x}^*(t \mid D_{ij_x}=x)}\,dx}{\int_{\cD_i}{K_h(x)}\,dx} &= \frac{\int_{\cD_i}{K_h(x)\big[f_i^*(t)+(f_i^*)'(t)x+\frac{(f_i^*)''(t)}{2}x^2\big]}\,dx}{\int_{\cD_i}{K_h(x)}\,dx} + \textit{O}(h^2)\\
    &=f_i^*(t)+\frac{\int_{\cD_i}{K_h(x)\big[(f_i^*)'(t)x+\frac{(f_i^*)''(t)}{2}x^2\big]}\,dx}{\int_{\cD_i}{K_h(x)}\,dx} + \textit{O}(h^2).
\end{align*}
Under assumption \ref{A7} and the condition that $\cD_i$ is finite, we have that for some constant $c>0$,
\begin{align*}
   &\brac*{\expect{f_i(t) \mid \pmb D}-f_i^*(t)}^2 \\
    &\leq \bigg(c\int_{\cD_i} |x|K_h(x) \,dx \bigg/ \int_{\cD_i} K_h(x) \,dx + c\int_{\cD_i} x^2K_h(x) \bigg/ \int_{\cD_i} K_h(x) \,dx\bigg)^2+o(1)\\
    &\rightarrow 0, \hspace{1cm}\text{as h} \rightarrow 0.
\end{align*}
\vspace{2mm}
Now for the variance term, by Assumption \ref{A8}, we have
\[
  \var[\bigg]{\sum_{j \neq i}[{K_h(D_{ij}){K_h(t_j-t)}}] \mid  \pmb D} \leq c'\sum_{j \neq i}\big[{K^2_h(D_{ij})\var{K_h(t_j-t) \mid \pmb D_j}}\big]
\]
Hence, as $h \gg m^{-1/2}$, by Lemma \ref{lem1}, Assumption \eqref{kernel} and the fact that $K(\cdot)$ is positive and bounded, we take Taylor expansion again and obtain that
\begin{align*}
    \var{f_i(t) \mid  \pmb D} &\leq c'm^{-1}\frac{\int_{\cD_i}K^2_h(x)\big[f_{j_x}^*(t \mid D_{ij_x}=x)\big(R(K)/h-f_{j_x}^*(t \mid D_{ij_x}=x)\big)\big] \,dx +O(1)}{(\int_{\cD_i}K_h(x) \,dx)^2}\\
    &\leq c'm^{-1}\frac{\int_{\cD_i}\frac{R(K)}{h}K^2_h(x)\big[f_i^*(t)+(f_i^*)'(t)x+\frac{(f_i^*)''(t)}{2}x^2\big] \,dx +O(1)}{(\int_{\cD_i}K_h(x) \,dx)^2}\\
    &\leq c'm^{-1}\frac{\int_{\cD_i}\frac{R(K)}{h}K^2_h(x)f_i^*(t)\,dx +O(h^{-1})}{(\int_{\cD_i}K_h(x) \,dx)^2}\\
    & \leq c''m^{-1}\frac{\int_{\cD_i}\frac{R(K)}{h}K^2_h(x)\,dx+O(h^{-1})}{(\int_{\cD_i}K_h(x) \,dx)^2}\\
    & \leq c''m^{-1}h^{-2}R(K)\frac{\int_{\mathbb{R}} K^2(y) \,dy}{(\int_{\cD_i}K_h(x) \,dx)^2}\\
    & \leq c'''m^{-1}h^{-2} = o(1),
\end{align*}
for some constant $c'',c'''>0$. Hence, as $\pmb D_i \rightarrow \cD_i$, combining the bias term and variance term, the consistency result is proved.
\end{proof}

\subsection{Proof of Corollary 1}
\begin{proof}
Recall that 
\begin{equation*}
L_i = \frac{(1-\pi_i)f_0(T_i)}{f_i(T_i)}.
\end{equation*}
Then based on the consistency results on $\pi_i$ and $f_i(t)$ in Propositions \ref{prop_pi} (with $\cN_i = \{j \in [m], j\neq i\}$ and $h\gg m^{-1}$) and \ref{prop_density}, together with the condition that 
$1/f_i^*(t)$ is bounded and $\pi_i^{\tau}\leq 1-\xi$, we have, uniformly for all $i \in [m]$,
\[
L_i = (1+o_{\pr}(1))L_i^*.
\]
Then 
by the condition that $\frac{1}{k^*}\sum_{i=1}^{k^*}L_{(i)}^*\leq \alpha_1 < \alpha <\alpha_2\leq \frac{1}{k^*+1}\sum_{i=1}^{k^*+1}L_{(i)}^*$ with probability tending to 1 and $\Var \left(L_{(k^*)}^*\right) = o(1)$, it yields that
\[
L_{(k)} = L_{(k^*)}^* + o_{\pr}(1) = \EE \cbrac*{L_{(k^*)}^*} + o_{\pr}(1).
\]
Then based on the definitions of $g_{i,+}^{-1}$ and $g_{i,-}^{-1}$, we have that 
\[
t^{*,+}_i = g_{i,+}^{-1}\left[\EE\left\{L_{(k^*)}^*\right\}\right] \text{ and } t^{*,-}_i = g_{i,-}^{-1}\left[\EE\left\{L_{(k^*)}^*\right\}\right],
\]
and that
\[
t^+_i = g_{i,+}^{-1}\left[{(1+o_{\pr}(1))L_{(k)}}\right] \text{ and } t^-_i = g_{i,-}^{-1}\left[{(1+o_{\pr}(1))L_{(k)}}\right],
\]
based on the condition that $\pi_i^{\tau}\leq 1-\xi$. Then because $g_{i,+}^{-1}$ and $g_{i,-}^{-1}$ have bounded first derivative, we have
\[
    t^+_i = t^{*,+}_i + o_{\pr}(1) \text{ and } t^-_i = t^{*,-}_i + o_{\pr}(1).
\]
By Assumption \ref{A7}, $f_0$ is bounded, then we obtain
\[
w_i = w_i^* + o_{\pr}(1),
\]
uniformly for all $i \in [m]$.
\end{proof}

\subsection{Proof of Theorem \ref{FDR1}}
\begin{proof}
The FDP of the oracle procedure at the thresholding level $t$ can be calculated by
\begin{align*}
    \text{FDP}(t)&=\frac{\sum_{i=1}^{m} \II\{P_i^{w^*} \leq t, \theta_i=0\}}{\max\{\sum_{i=1}^{m} \II\{P_i^{w^*} \leq t\},1\}}\\
    &=\frac{\sum_{i=1}^{m} \pr(P_i^{w^*} \leq t, \theta_i=0)}{\max\{\sum_{i=1}^{m} \II\{P_i^{w^*} \leq t\},1\}} \cdot \frac{\sum_{i=1}^{m} \II\{P_i^{w^*} \leq t, \theta_i=0\}}{\sum_{i=1}^{m} \pr(P_i^{w^*} \leq t, \theta_i=0)}\\
    &=\frac{\sum_{i=1}^{m}w_i^*(1-\pi_i^*)t}{\max\{\sum_{i=1}^{m} \II\{P_i^{w^*} \leq t\},1\}} \cdot \frac{\sum_{i=1}^{m} \II\{P_i^{w^*} \leq t, \theta_i=0\}}{\sum_{i=1}^{m} \pr(P_i^{w^*} \leq t, \theta_i=0)}\\
    &\leq \frac{\sum_{i=1}^{m}w_i^*(1-\pi_i^{\tau})t}{\max\{\sum_{i=1}^{m} \II\{P_i^{w^*} \leq t\},1\}} \cdot \frac{\sum_{i=1}^{m} \II\{P_i^{w^*} \leq t, \theta_i=0\}}{\sum_{i=1}^{m} \pr(P_i^{w^*} \leq t, \theta_i=0)}.
\end{align*}

Then by Steps 2 and 3 in the proofs of Theorem \ref{thm:FDR_idp} by replacing $w_i$'s with the true $w_i^*$'s, and together with the proofs of Theorem 2 in \cite{Caietal20}, by Assumption \ref{A9}, we have
\beas
\sup_{0\leq t\leq t_m}\Bigg{|} \frac{\sum_{i=1}^{m} \II\{Z_i^{w^*} \geq t, \theta_i=0\} - \sum_{i=1}^{m} \pr(Z_i^{w^*} \geq t, \theta_i=0)}{\sum_{i=1}^{m} \pr(Z_i^{w^*} \geq t, \theta_i=0)}\Bigg{|}\rightarrow 0,
\eeas
in probability.
Then the FDP and FDR are controlled and Theorem \ref{FDR1} is proved.
\end{proof}

\subsection{Proof of Theorem \ref{FDR2}}
\begin{proof}
Note that, the FDP of the data-driven procedure at the thresholding level $t$ can be calculated by
{
\begin{align*}
    FDP(t)&=\frac{\sum_{i=1}^{m} \II\{P_i^{w} \leq t, \theta_i=0\}}{\max[\sum_{i=1}^{m} \II\{P_i^{w} \leq t\},1]}\\
    &=\frac{\sum_{i=1}^{m} \pr(P_i^{w} \leq t \mid \theta_i=0 ,w_i)\pr(\theta_i=0)}{\max[\sum_{i=1}^{m} \II\{P_i^{w} \leq t\},1]} \cdot \frac{\sum_{i=1}^{m} \II\{P_i^{w} \leq t, \theta_i=0\}}{\sum_{i=1}^{m} \pr(P_i^{w} \leq t \mid \theta_i=0 ,w_i)\pr(\theta_i=0)},
\end{align*}
Define the event $A=[\{w_i\}_{i=1}^m: w_i = w_i^*+o(1)]$, then based on the result of Corollary \ref{cor_weight}, we have that $\pr(A)\rightarrow 1$. Next, we shall focus on the event $A$. For $\{w_i\}_{i=1}^m\in A$, uniformly for all $i \in [m]$,
\[
\pr(P_i^{w} \leq t| \theta_i=0, w_i, T_i>0) = w_i t =[1+o(1)]w_i^* t,
\]
uniformly for all $t$ defined in the range defined in Step 3 of Theorem \ref{thm:FDR_idp}. The same equality holds if we replace the condition $T_i>0$ by $T_i<0$ because the oracle quantity $w_i^*$ is fixed given the sign of $T_i$.
}
Then we have, uniformly for all $i \in [m]$,
\begin{align*}
\pr(P_i^{w} \leq t \mid \theta_i=0 ,w_i) 
&= \pr(P_i^{w} \leq t \mid \theta_i=0,w_i, T_i >0)\pr(T_i >0 \mid \theta_i=0,w_i) \\
&\textcolor{white}{=}+\pr(P_i^{w} \leq t \mid \theta_i=0,w_i, T_i <0)\pr(T_i <0 \mid \theta_i=0,w_i) \\
&= [1+o(1)]w_i^* t \brac*{\pr(T_i >0 \mid \theta_i=0,w_i) +\pr(T_i <0 \mid \theta_i=0,w_i)} \\
&= [1+o(1)]w_i^*t,
\end{align*}
which implies that 
\begin{align*}
    \pr(P_i^{w} \leq t \mid \theta_i=0 ,w_i)\pr(\theta_i=0) = [1+o(1)]w_i^*(1-\pi_i^*)t \leq [1+o(1)]w_i^*(1-\pi_i^{\tau})t.
\end{align*}
Thus, based on the results of Proposition \ref{prop_pi} and Corollary \ref{cor_weight} and proofs of Theorems \ref{thm:FDR_idp} and \ref{FDR1}, we obtain that the oracle-assisted weight produces a more conservative procedure asymptotically. This concludes the proof of Theorem \ref{FDR2}.
\end{proof}

\end{document}